\documentclass[12pt,preprint]{aastex}

\newcommand{\lxlbol}{$L_{\rm x}/L_{\rm bol}$}
\newcommand{\lglxlbol}{$\log{(L_{\rm x}/L_{\rm bol})}$}
\newcommand{\lx}{$L_{\rm x}$}
\newcommand{\lbol}{$L_{\rm bol}$}

\slugcomment{Accepted to ApJSuppl, COUP Special Issue}

\shorttitle{X-ray emission from early-type stars in the ONC}
\shortauthors{Stelzer et al.}

\begin{document}

\title{X-ray emission from early-type stars \\ in the Orion Nebula Cluster}

\author{B. Stelzer\altaffilmark{1,2}, E. Flaccomio\altaffilmark{2}, T. Montmerle\altaffilmark{3},
G. Micela\altaffilmark{2}, S.
Sciortino\altaffilmark{2}, F. Favata\altaffilmark{4}, T.
Preibisch\altaffilmark{5} \& E. D. Feigelson\altaffilmark{6}}

\email{stelzer@astropa.unipa.it}

\altaffiltext{1}{Dipartimento di Scienze Fisiche ed Astronomiche,
Universit\`a di Palermo, Piazza del Parlamento 1, I-90134 Palermo,
Italy}

\altaffiltext{2}{INAF - Osservatorio Astronomico di Palermo,
Piazza del Parlamento 1, I-90134 Palermo, Italy}

\altaffiltext{3}{Laboratoire d'Astrophysique de Grenoble,
Universit\`e Joseph-Fourier, F-38041 Grenoble, France}

\altaffiltext{4}{Astrophysics Division - Research and Science
Support Department of ESA, ESTEC, Postbus 299, 2200 AG Noordwijk,
The Netherlands}

\altaffiltext{5}{Max-Planck Institut f\"ur Radioastronomie, Auf
dem H\"ugel 69, D-53121 Bonn, Germany}

\altaffiltext{6}{Department of Astronomy \& Astrophysics,
Pennsylvania State University, University Park PA 16802, USA}

\begin{abstract}
The X-ray properties of twenty $\sim$1\,Myr old O, B, and A stars
of the Orion Trapezium are examined with data from the {\em
Chandra} Orion Ultradeep Project (COUP). On the basis of simple
theories for X-ray emission, we define two classes separated at
spectral type B4: hotter stars have strong winds that may give
rise to X-ray emission in small- or large-scale wind shocks, and
cooler stars that should be X-ray dark due to their weaker winds
and absence of outer convection zones where dynamos can generate
magnetic fields.  Emission by late-type magnetically active
companions may be present in either class.

Sixteen of the $20$ stars are detected with a wide range of X-ray luminosities, 
$\log{L_{\rm x}}\,{\rm [erg\,s^{-1}]} \sim 29 - 33$ and
X-ray efficiencies \lglxlbol$\, \sim -4$ to $-8$. 
Only two stars, $\theta^1$\,Ori\,D (B0.5) and NU\,Ori (B1), 
show exclusively the constant  
soft-spectrum emission at \lglxlbol$\, \sim -7$ expected from 
the standard model involving many small shocks in an unmagnetized
radiatively accelerated wind. 
Most of the other massive O7-B3 stars exhibit some combination 
of soft-spectrum wind emission, hard-spectrum flaring, and/or rotational
modulation indicating large-scale inhomogeneity.   
Magnetic confinement of winds with large-scale shocks 
can be invoked to explain these phenomena. This is supported in some cases
by non-thermal radio emission and/or chemical peculiarities, 
or direct detection of the magnetic field ($\theta^1$\,Ori\,C). 

Most of the stars in the weak-wind class exhibit
X-ray flares and $\log{L_{\rm x}} < 31$\,erg s$^{-1}$, consistent
with magnetic activity from known or unseen low-mass companions.
In most cases, the X-ray spectra can be interpreted in terms of a
two-temperature plasma model with a soft component of $3-10$\,MK
and a hard component up to $40$\,MK.  All non-detections belong to
the weak-wind class.  A group of stars exhibit hybrid properties
-- flare-like behavior superimposed on a constant component with
$\log{L_{\rm x}} \sim 32$\,erg s$^{-1}$ --  which suggest both
magnetic activity and wind emission.
\end{abstract}


\section{Introduction}

Among normal stars, X-ray emission is often strongest in the
youngest stellar populations.  X-ray emission from OB stars was first
reported from observations with the $Einstein$ observatory
\citep{Harnden79.1, Pallavicini81.1},
and dozens were detected with the {\em ROSAT} All-Sky Survey with 
characteristic efficiency \lglxlbol$\, \sim -7$ \citep{Berghoefer97.1}. 
Their soft and nearly constant emission was attributed to the integrated
emission from a myriad of small shocks due to instabilities in the 
radiatively-driven outflow 
\citep[see][for a review of the wind properties of massive
stars]{Lucy80.1, Owocki99.1, Kudritzki00.1}. 
Their soft spectrum and mildly variable X-ray emission 
stood in contrast to the hard and often wildly variable X-ray emission of
lower mass T Tauri stars. The variability of these latter stars is 
attributed to flares resulting 
from reconnection events in large magnetic loops as in the Sun and
other magnetically active stars \citep{Feigelson99.1, Favata03.1}.
Thus, early-type stars were viewed to radiate X-rays via a thermal plasma
produced by a {\it thermal} heating mechanism (wind shocks),
while late-type stars radiate via a thermal plasma produced by a {\it
non-thermal} heating mechanism (magnetic reconnection flares).

Recent studies by telescopes with sensitivity to hard X-rays and
high-resolution spectral capabilities have thrown this simple view 
into confusion.  The brightest star in the Orion Trapezium cluster,
the O7 star $\theta^1$\,Ori\,C, exhibits a hard spectral component and a
strong rotational modulation inconsistent with spherically symmetric wind
emission. 
The Trapezium O9.5 star $\theta^2$\,Ori\,A 
showed a powerful flare superposed on wind-like emission \citep{Feigelson02.1}. 
Spectral lines in these and other O stars are broadened
indicating production in the wind acceleration zone, but the line
strengths, shapes and absorption often differ from those predicted
by the standard wind-shock model \citep[e.g.][]{Miller02.1, Schulz03.1, Cohen03.1}. 
Some of these effects can be attributed to the confinement, and resulting large-scale
shocks, of the wind by a strong stellar magnetic field \citep{Babel97.1, UdDoula02.1, Schulz03.1}. 
This is supported by the detections of nonthermal radio
emission and atmospheric chemical peculiarities in some of these stars. 

The X-ray properties of intermediate-mass late B and A stars have 
also been confusing, but for different reasons. 
With neither powerful fast winds, nor outer convection zones, no X-rays were
expected. Nonetheless, a substantial number are detected with a
wide range of X-ray luminosities \citep[e.g.][]{Schmitt85.1,
Caillault89.1, Simon95.1, Cohen97.1}. In some cases, this emission
can be attributed to the presence of low-mass, late-type star
companions to the intermediate-mass stars.  This was directly
demonstrated by resolving several visual binaries with the
$Chandra$ $X-ray$ $Observatory$ \citep{Stelzer03.1, Behar04}.  But
other cases are less clear: some X-ray sources appear to coincide
precisely with the B star \citep{Stelzer04.1}, and the youngest A
and B stars (Herbig Ae/Be stars) sometimes outshine their lower
mass companions \citep{Feigelson03.1}, perhaps due to star-disk
magnetic interactions \citep{Hamaguchi05.1}.

These issues can be fruitfully addressed in the Orion Nebula
Cluster (ONC) where a roughly coeval population of young stars,
covering a variety of early spectral types, can be observed in a
single X-ray exposure.  With its high spatial resolution and
excellent astrometry, {\em Chandra} is the best instrument for a
study of this rich and crowded field.  Results from $Chandra$
observations carried out in 1999 and 2000 already showed that most
OBA stars in the ONC are X-ray emitters\footnote{Following
\citet{Getman05.1} and other COUP studies, $L_t$ is the X-ray
luminosity in the `total' $0.5-8$ keV band assuming a distance of
450 pc to the ONC. We later refer to $L_s$ in the `soft' $0.5-2$
keV band, and $L_{t,c}$ where the total band luminosity is
corrected for absorption.} above $\log L_t \simeq 28$
erg\,s$^{-1}$ \citep{Feigelson02.1, Flaccomio03.1}. The
luminosities and variability of some OB stars are 
consistent with wind emission. Similar to findings from earlier
X-ray observations, the emission levels are seen to scatter around the canonical 
$L_{\rm x}/L_{\rm bol} \simeq 10^{-7}$ relation. This relation has no immediate 
physical interpretation but could be recovered in a scaling analysis taking account of 
the detailed wind-shock structure and wind opacity \citep{Owocki99.1}. 
The large scatter of the data around the canonical value has remained unexplained,
but may be related to the fact that $L_{\rm x}$ and $L_{\rm bol}$ are coupled to each
other only indirectly via the wind density parameter $M_\odot/v_\infty$. 
Contrary to expectation, the $\simeq 30$ M$_\odot$  massive
Trapezium star $\theta^2$\,Ori\,A exhibited rapid high-amplitude
variability characteristic of magnetic flares \citep{Feigelson02.1}.
Together with other works on early-type stars, these results hint
that the X-ray emission mechanism in hot stars is more complex than the
simple wind-shock picture outlined above, suggesting in particular
that transient magnetic phenomena may be important even in massive
stars. In this paper, we seek to clarify these issues and
elucidate the respective roles of wind shocks and magnetic
activity in the X-ray emission of early-type stars.

Our study of early-type stars in the ONC is based on the
nearly-continuous 9.7\,day observation of the Orion Nebula made
with $Chandra$ in 2003. Known as the $Chandra$ Orion Ultradeep
Project (COUP), this is the most comprehensive X-ray study of a young
stellar cluster performed to date. The observation, data analysis,
source lists and characteristics are described in detail by
\citet{Getman05.1}. After describing the sample and presenting its
basic X-ray properties (\S 2), we examine their X-ray spectra (\S
3) and various types of variability (\S 4). Section 5 with
Appendix\,A is dedicated to multiple systems, \S 6 to systems
with disks, and Appendix\,B to a new X-ray detection of an
early-type Trapezium star in a crowded region of the field. In the
discussion (\S 7), we confront different possible emission
mechanisms for our targets.

\section{Early-type stars in the COUP}\label{sect:sample}

\subsection{Classification \label{sect:def_class}}

In order to investigate whether or not magnetic phenomena are
involved in the X-ray emission mechanisms for the early-type stars
of the ONC, we first divide the whole COUP sample of ONC members
confirmed by optical spectroscopy (`optical' COUP sample) into
three groups distinguished by the simplest emission mechanism
associated as a class with their spectral types (\S 1): the
hottest stars dominated by wind shocks, the cooler stars dominated
by magnetic activity, and an intermediate `blank zone'. In this last
class no X-rays are expected, 
because the stars have no, or only a very shallow
(outer) convective zone \citep[at most a few percent of the
stellar radius;][]{Siess00.1}, hence no magnetic activity, and
have weak, slow radiative winds or no wind at all. 
We seek a quantitative basis for dividing these classes here.

In hot massive stars, a radiative force is exerted on heavy atoms
in their atmospheres by photospheric ultraviolet photons,
transmitting outward momentum to the entire atmosphere via Coulomb
collisions \citep[e.g.,][]{Kudritzki00.1}. When the radiative
acceleration $g_{\rm rad}$ of the gas exceeds the local gravity
$g_\star$, a wind is triggered with net acceleration $g_{\rm rad}
- g_\star$. The wind velocity increases up to the sonic point,
reaching a ballistic regime with ``terminal velocity'' $v_\infty >
v_{\rm esc}$, where $v_{\rm esc}$ is the escape velocity of the
star. The development of a radiatively accelerated wind thus
depends both on a star's radiation field indicated by its spectral
type (or $T_{\rm eff}$) and its gravity $g_\star$. Metallicity
should also play a role but is not considered here, since all
stars in our sample are assumed to have the same metallicity.

Early models for radiatively driven winds 
\citep[e.g.,][]{Castor75.1, Abbott79.1} gave a
simple, continuous relationship between $v_\infty$, the mass-loss
rate $\dot{M}$, and $T_{\rm eff}$, so that the transition between
slow winds and fast winds was gradual. However, observations
(particularly in the X-ray band) suggest a sharper transition both
in $v_\infty$ and $\dot{M}$ around mid-B spectral types. Taking
into account the Doppler shifts and shadowing of the photospheric
lines, \citet{Babel96.1} calculated the locus of the stationary
wind regime in the two-dimensional $g-T_{\rm eff}$ plane. He found
that for low temperatures (e.g., $T_{\rm eff} < 10,000$\,K), only
weak-gravity stars can have winds, whereas on the main-sequence,
for high temperatures ($T_{\rm eff} > 20,000$\,K), winds can exist
even for high-gravity stars. In practice, the transition between
the weak-wind and fast-wind regimes near the main-sequence takes
place for $T_{\rm eff} \simeq 14,000 - 18,000$\,K; i.e., between
spectral types B3 and B5 on the temperature scale of
\citet{Hillenbrand04.1}.

This choice of boundary between strong and weak winds is supported
by two lines of evidence. First, the dependence of
$v_\infty/v_{\rm esc}$ on $T_{\rm eff}$ saturates at
$v_\infty/v_{\rm esc} \sim 3$ for $T_{\rm eff} \geq 20,000$\,K
\citep{Kudritzki00.1}.  Second, a simple model of X-ray emission
from wind shocks shows a break from $L_x \propto L_{bol}^1$ to $L_x
\propto L_{bol}^3$ around mid-B stars \citep{Owocki99.1},
basically due to a rapid change in the opacity of the winds to
 UV and X-ray photons. 

We therefore classify our sample in two groups based on their
spectral type range and expected source of X-ray emission:
\begin{enumerate}

\item stars with spectral type O through B3 that have strong,
optically thick winds (SW),

\item stars with spectral type B5 through A9 that have weak,
optically thin winds (WW),

\end{enumerate}

\noindent For comparison, we will also make use of the properties of cool, 
convective stars with spectral type F0 and later that are magnetically 
active (MA).
This physical classification will help us identify and interpret
deviations from the simple model predictions outlined in \S 1.

\begin{figure}
\begin{center}
\resizebox{16cm}{!}{\includegraphics{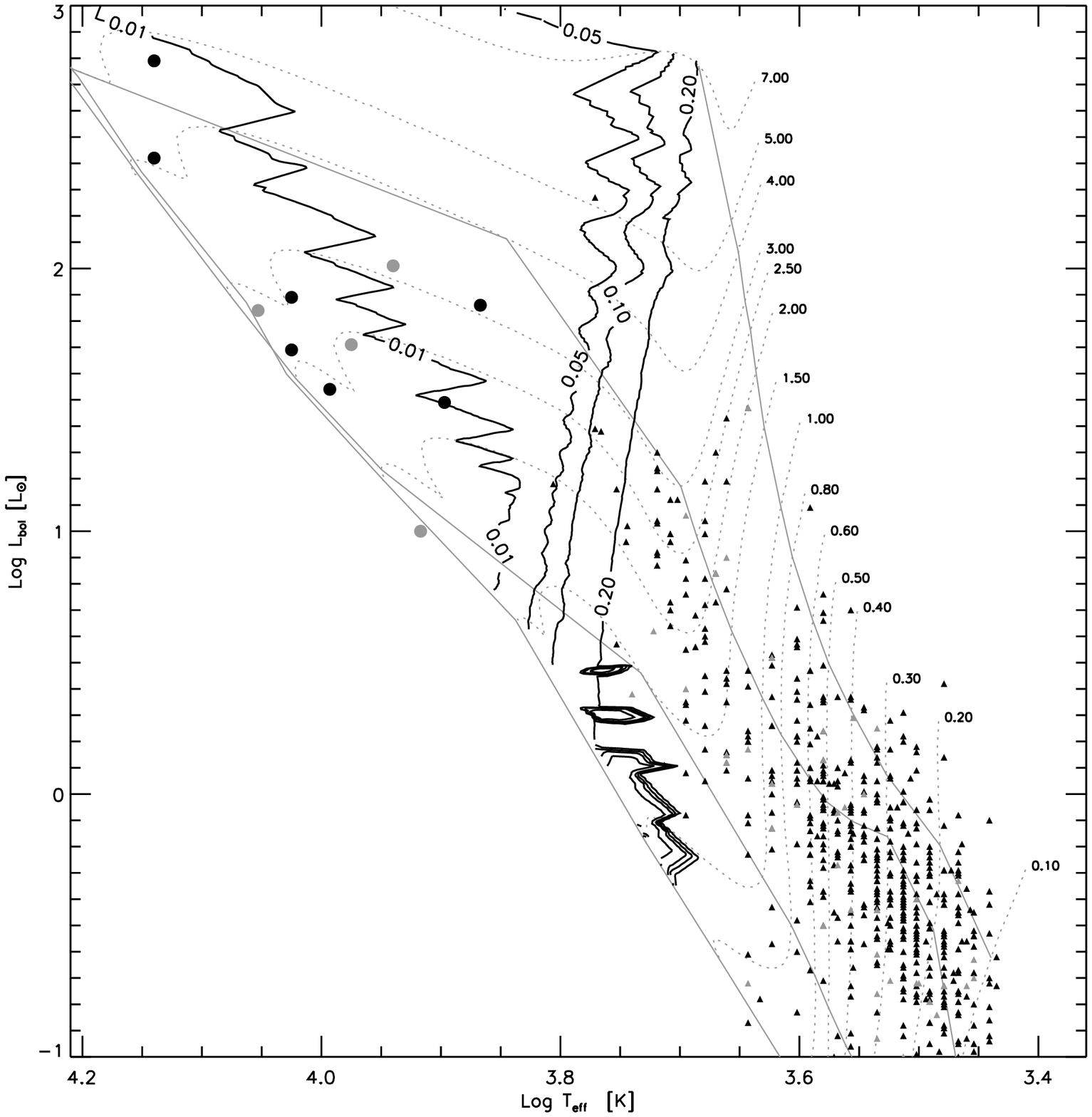}}
\caption{Hertzsprung-Russell diagram with pre-main sequence models
from \protect\citet{Siess00.1}.  Dotted grey curves show isomass
evolutionary tracks, solid grey curves show isochrones, and solid
black jagged contours indicate the size of the convective envelopes 
labeled in units of the stellar radius. Symbols show 
spectroscopically confirmed ONC stars observed in the COUP: 
{\em triangles} are magnetically active (MA) stars, and {\em circles} are 
weak wind (WW) stars. The strong wind (SW) stars are located outside 
the plotted range to the left. Grey symbols indicate stars not
detected in the COUP.} 
\label{fig:hrd_env}
\end{center}
\end{figure}

The coincidence of the lower boundary of the WW stars with the 
appearance of outer convection zones in cool stars is demonstrated in 
Fig.~\ref{fig:hrd_env}. It shows the Hertzsprung-Russell diagram 
with evolutionary interior models by \citet{Siess00.1}. The
jagged curves denote the extent of the convective envelope as a
fraction of the stellar radius.  
The data points show the `optical COUP sample' of  
\citet{Hillenbrand97.1} (henceforth H97). In all WW stars (shown as large circles),  
the convective envelopes comprises at most $5$\,\% of the stellar radius. 

Roughly two thirds ($21/32$) of the OBA stars in the ONC are in the field-of-view
of the COUP.
One star, JW\,602, has poorly
characterized optical properties (the spectral type is classified
as `B:') and is discarded. For the remaining 20 stars, 9 fall in
the SW sample and $11$ in the WW sample\footnote{Two of the WW stars with late-B
spectral type have unknown membership probabilities:
$\theta^1$\,Ori\,E and $\theta^1$\,Ori\,A. Since B stars are
intrinsically rare, we consider it unlikely that these stars are
not members of the ONC.}. 

Table~\ref{tab:sample} summarizes stellar parameters for the 20
COUP OBA stars, in order of decreasing effective temperature. We
adopt the $V$ magnitude, $A_{\rm V}$, and $L_{\rm bol}$ from
\citet{Getman05.1}, who had extracted the parameters from the
original data by H97. When a range of spectral types are present
in the literature, a spectral type was assigned corresponding to
an estimated average of the group (Hillenbrand, unpublished) and
converted to an effective temperature using the scale of
\citet{Hillenbrand04.1}. Given the possibility of yet unresolved
MA late-type companions, we searched the literature for
multiplicity of our targets which is summarized in columns $7-9$.
Finally, in order to investigate the possibility of magnetic
star-disk interactions, as in young protostars and T Tauri stars
\citep[e.g.,][]{Montmerle00.1, Bouvier99.1}, we provide in column
10 flags for stars with infrared (IR) excess indicating the
presence of circumstellar material.

\subsection{X-ray properties}

The time-averaged X-ray properties for the $20$ stars integrated
over the 9.7-day COUP exposure are presented in
Table~\ref{tab:xparams}. Except for the hardness ratio (column 9), 
the values are reproduced from the tables of \citet{Getman05.1}
who give details of their derivation.  The first three columns
give the COUP sequence number, the optical counterpart from
Table~\ref{tab:sample}, and the spectral type. Following columns
provide the off-axis angle and the offset between X-ray source and
the near-IR stellar positions. The net source counts in the broad
energy band ($0.5-8.0$\,keV) after background subtraction and
upper limits for undetected sources are given in column 6.

Four of the twenty targets are heavily piled-up in the ACIS CCD
detector: COUP\,809 ($\theta^1$\,Ori\,C), 1232
($\theta^2$\,Ori\,A), 745 ($\theta^1$\,Ori\,A), and 732
($\theta^1$\,Ori\,E). In these cases, only $1-4$\,\% of the
photons incident on the detector are extracted from an annular
region around the piled-up core of the point spread function (PSF). 
The spectral and luminosity analysis takes this into account in a
self-consistent manner. 
The luminosity values obtained for heavily piled-up sources are
subject to systematic uncertainties of order $\pm 20$\,\%. 
The treatment of piled-up COUP sources is described in \S\,6 of \citet{Getman05.1}.

All O to mid-B (SW) stars are detected, including the B2 star
Par\,1772, a faint X-ray source in a crowded region which had not
been detected in previous Orion Nebula X-ray studies (see
Appendix\,A and Fig.~\ref{fig:crowded_images} for details). Four
mid-B to A (WW) stars are undetected: $\theta^1$\,Ori\,F (B8),
JW\,531 (A1), JW\,108 (A3), and JW\,608 (A5).  
None of these stars has exhibited X-ray emission in previous observations.
Upper limits to their X-ray luminosities were computed with
XSPEC from the counts given in Table~\ref{tab:xparams}, 
assuming an isothermal plasma with energy $kT = 1$ keV,
abundance of $0.3$\,solar, and absorbing column derived from the visual absorptions given 
in Table~\ref{tab:sample} according to $N_H$[cm$^{-2}] =
1.6~10^{21}$ A$_V$[mag] \citep{Vuong03.1}. 

We also list the PSF fraction included in
the photon extraction area (column 7), and the effective exposure
time (column 8) at that location in the ACIS detector. The PSF
fraction is $\sim 90$\,\% for most COUP sources, but is lower when
the source suffers from pile-up or has a very close neighbor. In
such cases, special photon extraction regions have been defined.

The hardness ratio in column 10 is defined as $H/S$, where $H$ is
the number of counts in the hard band ($E>1.8$\,keV) and $S$ is
the number of counts in the soft band ($E<1.8$\,keV). 
This differs from the hardness ratios defined by \citet{Getman05.1}. 
Our choice of the boundary between the hard and soft band is near to the peak
of the observed spectrum of our targets, and therefore guarantees
a significant fraction of the total collected source photons in
each of the bands. 
In the last column, we provide the number of the respective X-ray
source in the previous {\em Chandra} ACIS observation 
\citep[][ hereafter FBG02]{Feigelson02.1}.

\section{Spectral Analysis}\label{sect:spec}

We adopt the procedure to extract the X-ray photons 
described by \citet{Getman05.1}, but we conduct a more
detailed spectral analysis than their automated fitting procedure,
which did not take in account visual absorptions, unusual abundance
or temperature distributions, or variability.  In addition, as
illustrated in their Figure~6, their fits may give incorrect
absorption column densities. Our resulting spectral fits to
thermal plasma models are summarized in Table~\ref{tab:spec}.
X-ray fluxes and luminosities were computed from the spectral fits
using the {\it XSPEC flux} command assuming a distance of
$450$\,pc. We list both the absorbed ($L_{\rm t}$) and the
absorption-corrected ($L_{\rm t,c}$) luminosities in the
$0.5-8$\,keV broad band.

Spectral modeling was first performed based on a one- or
two-temperature plasma in collisional ionization equilibrium using
the MEKAL code \citep{Mewe85.1, Mewe95.1}.  Elemental abundances
were fixed at $Z=0.3$~Z$_\odot$  with abundance ratios matching
those of \citet{Anders89.1}. This choice of subsolar abundances
follows \citet{Getman05.1}, and has been proven appropriate in
X-ray studies of various star forming regions
\citep[e.g.][]{Nakajima03.1, Feigelson02.1, Preibisch03.1}. 
The spectra of over half of the ONC early-type stars 
are well-fit by an absorbed, two-temperature
(2-T) plasma subject to soft X-ray absorption by a cold
intervening medium with column density $N_{\rm H}$. 
As noted by Getman et al., in some sources the spectral models 
are not unambiguous: a large uncertainty is introduced by the choice of $N_{\rm H}$,
resulting in either very cool and heavily absorbed plasma or
somewhat hotter plasma subject to light absorption. We have also
tested  models with $N_{\rm H}$ fixed on the value expected from
the \citet{Vuong03.1} extinction law, but found after visual
inspection of the best fits that the models with free column
density are more adequate. 

In about half of the sample, the spectral parameters resulting
from our analysis are almost identical to those presented in
\citet{Getman05.1}. Discrepancies for the remaining cases can be
attributed to two causes.  First, Getman et al.\ sometimes give
parameters of a 1-T model while, to achieve a uniform description
of our sample, we prefer the 2-T models.  The exception is
COUP\,995 where the 2-T fits result in lower temperatures and
unreasonably high luminosity. Second, the fitting procedure
encounters a local rather than global minimum in one of the
studies.  In these cases, we retain the model which both gives a
satisfactory fit ($\chi^2_{\rm red} \sim 1$) and has absorption
consistent to expected ONC values.  That is, the derived column
density is neither extremely low ($\log{N_{\rm H}}\,{\rm
[cm^{-2}]} \leq 20$) nor extremely high ($\log{N_{\rm H}}\,{\rm
[cm^{-2}]} \geq 22$).

In 4 of the brightest COUP hot stars, no 2-T model provides an
adequate description of the observed spectrum.  COUP\,732
($\theta^1$\,Ori\,E), 745 ($\theta^1$\,Ori\,A), and 809
($\theta^1$\,Ori\,C), which all suffer strongly from photon pile-up
in the ACIS detector, show a high-energy excess which calls for
the inclusion of a third high-temperature component in the model.
This high-temperature component does not significantly affect the
derived luminosity.  For $\theta^1$\,Ori\,C (COUP\,809), the
brightest X-ray source in the field, substantial discrepancies
between best-fit model and data are left, 
which may be related to residual pile-up in the extracted area
and/or uncertainties in the response for the annular extraction region.  
Since our
modelling did not improve the result for COUP\,809, we use the
spectral parameters given by \citet{Getman05.1}, but caution that
the luminosity might be overestimated (see \S~\ref{subsect:809}).

\begin{figure}
\begin{center}
\resizebox{16cm}{!}{\includegraphics{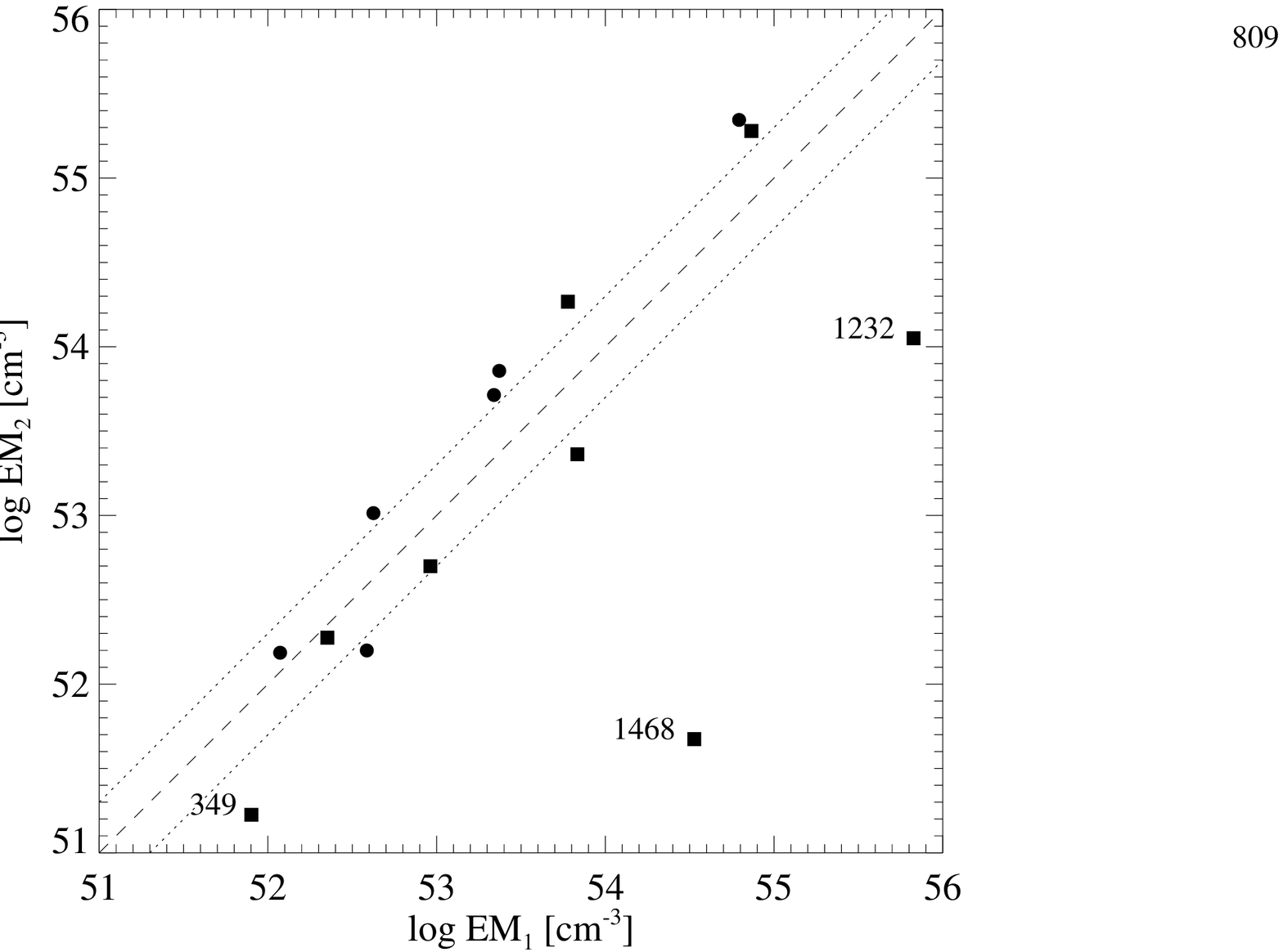}} \caption{Relation 
between the emission measures from the two components of the 
spectral fits from Table~\ref{tab:spec}. For 3-temperature models, we 
plot the $EM$ corresponding to the two lower temperatures which 
dominate the emission. The region included within the dotted lines 
marks differences of a factor two and lower. Squares are SW stars,
and circles are WW stars. Three outliers are 
labelled by their COUP source numbers.} \label{fig:em}
\end{center}
\end{figure}

The emission measures for the soft ($EM_1$) and hard ($EM_2$)
components of the 2-T fits are compared in Figure~\ref{fig:em}.
There is a general tendency of the WW stars to show $EM_2/EM_1>1$,
and of the SW stars to show $EM_2/EM_1<1$.  Some SW stars --
$\theta^2$\,Ori A, NU\,Ori and to lesser extent Par\,1772 -- are very
strongly dominated by the soft component.

In addition to this plasma modelling procedure, we examined a
simple model-independent spectral measure using the hardness
ratio.  The median values of $HR$ for the three sub-samples are: 
$\langle HR \rangle_{\rm MA} = 0.37$, $\langle HR
\rangle_{\rm WW} = 0.29$, and $\langle HR \rangle_{\rm SW} = 0.18$.
Nonparametric two-sample tests indicate that the difference in
$HR$ distributions for SW stars and MA stars is significant at the
$P<1$\% confidence level, while WW and MA stars are statistically
indistinguishable.

\section{Variability}\label{sect:var}

\subsection{Statistical tests}\label{subsect:var_stat}

Variability is a characteristic property of magnetically active low-mass 
pre-main sequence stars \citep{Montmerle83.1}. 
Strong variations of their X-ray output are observed on short timescales (minutes to hours),
while longterm variability is less pronounced and may be the result of
catching the star during different phases of short-term events. 
Studies of X-ray variability on short timescales in hot wind-driven stars 
are practially absent in the literature. But comparing different sets of 
observations for a given star has provided little or no evidence for long-term 
variations \citep{Berghoefer97.1}, leading to the conviction that hot stars show no
X-ray variability.  
In this vein, variability can be an
important diagnostic for discriminating the origin of the X-ray
emission in wind shocks $vs.$\/ magnetic activity. 
On the other hand, to our knowledge no study has been performed where the variability 
of hot and cool stars was directly compared in a consistent manner.
This is the purpose of this section. 

The observable variable phenomena include flares of different morphology,
rotational modulation, and irregular variations on short and long
timescales. For subsamples of COUP stars detailed variability studies 
are presented by \citet{Favata05.1, Flaccomio05.1, Wolk05.1}. 
Classification of a given event into any of 
the above mentioned categories of variability is not straightforward. 
Especially the definition of a flare, generally viewed as an event
with rapid rise and spectral hardening followed by decay and
spectral softening, escapes consensus on its quantitative
description throughout the literature. 

We apply here a method described by \citet{Wolk05.1} for analysis
of flare-like variability in COUP MA sources.  It is based on
Maximum Likelihood Blocks (MLB), a modification of Bayesian Blocks
developed by \citet{Scargle98.1}, in which the X-ray lightcurve is
divided into contiguous segments of constant signal. The
segmentation algorithm has two free paramenters: the minimum
number of counts accepted in a given segment and the significance,
defined as the probability that a constant lightcurve is
spuriously segmented due to random fluctuations. In this study of
OBA COUP stars, we set a minimum segment of $20$ counts and a
significance for change points of $95$\,\%.

The results of the variability tests are summarized in
Table~\ref{tab:variab}. Stars are sorted in order of decreasing
effective temperature.  Columns 4 and~5 flag stars variable
according to a simple Kolmogorov-Smirnov (KS) test and the MLB
test. The KS-test shows $12$ out of $16$ targets are variable at
the 95\,\% confidence level.  The MLB test indicates $14$ of our
targets can not be described with a single block and thus are
likely variable.

\begin{figure}
\begin{center}
\resizebox{16cm}{!}{\includegraphics{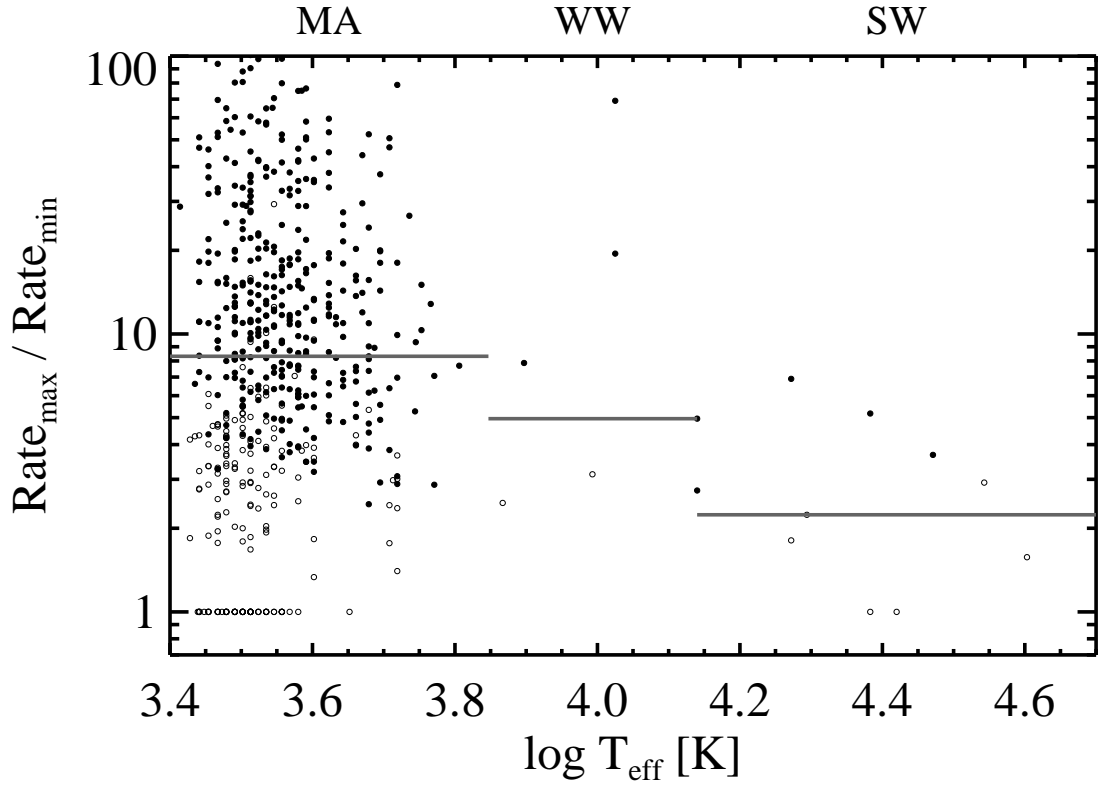}} \caption{Ratio of the 
maximum to minimum count rate blocks in the MLB test for all sources 
in the COUP optical sample plotted versus the effective temperature. 
If the maximum count rate corresponds to a flare the plotting symbol 
is filled. The thick grey lines mark the median values for these 
ratios in each class: MA = magnetically active, WW = weak-wind, and 
SW = strong-wind.} \label{fig:varampl}
\end{center}
\end{figure}

The ML blocks are also used to define a `characteristic level' and
flares, represented by periods of `very elevated' count rate and
rapid flux change as quantified by the derivative of the
lightcurve \citep{Wolk05.1}.  A total of $15$ flares are
identified, distributed among the $20$ targets as summarized in
column 6 of Table~\ref{tab:variab}. Columns 7 and~8 give the
fraction of the time spent in the `characteristic level' ($t_{\rm
char}/t_{\rm total}$) and the ratio between the count rate of the
highest segment and the `characteristic level' ($R_{\rm
peak}/R_{\rm char}$). In Figure~\ref{fig:varampl}, the amplitude
of the lightcurve variations, $R_{\rm max}/R_{\rm min}$ is
compared to that of COUP low-mass MA stars. Stars for which the
maximum count rate corresponds to a flare are distinguished by
filled plotting symbols. The amplitudes clearly indicate that SW
stars are less strongly variable than MA stars.

\begin{figure}
\begin{center}
\resizebox{16cm}{!}{\includegraphics{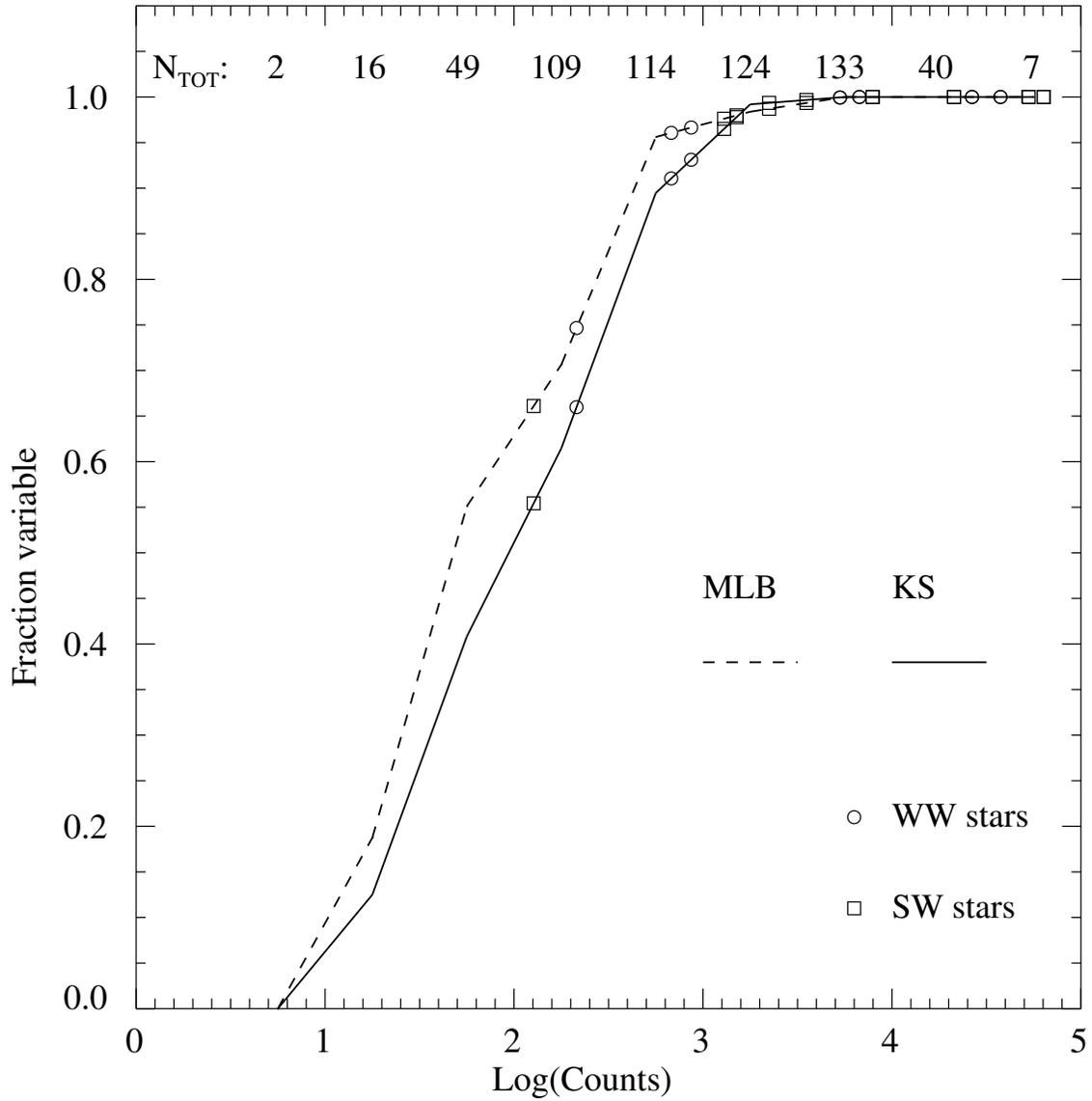}} \caption{Fraction of 
COUP sources which are X-ray variable as a function of source 
brightness. The curves show the distribution for MA stars in 
the COUP according to Kolmogorov-Smirnov and Maximum Likelihood 
Blocks criteria. The individual hot SW and WW stars are plotted as 
open boxes and circles, respectively. } \label{fig:varfrac}
\end{center}
\end{figure}

The sensitivity to the detection of variability depends on the
signal-to-noise. Fig.~\ref{fig:varfrac} shows the fraction of
variable stars for different count levels. The curves represent
COUP MA stars. The solid line is for the KS-test and the dashed
line for the MLB test. The total number of stars in each bin is
indicated on top of the graph. These data demonstrate that nearly all
X-ray-bright low-mass stars are found to be variable. 
An equivalent study for the SW and WW stars is prohibited by the small
number of objects in our sample. Instead, we place the SW
and WW stars on the MLB and KS curves for MA stars, at the position 
corresponding to their observed counts. From this comparison,  
we find that 96\% of WW and SW stars are expected to be variable. 
For the WW, this is consistent with the observed fraction of $100$\% 
variable stars.
The fraction of variable WW stars is thus similar to that of MA stars. 
But for the SW stars, the expected variable fraction is somewhat higher 
than the observed fraction of $78$\% variable stars (see Table~\ref{tab:variab}). 
A deficiency of SW variability is also suggested by the fact that
only $4$ flares appear in $9$ SW stars compared to $11$ flares in $7$
detected WW stars.

\subsection{Flares}\label{subsect:flares}

The lightcurves of all detected targets are shown in
Fig.~\ref{fig:lcs} together with the time evolution of the
hardness ratio evaluated in each segment of the MLB test.
Signatures of heating during phases of increased intensity are
obvious support for the interpretation of these episodes as magnetic
reconnection flares. In the following, the characteristics of the
flares on SW and WW stars are compared to the flares on solar-mass
COUP stars presented by \citet{Wolk05.1} which uses the same MLB
and flare characterization methodology. Since only $4$ flares were
observed on SW stars, we did not compute statistics for this group
alone, but rather examine how the derived values for WW stars
change if we combine the sample with the SW stars. Spectral
analysis is based on two-temperature plasma fits of flare segments
(see \S~\ref{sect:spec}) for the 12 (9) flares in the SW+WW (WW)
samples where $>$50\,counts were collected.  The results of this
comparison are summarized in Table~\ref{tab:flare_comp}.

A histogram showing the distribution of flare durations for the WW
stars is given in Figure~\ref{fig:histo_flareduration}. Although
the median of the flare duration on WW stars is only about $1/3$
of the value for solar analogs, the distribution seems to be
roughly in agreement with the $0.1 \times e^{(-t/100\,ks)}$
relation that describes the distribution for solar-type stars.
Following the arguments of \citet{Wolk05.1}, we derive a flare
frequency of $1$ flare every 710\,ks for WW stars which is similar
to the value of $1$ flare every $640$\,ks obtained for ONC solar
analogs.  For SW stars, the flare frequency is three times lower
at $1$ flare every $\sim 2200$\,ks.

The median flare luminosity and energy of WW stars is similar to
the respective values for solar-type stars, while in the total
sample of SW plus WW stars both the flare luminosity and energy
are increased by almost $1$ dex. A possible interpretation is   
that the flares of SW stars are $not$ drawn from the population 
of COUP MA stars, but are the result of a different emission mechanism. 
Since the sudden energy release observed in flares points at the 
presence of magnetic fields one possible explanation for the flares 
on SW stars is that they arise from magnetically mediated events
in the strong winds. Note, however that the flares on SW stars 
are not particularly strong with respect to the characteristic level; 
their high luminosities are due to the overall higher emission level 
of SW stars with respect to the WW and solar-type stars.
The different quiescent luminosity of the SW stars and WW stars
 may give rise to a bias, because smaller flares are more difficult
to detect on brighter stars. Indeed, if we examine the individual events, 
the strongest flares on SW stars are comparable to the strongest flares 
on WW stars and solar-type stars, at odds with the above 
mentioned interpretation. 
A thorough quantitative comparison of COUP spectra/variability 
and the MCWS model can be obtained with help of simulations, but it is
beyond the scope of this paper and will be addressed in a future study.

\subsection{Rotational Modulation}\label{subsect:rotmod}

Rotational modulation of the X-ray signal is expected if the X-ray
emission is concentrated in long-lived inhomogeneous magnetic
field structures, such as active regions on a MA star or the 
magnetosphere in a magnetically confined wind shock (MCWS). 
However, since the X-ray emitting plasma is optically thin, rotational
modulation may exist only as a result of asymmetries along the
line of sight: stellar occultation (provided the emitting volume is not
much larger than the star), or non-spherically symmetric absorbing
material such as the `cooling disk' of the MCWS model.

In contrast to the optical band where rotational patterns related to
active regions are readily observed in many late-type stars,
reports on rotational modulation in the X-ray band have been
scarce due to inadequate time coverage of past observations.  The
rotation periods of active late-type stars are on the order of
$\sim 1-10$\,days, accessible to the 13.2 days spanned by the COUP
exposure but longer than previous X-ray exposures of pre-main
sequence stars.

\citet{Flaccomio05.1} detect rotational modulation in $\simeq 10$\,\% 
of a subgroup of $>200$ COUP stars examined systematically, 
which definitely is the biggest sample studied in this respect to date. 
They compute the Lomb Normalized Periodogram
\citep{Scargle82.1} on binned COUP lightcurves. The peak
frequencies in the periodogram are then identified and a False
Alarm Probability (FAP) is computed using Monte Carlo techniques
allowing for correlated noise.  The analysis is made both on the
original data and on lightcurves trimmed of bright flares. We now
apply these methods to the ONC early-type star sample.

We examine the COUP lightcurves of the three stars in the SW and WW
sample which have photometrically measured rotation periods:
$\theta^1$\,Ori\,C, JW\,660, and JW\,165. Two additional stars,
JW\,831 and JW\,531, were surveyed for starspot induced
photometric variability by \citet{Choi96.1}, but no period was
found.  
At the present stage we discuss only stars with known optical period,
because the search for X-ray
periods is complicated by flares and other types of variability. 
A systematic analysis of the whole COUP sample with known optical
period \citep{Flaccomio05.1} 
has shown that sometimes the X-ray period does not coincide with the optical one, 
suggesting that the complex X-ray lightcurves can result in spurious periods. 
Our results for the three stars analysed here are summarized in Table~\ref{tab:rotmod}, and
phase-folded lightcurves and hardness ratios can be found in
Fig.~\ref{fig:lcs_folded}.  
The median count rate
and the amplitude (columns 5 and~6 of Table~\ref{tab:rotmod}) were
computed from the folded lightcurve. The uncertainty of the
amplitude derives from the scatter of the data near maximum and
minimum, and thus takes into account the intensity changes from
one cycle to the other.

\subsubsection{$\theta^1$\,Ori\,C = COUP\,809}\label{subsubsect:theta1oric}

An optical period of $15.43$\,d was reported by \citet{Stahl93.1},
and also evidenced in a set of $10$ {\em ROSAT} X-ray observations
separated by $\sim 2$ days each, such spanning a little more than
one cycle \citep{Gagne97.1}. The same variability pattern is 
recovered in the COUP lightcurve (Fig.~\ref{fig:lcs_folded}). The
X-ray lightcurve is almost exactly in phase with the H$\alpha$
ephemeris measured $10$~years earlier, and the X-ray period agrees
remarkably well with the optical period. The X-ray hardness ratio
varies out of phase with the intensity by roughly half a cycle.
While there is a concern that incomplete correction for photon
pile-up in the ACIS detector may be responsible for the
variation of the hardness ratio, we believe this is not occurring. 
If this were the case, one would expect a correlation
with the lightcurve and not an anti-correlation because the
increased pile-up during higher intensities would shift photons
into the high-energy portion of the spectrum.

Astrophysical discussion of this source in the context of detailed
evidence for an inclined dipole field in $\theta^1$\,Ori\,C and
magnetically trapped wind plasma \citep[e.g.][]{Babel97.1,
Donati02.1, Schulz03.1, Smith05.1} will be the subject 
of a forthcoming study.

\subsubsection{JW\,660 = COUP\,1116}\label{subsubsect:jw660}

A rotation period of $6.15$\,d was reported by \citet{Choi96.1}
and \citet{Herbst00.1} for this B2.5 star, although the cause of a
spot pattern on an early-B star is not clear.  We find two
episodes of enhanced count rates in the COUP lightcurve, which may
define an X-ray period of $6.67$\,d (Fig.~\ref{fig:lcs_folded}).
The amplitude of the folded lightcurve is poorly defined due to
the different peak intensity in the two consecutive cycles.  It is
also possible that the two features are long-duration flares
unrelated to any rotating structures.  The hardness ratio varies
in phase with the brightness of the source. Spectral hardening
during phases of higher intensity is a typical signature of
magnetic activity and is opposite to the pattern seen in
$\theta^1$\,Ori\,C.

More detailed information on the run of temperature and emission
measure is obtained from time-resolved spectroscopy. The spectrum
of JW\,660 can be described adequately by a two-temperature
thermal model with comparatively high temperature at all phases of
the presumed rotation cycle. These fits require an absorbing
column somewhat higher than the value inferred from the optical
extinction. During the maximum of the cycle the higher temperature
is slightly enhanced and the emission measure shifted towards the
hot component, in agreement with the evolution of the hardness
ratio. A feature from highly ionized iron at $\sim 6.7$\,keV is
pronounced during the phase of maximum brightness. Its appearance
results from improved statistics rather than a true change in the
spectral shape.

\subsubsection{JW\,165 = COUP\,113}\label{subsubsect:jw165}

A photometric period of $5.77$\,d was reported by
\citet{Herbst00.1} for this A7 star, albeit based on a single
observing season. The period search on the COUP data results in a
period of $4.6$\,d, but the folded X-ray lightcurve is rather
noisy (Fig.~\ref{fig:lcs_folded}).
Spectral fitting suggests a soft X-ray absorption inconsistent
with the visual extinction.

\section{Multiple Systems}\label{sect:binaries}

About half of our sample has been carefully examined for
multiplicity using traditional spectroscopic and newer speckle and
adaptive optics imaging techniques, as indicated in columns $7-9$
of Table~\ref{tab:sample}. High-multiplicity visual systems are
common among the ONC OB stars, and we summarize known components
of our sample stars in Table~\ref{tab:binaries}. In all cases the
primary falls into the SW category, and most of the secondaries
seem to be WW stars rather than MA stars. However, the mass estimates
are often based on photometry and have large uncertainties. We
also caution that additional, faint lower-mass companions may be
present which could contribute to the X-ray emission.  Recall that
1\arcsec\/ separation corresponds to a projected distance of
440\,AU. Five well-characterized systems are presented in Appendix\,A 
where we discuss the characteristics of their X-ray emission
with the aim to identify the X-ray emitting component.

Only one of these multiple systems is resolved in the COUP: the
quintet $\theta^1\,$Ori\,B consists of two X-ray sources
(Fig.~\ref{fig:binary_images}, see Apppendix\,B for details).
$\theta^1$Ori\,B\,East (COUP\,778) emits $\log L_{\rm
t,c}=30.1$ erg s$^{-1}$ with an unusually soft spectrum and no
strong variability.  This source can be plausibly attributed to
wind shocks of the B2.5 primary. In contrast,
$\theta^1$\,Ori\,B\,West (COUP\,766) has all the
characteristics of a MA lower-mass member of the
Orion population with several flares.  It is not clear which of
the components of the $\theta^1$\,Ori\,B\,West binary is the
principal X-ray emitter.

\section{Stars with evidence for disks}\label{sect:haebe}

Our sample includes four stars listed in the catalog of Herbig Ae/Be
stars by \citet{The94.1} in their Table\,5 labelled `non emission
line early-type shell stars and young stellar candidates'. Stars
in this table lack emission line signatures of accretion, thus do
not satisfy the classical definition of HAeBe stars as emission
line objects, but IR photometric excesses were seen with
$IRAS$. A dust shell is identified around one additional star of our sample
\citep[][ see flag for IRAS excess stars in the last column of Table~\ref{tab:sample}]{Tovmassian97.1}. 
The $IRAS$ measurements indicate that a
substantial amount of the luminosity of these stars is irradiated
in the far-IR, probably related to dust emission at
$100-1000$\,AU from the star.  Par\,1772 and V361\,Ori are
discussed by \citet{Manoj02.1} who estimate dust masses between
those of typical HAeBe stars and of Vega-like stars. The absence
of emission features in the spectrum of Par\,1772 and JW\,153 was
confirmed by \citet{Boehm95.1}, thereby showing that the accretion
phase has ended in these objects.

Figure~\ref{fig:jhk} shows the near-IR color-color diagrams
for our sample.  For those stars detected in the COUP  the $JHKL$
magnitudes are reported by \citet{Getman05.1} while for the
nondetections we compiled IR photometry from the 2\,MASS
database and the literature \citep{Hillenbrand98.1, Carpenter01.1,
Muench02.1}. No strong indications for excess emission above the
photosphere is seen in the $JHK$-diagram, but several stars have
$L$-band excesses. 
A comparison
to the intrinsic IR colors of a late-type star shows that the redder
color expected from the presence of a hidden cool star can not account
for the observed excesses. We conclude that the near-IR excesses must be
ascribed to circumstellar material. 
We note that the five $IRAS$ excess stars do not show stronger $K-L$
excess than the remaining sample. This suggests they have
evolved disks where the inner, hotter part has been cleared.

We find no link between X-ray properties and the presence or
absence of an IR excess. For example, a wide spread $-8 <
\log{(L_{\rm x}/L_{\rm bol})} < -5$ is seen among the stars with
evidence for disks, which does not favor a common origin of their
X-rays (Figure~\ref{fig:lxlbol_lbol}). One of the five $IRAS$
excess stars, JW~531, remains undetected in the COUP.

\section{Discussion}

\subsection{COUP results in perspective}\label{subsect:disc_coup}

The COUP is the most sensitive X-ray image ever obtained of a rich
young ($\sim 1$\,Myr old) stellar cluster with an initial mass
function extending up to $\simeq 45$\,M$_\odot$, providing the
opportunity to study the X-ray emission of a coeval sample of
early-type stars on a long continuous time baseline. The X-ray
emission mechanisms of early-type stars are highly disputed (\S\,1). In
the hotter stars, our SW sample, the stellar winds most likely
play a crucial role. About half of the SW sample is dominated
by soft X-rays that are likely generated by a myriad
of small shocks produced by instabilities in their line-driven
winds \citep{Lucy80.1, Owocki99.1}. The other half, mostly the B1 to B3 type
stars, the cooler stars in the SW sample, also show substantial hard
emission, unexplained within the wind-shock model. 
This emission may indicate that magnetic fields are involved. 

Direct observational evidence for the existence of
magnetic fields on early-type stars is scarce and mostly
restricted to the class of CP stars \citep{Landstreet92.1}. On the
other hand, magnetic fields are often held responsible for
imposing the cyclic variability of wind properties observed in the
UV \citep[e.g.][]{Kaper99.1, deJong01.1} and for the non-thermal
radio emission seen in some O stars \citep{Abbott84.1,
VanLoo04.1}.  While traditionally magnetic fields on hot stars are
believed to be fossil remnants of the star formation process,
numerical modelling has shown that dynamos may operate in their
convective cores \citep{Charbonneau01.1}. However, the theory has
difficulties transporting magnetic flux throughout the thick
radiative layer to the stellar surface \citep{MacGregor03.1}.
Hybrid models including winds disturbed in large-scale
magnetospheres have been developed to explain both radio and X-ray
emission from early-type stars \citep{Trigilio04.1}.

These models have two implications for the mechanism of X-ray
production.  First, X-ray emission may be produced as a result of
magnetic reconnection in the current sheet that forms in the
equatorial plane where gas pressure opens the field lines. This
model explains the observed non-thermal radio emission of magnetic
CP stars \citep{Linsky92.1}, and \citet{Usov92.1} have estimated
that the expected X-ray luminosities are $\log{L_{\rm x}} \sim
31-34\,{\rm erg\,s^{-1}}$. Therefore, gas heating by magnetic
reconnection or other instabilities at the
equatorial current sheet from a dipolar field of an early-type
star can account for X-ray emission from radiative stars; i.e.,
both SW and WW stars. Second, in the magnetically confined wind
shock (MCWS) model \citep{Babel97.1}, the major effect of the
magnetosphere is to confine and channel the wind to the equatorial
plane where a quasi-stationary large-scale shock forms. 
The cooling of shocked material results in the formation of a 
`cooling disk', partially absorbing the emitted X-rays and periodically
modulating the observed X-rays. This model
can explain the high temperatures and unusual X-ray line shapes
seen in $\theta^1$ Ori C \citep{Schulz03.1, Gagne05.1}.

In less luminous WW stars, the wind velocities are smaller than in
SW stars. Production of X-ray emission in small-scale wind shocks
becomes questionable although the MCWS model may still apply;
indeed, this model was first devised for the magnetic Ap star, 
IQ\,Aur \citep{Babel97.1}.  Perhaps the most common explanation
for the X-ray emission from WW stars -- applicable to the weaker
SW stars as well -- invokes unknown or unresolved late-type
companions (separations smaller than $\sim$\,100\,AU). 
Such objects would by their nature be magnetically
active (MA), and produce relatively hard, highly variable X-ray
emission with typical luminosity between $\log{L_{\rm x}} \simeq
28-31\,{\rm erg\,s^{-1}}$.

Distinguishing between the different possible emission scenarios
based on observational facts is not straightforward. In this
section, we use the results from the COUP to test our {\it a
priori} division of the early-type stars in the ONC into SW and WW 
stars, and we provide a tentative classification of all
targets. This classification scheme involves three criteria --
X-ray luminosity, spectrum, and variability -- and is summarized
in Table~\ref{tab:class}.

\subsection{Classification of ONC early-type stars X-ray emission}

We consider first Fig.~\ref{fig:lxlbol_lbol} which plots 
the $L_{\rm x}/L_{\rm bol}$ ratio against the
effective temperature for the SW and WW stars in the COUP.
For both groups of stars, a wide spread of \lxlbol~ ratios is seen
without dependence on $T_{\rm eff}$.  
The WW stars show a very wide spread
around $\log{(L_{\rm x}/L_{\rm bol})} \sim -5$ to $-6$. 
The SW stars do not cluster near
the empirical relation \lglxlbol~$\simeq 10^{-7}$ reported in
earlier observations of X-ray emission from hot stars and
interpreted in terms of the standard wind-shock model
\citep{Sciortino90.1, Berghoefer97.1}.  They scatter by more than
two dex around this level.    

Nonetheless, we use this $L_{\rm x} - L_{\rm bol}$ scaling relation to help
classify and understand the X-ray emission of ONC early-type stars.  
We compare the observed luminosities $L_{\rm t,c}$ to the `predicted' luminosity
$L_{\rm x,p} = 10^{-7}~L_{\rm bol}$ for stars of the SW class. For
the WW class, we use a `predicted' luminosity $L_{\rm x,p} =
10^{-4}~L_{\rm bol}$ which is characteristic of lower-mass MA
stars \citep{Preibisch05.1}.  These levels are shown as dotted
lines in Figure~\ref{fig:lxlbol_lbol}.  Since MA companions are
likely to have $L_{\rm bol}$ lower than the primary, this $L_{\rm
x,p}$ is actually a lower limit. In column~4 of
Table~\ref{tab:class}, we distinguish the following situations:
\begin{description}

\item If $L_{\rm t,c}/L_{\rm x,p} < 10$, the emission mechanism
defined above for the SW and WW stars respectively is considered
to be the dominating mechanism. We denote this mechanism in
column~4 by ``W'' for wind-shock and ``T'' for T Tauri-like
magnetic activity.   Note that, in the latter case, as an alternative
to a MA late-type companion, the emission
might be due to a magnetic reconnection process in a wind. 

\item If $L_{\rm t,c}/L_{\rm x,p} > 10$, a different or additional
mechanism is considered to be present. We denote this situation by
a ``$>$'' sign in column~4.

\end{description}

In column~5 of Table~\ref{tab:class}, we summarize the result of
the X-ray spectral analysis (\S 3) for all detected stars. We
consider the X-ray spectrum characteristic for small-scale wind
shocks if the soft component (with $T \sim 2.5$\,MK) of the
two-temperature fits dominates the emission measure (see
Figure~\ref{fig:em}), or if a hard component is altogether absent.
Such sources are denoted by ``W''.  We see a clear pattern that
such soft-dominated spectra are seen only in some SW stars and not
in any WW stars.

Column~6 tabulates results of the variability analysis (\S 4)
based on the COUP light curves. Constant or weakly variable light
curves characteristic of small-scale wind shocks are denoted
``W''. Flare-like high-amplitude variability indicates magnetic
reconnection (``T''), which might be either from lower-mass MA
T Tauri companions or from large-scale shock events in a magnetically
confined wind. Smooth lightcurves (``S'') indicate non-flare-like
variability such as periodic modulations with moderate amplitudes 
(factors $\sim\,2-3$).

In column~7, we categorize the observational indications for the
presence of circumstellar material (\S 6). Recall that no
emission-line stars (= HAeBe stars) have been identified in the
ONC, but some evidence for disks is seen in the near-IR
color-color diagrams or in far-IR fluxes. Finally, column~8
collects observational evidence for magnetic fields
including rotational modulations (RM), non-thermal cm radio
emission (NT), chemically peculiar characteristics (CP), or direct
indicators such as by Zeeman or polarization measurements (B).

Some immediate conclusions may be drawn from Table 8 and the
underlying findings presented throughout this paper:
\begin{enumerate}

\item There is a clear distinction between the higher-mass SW and
all WW stars in terms of their X-ray properties. This justifies
{\it a posteriori} our adopted division between SW and WW
classification.  However, the location of the transition between
the two classes remains ill-defined and may be earlier than the B4
boundary adopted in \S~\ref{sect:def_class}.  Some B0$-$B3 stars
have luminosities consistent with X-ray production in the wind,
but have spectra and variability characteristics of magnetic
activity.

\item In the more massive SW stars with spectral types B0$-$O7,
small-scale wind shocks predicted by the original model of
\citet{Lucy80.1} appear to be the dominant X-ray emission
mechanism.  But even here, magnetic processes seem to be present; e.g.,
to explain the powerful $L_{\rm x} \sim 10^{32}$ erg s$^{-1}$ COUP flare
in $\theta^1$\,Ori\,A (Figure~\ref{fig:lcs}), and a similar flare in
$\theta^2$\,Ori\,A seen in the earlier $Chandra$ study of
\citet{Feigelson02.1}.  If one includes the rotational modulation of 
$\theta^1$ Ori C and the X-ray independent indicators of magnetic
fields, all three of the most massive Trapezium stars --
$\theta^1$\,Ori\,C, $\theta^2$\,Ori\,A and $\theta^1$\,Ori\,A -- show
hybrid wind/magnetic characteristics; see \citet{Schulz03.1} 
for a discussion of these hybrid systems in the Orion Nebula hot stars. 
These stars are promising candidates for the MCWS model.

\item The less massive SW stars similarly show hybrid
characteristics combining winds and magnetic activity.  Here the
X-ray luminosities are lower, $L_{\rm x} \simeq 29-31$ erg s$^{-1}$, 
as expected for lower-mass T Tauri companions, and
evidence for magnetic fields is found in some but not all objects. 
Interestingly, most SW stars with emission above the canonical line of \lglxlbol$=-7$ 
are known to have lower-mass companions that are unresolved in X-rays. 

\item In the WW class, the X-ray emission is either undetected
with very sensitive limits ($L_{\rm x}/L_{\rm bol} < 10^{-7}$), or is
detected with properties associated with magnetic activity from a late-type
companion.

\item Objects with circumstellar disk indicators show no evidence for
a common X-ray emission mechanism. 
We conclude that disks do not play a significant
role in the X-ray emission of the early-type ONC stars.

\end{enumerate}

We will now discuss the sample stars in more detail, starting with
the cooler intermediate-mass WW stars and proceeding to the SW
stars.

\subsection{WW stars}

The situation for the ONC WW stars (i.e., spectral types B5 and
later) seems fairly straightforward (Table~\ref{tab:class}). Out
of 11 stars, 4 are not detected with upper limits several orders
of magnitude below the activity level and clearly lower than the
$L_{\rm x}$ of all detected WW stars (see
Fig.~\ref{fig:lxlbol_lbol}). A bimodal distribution of X-ray
luminosities for B- and A-type stars has been suggested in other
studies \citep{Preibisch01.1, Feigelson02.1, Stelzer03.1}, 
explained by zero emission by the intermediate-mass stars and
emission by lower-mass MA companions when present.  Our data can
not distinguish between two separate populations and a single
population with a very wide dispersion.

The X-ray luminosities of the majority of the detected WW stars
range between $\log{L_{\rm x}}\,[{\rm erg\,s^{-1}}] \sim 29-31$
erg s$^{-1}$, typical of ONC late-type MA stars
\citep{Feigelson05.1, Preibisch05.1}, and therefore consistent
with the unresolved binary companion hypothesis.  Our study of
their variability in general and flaring in particular, 
and their spectral hardness support
the identification of WW X-ray emission with late-type MA
companions.  
Not enough is known about $\theta^1$\,Ori\,BW to assign an X-ray emission 
mechanism, although the flare-like variability suggests the possible presence
of an MA companion. 
Remarkably, none of the WW stars in the COUP sample
is a known multiple; the multiple systems discussed in Appendix~A
are all SW stars. However, this is most likely due to a lack of systematic
binarity searches in the WW population.

The WW star $\theta^1$\,Ori\,E (B5) has unique properties in the
COUP data. It has a much higher time-averaged X-ray luminosity
than the other WW stars, $\log{L_{\rm t,c}}=32.4$ erg s$^{-1}$ and
$\log(L_{\rm t,c}/L_{\rm bol}) = -3.6$ (Table~3). 
Its \lglxlbol\, level is several orders of
magnitude above any standard wind-shock prediction. 
This star also shows evidence for non-thermal radio emission
\citep{Felli93.1}, and \citet{Schulz03.1} found plasma
temperatures ranging from $4$ to $47$\,MK in a study of $Chandra$ 
grating observations.
Therefore, it may have an extended magnetosphere,
able to confine its weak wind within the MCWS framework. 
$\theta^1$\,Ori\,E has also shown a flare, which -- if attributed
to a late-type MA companion -- would be one of the most 
powerful flares seen in any T Tauri star \citep{Grosso04.1, Favata05.1}. 
Altogether, the COUP properties of $\theta^1$\,Ori\,E are not easily explained in
a coherent fashion.

The X-ray properties of the A7 star JW\,165 are also worth noting.
Its smooth lightcurve is consistent with both rotational modulation
or slow flares. Its late spectral type 
implies that its wind must be very weak. For comparison, the
calculations by \citet{Babel97.1} for IQ Aur, a hotter A0 star,
give $\dot{M} \sim 10^{-10} M_\odot~{\rm yr}^{-1}$. JW\,165 may
have an unseen magnetosphere, although in this case its mass-loss
rate should be $\dot{M} > 10^{-12} M_\odot~{\rm yr}^{-1}$ since it
is not known to be CP.  Perhaps more likely, it has a MA T Tauri
companion responsible for the emission and peculiar 
variability observed during the COUP exposure.

\subsection{SW stars}

Based on the emission diagnostics in column~4 of
Table~\ref{tab:class}, the SW stars can be subdivided in two
groups: (i) those for which the X-ray luminosity is consistent
with small-scale wind shock emission, and (ii) those which are
overluminous by one or two orders of magnitude with respect to the
corresponding predictions.  With the exception of JW 660, all
stars in the B0.5-B3 range fall in the ``W'' category according to
their X-ray luminosity. 
However, a MA companion is suggested for
$\theta^2$\,Ori\,B. Here the \lglxlbol\, level is compatible with
the prediction for wind-shock emission, but in absolute terms 
$L_{\rm x}$ is low enough to originate from a late-type T Tauri star.

JW 660 (B3) is one of three stars showing evidence for rotational
modulation in the COUP.  The time-evolution of its hardness ratio
shows a correlation with the X-ray brightness, opposite to the 
{\it anti}correlation found in $\theta^1$ Ori C where the MCWS mechanism
appears to dominate. Rotational modulation of the X-ray emission
can result from the eclipse of a large magnetic loop by the star
\citep{Stelzer99.1}, and the joint modulation of intensity and
temperature is typical for magnetic activity phenomena.  However,
we can not exclude applying the MCWS scenario to JW\,660, as this
model does not make predictions on the variation of hardness with
phase. It would be very valuable to search JW\,660 for a large-scale
magnetic field.

The remaining stars (B0-07) are the most massive in the ONC. They
are all very luminous with $\log{L_{\rm t,c}} > 32$ erg s$^{-1}$, 
about one order of magnitude or more above the value ``predicted'' for
small-scale wind shocks. Next we discuss these stars in turn.

The most extreme case is obviously
$\theta^1$\,Ori\,C with $\log{L_{\rm t,c}} \sim 33.3$ erg
s$^{-1}$. 
{\em ROSAT} discovered an X-ray
modulation with the same period as the optical period \citep[$P =
15.4$d;][]{Gagne97.1}. To explain its extreme X-ray luminosity and
rotational modulation, \citet{Babel97.2} applied a MCWS model
originally devised for the X-ray emission of the Ap star IQ Aur
\citep{Babel97.1}. 
For $\theta^1$\,Ori\,C, the MCWS model postulates a magnetic field of order 400 G buried in the
wind and confining its dense, low-velocity layers.  Because the
outer layers of the wind were predicted to be open and sitting
above the magnetically confined layers, such a magnetic field was
difficult to detect.  However, this was achieved by
\citet{Donati02.1} who found the predicted field via a
rotationally modulated Zeeman effect.
The COUP data enables for the first time a continuous study of X-ray spectral 
changes along the rotation cycle of $\theta^1$\,Ori\,C. 
The anti-correlation between source luminosity and
hardness is unusual and detailed modelling is needed. At present,
we speculate that the explanation for this anti-correlation results from particular 
viewing angles of the X-ray absorbing `cooling disk' predicted by the MCWS model.
In a recent study based on high-resolution spectroscopic
X-ray observations of $\theta^1$\,Ori\,C \citet{Gagne05.1} reported 
a small increase of the column density during phases when the disk is viewed edge-on,
corresponding to the X-ray minimum. This is consistent with the increase in spectral hardness
we observed during the COUP. 

The second-most hot star, $\theta^2$\,Ori\,A (O9.5), can be accommodated in the class of
small-scale wind shock emitters according to its characteristics
shown during the COUP, but dramatic flares were seen 
in the earlier $Chandra$ observation described by \citet{Feigelson02.1}.
These flares can be caused by magnetic reconnection in the wind
(which implies that this star be magnetic), or by an unusually active
unseen MA companion. 

$\theta^1$\,Ori\,A similarly shows hybrid characteristics.  
This star shows evidence for
magnetic fields from its non-thermal radio emission
\citep{Felli93.1}, and an X-ray spectroscopic study shows plasma
with temperatures up to $43$\,MK \citep{Schulz03.1}. 
With its flare-like behavior, it reaches X-ray
luminosities unusually high for a low-mass MA companion.  We
speculate that its magnetic field is too weak for the wind to be
confined, but that strong reconnection events take place along the
wind. If the star rotates, the wind flow will, like the solar
wind, take the shape of an Archmedean spiral past the Alfv\'en
radius. In such a configuration, ``corotating interaction
regions'' (CIRs) analogous to those in the solar wind \citep[e.g.,
][]{Kissmann04.1}, or in the wind of the young pole-on B star AB
Aur \citep{Catala99.1}, may exist. These CIRs could trigger
reconnection events, such as proposed by \citet{Usov92.1}, to
which the flare-like events observed on $\theta^1$\,Ori\,A by COUP
might be related.  A direct search for magnetic fields on the
stellar surface would be valuable.
We recall, however, that $\theta^1$\,Ori\,A has two known
lower-mass companions, and possibly others still undetected.
The observed X-rays could arise from one extraordinarily
MA companion, similar perhaps to LkH$\alpha$~312
seen elsewhere in the Orion molecular cloud \citep{Grosso04.1}.

\section{Conclusions}

The COUP observation has permitted an in-depth study of the X-ray
emission of the early-type stars of the ONC with the unprecedented
exposure covering most of 13.2 contiguous days.  We examine a sample
of $20$ coeval O, B and A stars.  
We address the X-ray properties and underlying astrophysical
processes of two subclasses: 
$9$ hotter members with spectral types O7$-$B3, 
designated `strong wind' or SW stars, and 
$11$ cooler members with spectral types B5$-$A9, 
designated `weak wind' or WW stars.
We also investigate the effects of multiplicity, disks, and independent
evidence for stellar magnetic fields.

Addressing the mystery of X-ray emission from intermediate-mass
late-B and A stars, we find an extraordinary range in luminosities
ranging from nondetections with $\log{L_{\rm X}} < 27.6$ erg s$^{-1}$
comparable to the quiet Sun to violently flaring sources up to
$\log{L_{\rm X}} \simeq 31$ erg s$^{-1}$.  Based on flaring
lightcurves and hard spectra, the detected WW stars are very
likely associated with unseen late-type companions.  The case of
$\theta^1$\,Ori\,E (B5) is unusual with strong constant X-ray
luminosity around $L_{\rm X} \simeq 2 \times 10^{32}$ erg s$^{-1}$
and flares of comparable intensity.  Exhibiting non-thermal radio emission, this may be a
case for MCWS emission in a mid-B star.

The situation with the Trapezium SW OB stars is rich and intriguing.
The X-ray emission gives little
support for the long-standing claim that \lxlbol\, is constant
around $\simeq 10^{-7}$ for OB stars. While a rough trend between
\lx\, and \lbol\, is seen, three orders of magnitude scatter in
\lx\, can be present at a given \lbol\footnote{We suggest that the
previous evidence, particularly the {\em ROSAT} All-Sky Survey of
OB stars \citep{Berghoefer97.1}, suffered a distance-related
selection bias and missed the lower luminosity stars.}.
We note that this result is limited by the Orion population to stars
of spectral type O7 and later, and we can not say whether a constant
\lxlbol\, is present for the most luminous early-O stars. 

The X-ray properties from the
majority of the SW stars exhibit hybrid characteristics.  
Only $\theta^1$\,Ori\,D (B0.5) and NU\,Ori (B1), and perhaps the
faint Par\,1772 (B2) and $\theta^1$\,Ori\,B\,East (B3), exclusively show the 
roughly constant, soft spectrum emission attributable to the traditional
Lucy-White model of small-scale wind shocks.  
Most show combinations of soft wind emission, hard-spectrum short-lived
flares, and/or rotational modulation. If the flares of $\theta^2$\,Ori\,A from
a previous {\em Chandra} observation are included, all three of the hottest Trapezium
stars -- $\theta^1$\,Ori\,C (O7), $\theta^2$\,Ori\,A (O9.5) and $\theta^1$\,Ori\,A (B0) 
-- show these hybrid characteristics.
Together with independent indicators of magnetic fields, the COUP evidence
strongly supports the MCWS model where the harder and variable
X-rays are produced by magnetically-mediated large-scale shocks.

The eleven cooler WW stars lack both strong ultraviolet emission
to drive powerful winds and outer convection zones to drive a
solar-like dynamo.  They are thus predicted to be X-ray-quiet.  Yet
seven of these stars are detected in the COUP image and together
they exhibit a wide range of X-ray luminosities from 
$\log{L_{\rm x}} < 27.6\,{\rm erg\,s^{-1}}$ comparable to the contemporary Sun 
to flares up to $\log{L_{\rm x}} \sim 31\,{\rm erg\,s^{-1}}$.  
Based on their flaring lightcurves and hard spectra, very
similar to COUP properties of lower-mass stars in the ONC,  
we believe that the WW X-ray emission is most likely
produced by late-type companions.  
In one case, $\theta^1$\,Ori\,B, the high-multiplicity system
is resolved into two X-ray sources; one component seems to be a soft-spectrum 
wind source and the other one a flaring MA lower-mass companion. The
COUP study uncovered a similar multiple system, heavily obscured
by the Orion Molecular Cloud: the hot Becklin-Neugebauer Object
has a faint X-ray source that is nearly overwhelmed by a much
brighter, previously unknown low-mass companion
\citep{Grosso05.1}.

Our boundary at spectral type B4 between SW and WW stars is probably
not a strict divide between wind and companion X-ray emission in all
stars.  The case of $\theta^1$\,Ori\,E at B5 is unusual, showing strong
constant X-ray emission around \lx$\sim 2\,10^{32}\,{\rm erg\,s^{-1}}$ and flares of
comparable intensity.  It may represent a case for MCWS emission in
a mid-B star. 

We emerge from the present study with an increasing recognition of
the important role played by magnetic fields in massive stellar
systems.  In many cases, the X-ray properties no longer follow the
simple ideas of small-scale shocks from unstable radiatively
accelerated winds.  Violent reconnection events, either in the
winds of the massive primary or from MA
orbiting late-type companions, are frequently present.  Rotational
modulation of the X-ray signal is not infrequent.  The high
fraction of OB stars which are suggested from X-ray emission to
have magnetically confined winds is much higher than the few
percent of B stars possessing optically detected magnetic surfaces
\citep{Landstreet03.1}.  X-rays thus provide a powerful diagnostic
for OB magnetospheres and their interactions with radiatively
accelerated winds.  The evidence provided here concerning the
prevalence of strong magnetic fields in the hotter O7$-$B0 stars,
where optical detection techniques are difficult, may be
particularly important. 

It is now necessary to search for unresolved late-type companions
down to distances $\sim$ 10 AU, as is currently possible with
high-angular resolution optical and IR methods. On the other hand, our
study must encourage renewed efforts to search for magnetic fields in OB
stars, in particular in the Orion Trapezium.

\appendix

\section{Multiple stellar systems}\label{sect:multiple}

Here we present observational information on individual multiple systems in
order of decreasing spectral type of the primary component.

\subsection{$\theta^1\,$Ori\,C -- COUP\,809}\label{subsect:809}

$\theta^1\,$Ori\,C is a speckle binary with $0.037^{\prime\prime}$
separation (or 16 AU projected linear separation), 
composed of the $45\,M_\odot$ primary and a
$>6\,M_\odot$ secondary \citep{Schertl03.1}. It is the most
massive star in the Trapezium cluster, and also extraordinary in
terms of its X-ray properties: highest X-ray luminosity among the
COUP sources ($\log{L_{\rm t,c}}\,{\rm [erg/s]} \sim 33.3$),
rotational modulation of the X-ray emission in accordance with the
optical variations, hardness anti-correlated with the lightcurve
along the $15.4$\,d cycle. These pecularities make it likely 
that the X-ray emission originates on the O star.

Although we took carefully care of correcting for pile-up effects, 
the extracted spectrum of this extremely X-ray luminous source 
may still be affected by pile-up, so
that the spectral parameters and the luminosity are possibly
misestimated. 
\citet{Schulz01.1} derived $\log{L_{\rm x}}\,{\rm [erg\,s^{-1}]} \simeq 32.3$ 
from the zeroth order signal of an ACIS/HETG observation. 
But \citet{Gagne05.1} found $\log{L_{\rm x}}\,{\rm [erg\,s^{-1}]} \simeq 33$ 
from the same data set, in rough agreement with the COUP result. 
Furthermore, \citet{Flaccomio03.1} found $\log{L_{\rm x}}\,{\rm
[erg\,s^{-1}]} \simeq 33.2$, consistent with our COUP
result, from a {\em Chandra} HRC observation, which is free of
pile-up effects.

\subsection{$\theta^2\,$Ori\,A -- COUP\,1232}\label{subsect:1232}

This system is a hierarchical triple with a 25 M$_\odot$ O9.5
primary \citep{Preibisch99.1}, a $\simeq 9$ M$_\odot$ A2
spectroscopic companion \citep{Aikman74.1, Abt91.1, Morrell91.1}
and a $\simeq 7$ M$_\odot$ visual companion with separation
$0.38$\,\arcsec \citep{Mason98.1, Preibisch99.1}.  The X-ray emission
is unusually strong $\log L_{t,c} = 32.5$ erg s$^{-1}$ after
correction for absorption. The COUP source suffers heavy pileup
and only 1\% of the events are extracted for analysis.

Such high levels of activity are rarely seen in lower mass T Tauri
stars and, in such cases, are always associated with high
amplitude ($L_{\rm flare}/L_{\rm char} >> 10$), hard spectrum ($kT
\simeq 5-8$ keV) flares \citep[e.g.,][]{Tsuboi98.1, Grosso04.1,
Feigelson05.1, Favata05.1}.  However, $\theta^2\,$Ori\,A is only
weakly variable over the 13-day COUP observation: no flare is seen
and the luminosity ratio between the highest segment and the
characteristic level is $L_{\rm peak}/L_{\rm char} = 1.8$. 
In the earlier $Chandra$ ACIS study of the ONC of
\citet{Feigelson02.1}, $\theta^2\,$Ori\,A exhibited multiple
flares with peak $\log L_t \simeq 31.8$ erg s$^{-1}$ superposed on
a relatively constant component. This was the most dramatic
variation then known from an OB stellar system and can be
attributed either to rapidly changing large-scale shocks in a
magnetically confined stellar wind of the massive primary, or to
unusually powerful magnetic reconnection flares in a lower mass
companion.

The spectrum is well-fit by a two-temperature plasma dominated by
the soft component. The absorption column derived from the X-ray
spectrum is compatible with the optical extinction. We thus
believe the emission arises from the massive primary wind and not
a MA secondary.  Note that our individually constructed X-ray
spectral fit differs from that obtained by \citet{Getman05.1} from
a semi-automated procedure.  They reported a much lower absorbing
column density, and hence lower value of the absorption-corrected
$\log L_{\rm t,c}$.

\subsection{$\theta^1$\,Ori\,A -- COUP\,745}\label{subsect:745}

$\theta^1$\,Ori\,A is at least a triple system with a 16 M$_\odot$
B0.5 primary \citep{Schertl03.1}.  The nearer companion has an
eclipsing 65\,day orbit; models of the optical lightcurve suggests
a pre-main sequence $\sim 3$ M$_\odot$ $\sim$A0 star
\citep{Lloyd99.1}. The distant companion is a 4 M$_\odot$ late-B
star with $0.2$\,\arcsec\/ separation \citep{Petr98.1, Weigelt99.1,
Preibisch99.1}.  The system is a highly variable radio source with
spectral and polarization properties characteristic of non-thermal
flaring \citep{Felli93.1}.

$\theta^1$\,Ori\,A is a strong X-ray source which is affected by
photon pile-up in the ACIS detector. During the 13-day COUP
observation, it exhibited a nearly-constant `characteristic' level
of $\log L_{\rm t,c} \simeq 32.4$ erg s$^{-1}$ superposed with
several flares with peak luminosities around $31.9-32.6$ erg
s$^{-1}$.  A range of plasmas with energies from 0.7 to $>$2 keV
are present.

\subsection{NU\,Ori -- COUP\,1468}\label{subsect:1468}

NU Ori is at least a hierarchical triple binary with a
14\,M$_\odot$ B1 primary, a $\simeq 3$\,M$_\odot$ spectroscopic
companion with period of 8\,days \citep{Morrell91.1}, and a
$\simeq 1$\,M$_\odot$ companion with separation $0.47$\,\arcsec\,
\citep{Preibisch99.1}. The COUP X-ray luminosity is $\log L_{\rm
t,c} = 31.2$ erg s$^{-1}$.  This source was located 8\arcmin\/
off-axis where the $Chandra$ point spread function is degraded, so
the $0.4$\,\arcsec\/ offset between the COUP and primary component
positions can be reconciled with emission from any of the
components. However, the $L_{\rm x}/L_{\rm bol}$ level is that of
a typical wind X-ray source, and NU\,Ori showed no sign of
variability during the 13-day observation.  Together, these characteristics 
suggest that the observed emission comes from the primary and can be 
explained within the standard scenario of stellar winds.

\subsection{$\theta^1$\,Ori\,B -- COUP\,766+778}\label{subsect:778_766}

This system is at least a quintet. Near-IR images of the system are 
provided by \citet{Simon99.1} and by \citet{Schertl03.1}. 
The brightest member is the B2.5
component $\theta^1$ Ori B East with mass 6.3 M$_\odot$
\citep{Prato02.1}.  It is an eclipsing binary (hence the variable
star designation BM Ori). The secondary is a rapidly rotating
pre-main sequence star with large IR excess, estimated spectral
type of G2 \citep{Popper76.1, Antokhina89.1, Vitrichenko00.1}, and
mass of $\sim 2.7$\,${\rm M_\odot}$ \citep{Palla01.1}.
$\theta^1$\,Ori\,B West lies $0.94$\,\arcsec\/ to the southwest and is
a resolved binary with $0.14$\,\arcsec\/ separation. The estimates for
the masses of these two components range between
$4-1.6$\,M$_\odot$ and $3-0.7$\,M$_\odot$, respectively
\citep{Schertl03.1, Preibisch99.1}. The faintest known member of
the quintet has $K$-band magnitude 10.5 \citep{Simon99.1} with
estimated mass 0.2~M$_\odot$ \citep{Preibisch99.1}, and lies $\sim
0.6^{\prime\prime}$ northwest of $\theta^1$\,Ori\,B\,East.

$\theta^1$\,Ori\,B is the only visual multiple that is resolved
into multiple COUP sources, although several dozen COUP doubles
with separations $< 3$\arcsec\/ are seen in the MA population
\citep{Getman05.2}.  The COUP image of the region, shown in
Figure~\ref{fig:binary_images}, shows two X-ray sources:
$\theta^1$\,Ori\,B\,East (COUP\,778) and West (COUP\,766).
About 1/3 of the events from each source is retrieved with the
{\it ACIS Extract} procedure to reduce cross-contamination of the
point spread functions. However, some residual contamination of
the luminous flare in the brighter source~766 is seen in the
lightcurve of the weaker source~778 (Figure~\ref{fig:lcs}).

$\theta^1$Ori\,B\,East (COUP\,778) with $\log L_{\rm
t,c}=30.1$ erg s$^{-1}$ has an unusually soft spectrum without
strong variability.  The emission is thus plausibly attributed to
wind shocks of the B2.5 primary, rather than to reconnection
flares in its G2-type companion.

In contrast, $\theta^1$\,Ori\,B\,West (COUP\,766) has all the
characteristics of a magnetically active lower-mass member of the
Orion population \citep{Wolk05.1, Favata05.1}.  It shows several
flares including a very powerful one exhibiting rapid rise with a
hard spectrum, peak luminosity around $\log L_{\rm t} \simeq 31.5$ erg
s$^{-1}$, and slower decay over several hours with a softening
spectrum.  It is not clear which of the components of the
$\theta^1$\,Ori\,B\,West binary is the principal X-ray emitter.

\section{New X-ray Detection of Par\,1772 ( = COUP\,349)}\label{sect:crowded}

The high spatial resolution of {\em Chandra} combined with the
high sensitivity achieved during the extraordinary long exposure
of the COUP has brought forth a new X-ray detected B1.5 star,
Par\,1772.  It is sometimes considered a Herbig Be star due to
mid-IR excess in the $IRAS$ satellite data and optical
reddening.  However, it does not exhibit emission lines indicating
accretion nor near-IR photometric excess emission indicating a warm
inner disk, as seen in other Herbig Be stars \citep{Manoj02.1}. It
may have an annular disk. No companions were seen in speckle
imaging \citep{Preibisch99.1}.

In X-rays, the star was reported at a level of $\log L_{\rm s} \simeq
30.2$ erg s$^{-1}$ from $ROSAT$ HRI observations in 1991-92
\citep{Caillault94.1}.  However, the offset was reported to be
$7$\,\arcsec\/ so it seems likely that the association was confused by
the three COUP sources within $\simeq 10$\,\arcsec (see
Fig.~\ref{fig:crowded_images}). The star was not detected in
1999-2000 $Chandra$ ACIS or HRC observations \citep{Feigelson02.1,
Flaccomio03.1}. During the COUP 2003 exposure, it appeared as a
weak and roughly constant source with $\log L_{\rm t,c} = 28.9$
erg $^{-1}$.  Together with $\theta^2$ Ori B, it has one of the
lowest \lglxlbol\, values ($=-7.9$) ever seen in a
stellar X-ray source.  We can not determine whether the emission
arises from the massive primary or an unseen companion.

\acknowledgments

We thank N. Schulz for valuable comments on the manuscript. 
We appreciate helpful suggestions of the referee M. Gagn\'e. 
COUP is supported by $Chandra$ Guest Observer grand SAO GO3-4009A
(E.\ Feigelson, PI). B.S., E.F., G.M. and S.S. acknowledge
financial support from the Italian MIUR FRIN program and an INAF
program for the years 2002-04.  E.D.F. received support from the
$Chandra$ ACIS Team contract NAS8-38252.

\clearpage
\newpage

%

\begin{deluxetable}{lrrlrrcccc}
\tabletypesize{\small} \tablewidth{0pt} \tablecolumns{10}

\tablecaption{Stellar parameters of early-type stars in the COUP field of the ONC \label{tab:sample}}

\tablehead{ \colhead{ID} & \colhead{$V$} & \colhead{$A_V$} &
\colhead{SpTy} & \colhead{$\log T_{\rm eff}$} &
\colhead{$\log L_{\rm bol}$} & \multicolumn{3}{c}{Multiplicity} & \\
\cline{7-9}

& \colhead{mag} & \colhead{mag} && \colhead{K} &
\colhead{$L_\odot$} & \colhead{SB?\tablenotemark{a}} &
\colhead{VB?\tablenotemark{b}} &
\colhead{Optical?\tablenotemark{c}} &
\colhead{Disk?\tablenotemark{d}}
 }

\startdata
\hline
\multicolumn{10}{c}{`Strong-Wind' Sample} \\
\hline
$\theta^1$\,Ori\,C & $5.140$ & $1.74$ &   O7 & $ 4.603$ & $  5.40$ &  $-$     & $\times$ & \nodata  & \nodata  \\
$\theta^2$\,Ori\,A & $5.070$ & $1.12$ & O9.5 & $ 4.543$ & $  5.03$ & $\times$ & $\times$ & \nodata  & \nodata  \\
$\theta^1$\,Ori\,A & $6.730$ & $1.89$ &   B0 & $ 4.471$ & $  4.50$ & $\times$ & $\times$ & \nodata  & \nodata  \\
$\theta^1$\,Ori\,D & $6.700$ & $1.79$ & B0.5 & $ 4.420$ & $  4.35$ &  $-$     & $-$      & \nodata  & \nodata  \\
NU\,Ori            & $6.840$ & $2.09$ &   B1 & $ 4.383$ & $  4.33$ & $\times$ & $\times$ & \nodata  & $\times$ \\
$\theta^2$\,Ori\,B & $6.410$ & $0.73$ &   B1 & $ 4.383$ & $  3.96$ &  $-$     & $-$      & \nodata  & \nodata  \\
Par\,1772          & $8.370$ & $1.47$ &   B2 & $ 4.294$ & $  3.27$ & \nodata  & $-$      & \nodata  & $\times$ \\
JW\,660            & $9.620$ & $2.44$ &   B3 & $ 4.272$ & $  3.11$ & \nodata  & \nodata  & \nodata  & \nodata  \\
$\theta^1$\,Ori\,B & $7.260$ & $0.49$ &   B3 & $ 4.272$ & $  3.28$ & \nodata  & $\times$ & $\times$ & $?$      \\
\hline
\multicolumn{10}{c}{`Weak-Wind' Sample} \\
\hline
$\theta^2$\,Ori\,C & $8.240$ & $0.92$ &   B5 & $ 4.140$ & $  2.79$ &  $?$     & \nodata  & \nodata  & $\times$ \\
$\theta^1$\,Ori\,E & $-$     & $-$    &   B5 & $ 4.140$ & $  2.42$ & \nodata  & \nodata  & \nodata  & \nodata  \\
$\theta^1$\,Ori\,F & $-$     & $-$    &   B8 & $ 4.053$ & $  1.84$ & \nodata  & $-$      & \nodata  & \nodata  \\
JW\,197            & $10.16$ & $0.85$ &   B9 & $ 4.025$ & $  1.69$ &  $-$     & \nodata  & \nodata  & \nodata  \\
JW\,153            & $8.920$ & $0.13$ &   B9 & $ 4.025$ & $  1.89$ &  $-$     & \nodata  & \nodata  & $\times$ \\
JW\,831            & $9.470$ & $0.00$ &   A0 & $ 3.993$ & $  1.54$ &  $-$     & \nodata  & \nodata  & \nodata  \\
JW\,531            & $10.55$ & $1.56$ &   A1 & $ 3.975$ & $  1.71$ & \nodata  & \nodata  & \nodata  & $\times$ \\
JW\,108            & $10.19$ & $2.07$ &   A3 & $ 3.940$ & $  2.01$ &  $-$     & \nodata  & \nodata  & $?$      \\
JW\,608            & $11.89$ & $1.29$ &   A5 & $ 3.917$ & $  1.00$ & \nodata  & \nodata  & \nodata  & \nodata  \\
JW\,165            & $13.55$ & $4.20$ &   A7 & $ 3.897$ & $  1.49$ & \nodata  & \nodata  & \nodata  & \nodata  \\
JW\,599            & $19.20$ & $10.7$ &   A9 & $ 3.867$ & $  1.86$ & \nodata  & \nodata  & \nodata  & $?$      \\
\enddata

\tablenotetext{a}{Crosses indicate variable radial velocity or confirmed
spectroscopic binary, dashes indicate constant radial velocity \citep{Frost26.1, Abt91.1, Morrell91.1}.}

\tablenotetext{b}{Crosses indicate visual companions identified in speckle
surveys \citep{Weigelt99.1, Preibisch99.1, Petr98.1, Simon99.1,
Mason98.1}, dashes indicate negative results.}

\tablenotetext{c}{Optical binaries identified by eclipses are indicated by crosses
\citep{Antokhina89.1}.}

\tablenotetext{d}{Crosses mark HAeBe candidates \citep{The94.1,
Tovmassian97.1}.  `?' mark stars with $L$-band excess in the
near-IR color magnitude diagram (Figure~\ref{fig:jhk}).}

\end{deluxetable}

\clearpage
\newpage

\begin{deluxetable}{rllcrrrrcc}
\tabletypesize{\small} \tablewidth{0pt} \tablecolumns{10}

\tablecaption{COUP X-ray properties of early-type stars in the ONC
\label{tab:xparams}}

\tablehead{ \colhead{COUP} & \colhead{ID} & \colhead{SpTy} &
\colhead{Offaxis} & \colhead{$\Delta_{\rm ix}$} & \colhead{NetCts}
& \colhead{PSF}  & \colhead{Exp} & \colhead{HR} & \colhead{FBG02} \\

\colhead{\#}  &&& \colhead{\arcmin} & \colhead{\arcsec} &&
\colhead{frac} & \colhead{ks} && \# }

\startdata
\hline
\multicolumn{10}{c}{`Strong-Wind' Sample} \\
\hline
  809 & $\theta^1$\,Ori\,C &    O7 &  0.32 &  0.05 & 63444 & 0.01  &  834.7 & $0.26 \pm 0.00$ &  542  \\
 1232 & $\theta^2$\,Ori\,A &  O9.5 &  1.95 &  0.15 &  3525 & 0.01  &  827.6 & $0.09 \pm 0.01$ &  828  \\
  745 & $\theta^1$\,Ori\,A &    B0 &  0.53 &  0.13 & 21283 & 0.02  &  834.7 & $0.34 \pm 0.01$ &  498  \\
  869 & $\theta^1$\,Ori\,D &  B0.5 &  0.39 &  0.12 &  7901 & 0.87  &  832.9 & $0.04 \pm 0.00$ &  584  \\
 1468 & NU\,Ori            &    B1 &  8.41 &  0.40 &  2237 & 0.67  &  728.6 & $0.03 \pm 0.00$ &  996  \\
 1360 & $\theta^2$\,Ori\,B &    B1 &  2.69 &  0.15 &  1293 & 0.86  &  767.5 & $0.21 \pm 0.02$ &  916  \\
  349 & Par\,1772          &    B2 &  4.58 &  0.08 &   127 & 0.87  &  758.6 & $0.18 \pm 0.05$ &\nodata\\
 1116 & JW\,660            &    B3 &  2.13 &  0.07 & 52865 & 0.87  &  818.7 & $0.65 \pm 0.01$ &  746  \\
  778 & $\theta^1$\,Ori\,BE&    B3 &  0.60 &  0.13 &  1506 & 0.37  &  832.9 & $0.17 \pm 0.02$ &  519  \\
\hline
\multicolumn{10}{c}{`Weak-Wind' Sample} \\
\hline
 1473 & $\theta^2$\,Ori\,C &    B5 &  3.92 &  0.17 & 26525 & 0.86  &  763.9 & $0.38 \pm 0.01$ &  995  \\
  732 & $\theta^1$\,Ori\,E &    B5 &  0.59 &  0.11 & 37523 & 0.04  &  834.7 & $0.34 \pm 0.01$ &  495  \\
      & $\theta^1$\,Ori\,F &    B8 &\nodata&\nodata& $<29$ &\nodata&  822.5 & \nodata         &\nodata\\
  142 & JW\,197            &    B9 &  4.50 &  0.18 &  5311 & 0.87  &  811.7 & $0.29 \pm 0.01$ &  103  \\
  100 & JW\,153            &    B9 &  8.61 &  0.38 &   679 & 0.89  &  594.2 & $0.14 \pm 0.02$ &  ~70  \\
 1415 & JW\,831            &    A0 &  3.90 &  0.08 &   866 & 0.87  &  801.1 & $0.08 \pm 0.01$ &  952  \\
      & JW\,531            &    A1 &\nodata&\nodata& $<20$ &\nodata&  815.9 & \nodata         &\nodata\\
      & JW\,108            &  A3.5 &\nodata&\nodata& $<66$ &\nodata&  454.9 & \nodata         &\nodata\\
      & JW\,608            &    A5 &\nodata&\nodata& $<40$ &\nodata&  801.2 & \nodata         &\nodata\\
  113 & JW\,165            &    A7 &  9.44 &  0.32 &  6724 & 0.89  &  212.2 & $0.32 \pm 0.01$ &  ~81  \\
  995 & JW\,599            &    A9 &  6.20 &  0.16 &   214 & 0.87  &  767.5 & $0.75 \pm 0.13$ &  670  \\
\enddata

\end{deluxetable}

\clearpage
\newpage

~~

\begin{deluxetable}{rllcrrrccclccc}
\tabletypesize{\scriptsize}  \tablewidth{0pt} \tablecolumns{14}
\rotate

\tablecaption{COUP X-ray spectral parameters of early-type stars
in the ONC \label{tab:spec}}

\tablehead{
\colhead{COUP} & \colhead{ID} & \colhead{SpTy} &
\colhead{$\log N_{\rm H}$} & \colhead{$kT_1$} & \colhead{$kT_2$} &
\colhead{$kT_3$} & \colhead{$\log{EM_1}$} & \colhead{$\log{EM_2}$}
& \colhead{$\log{EM_3}$}& \colhead{$\chi^2_\nu$ (dof)} &
\colhead{$\log{L_{\rm t}}$} & \colhead{$\log{L_{\rm t,c}}$} &
\colhead{$\log{(L_{\rm t,c}/L_{\rm bol})}$} \\

\colhead{\#} &&& \colhead{cm$^{-2}$} & \colhead{keV} &
\colhead{keV}  & \colhead{keV}  & \colhead{cm$^{-3}$} &
\colhead{cm$^{-3}$} & \colhead{cm$^{-3}$} && \colhead{erg
s$^{-1}$} & \colhead{erg s$^{-1}$} &
}

\startdata
\multicolumn{14}{c}{`Strong-Wind' Sample} \\\hline
  809 & $\theta^1$\,Ori\,C & O7  & 21.2 &  0.59 &  1.38 &\nodata&  55.8 &  56.2 &\nodata&  5.59 ( 219) & 33.1 & 33.3 &   -5.7 \\
 1232 & $\theta^2$\,Ori\,A & O9.5& 21.5 &  0.23 &  2.45 &\nodata&  55.8 &  54.1 &\nodata&  0.83 (  44) & 31.8 & 32.5 &   -6.1 \\
  745 & $\theta^1$\,Ori\,A & B0  & 20.9 &  0.66 &  1.68 & $>$15 &  54.9 &  55.3 &  54.3 &  1.96 ( 153) & 32.3 & 32.4 &   -5.7 \\
  869 & $\theta^1$\,Ori\,D & B0.5& 21.5 &  0.23 &  0.58 &\nodata&  53.8 &  53.4 &\nodata&  2.05 (  61) & 30.2 & 30.7 &   -7.2 \\
 1468 & NU\,Ori            & B1  & 21.9 &  0.22 & $>$15 &\nodata&  54.5 &  51.7 &\nodata&  1.70 (  25) & 29.8 & 31.2 &   -6.7 \\
 1360 & $\theta^2$\,Ori\,B & B1  & 21.2 &  0.85 &  2.88 &\nodata&  52.4 &  52.3 &\nodata&  1.15 (  13) & 29.5 & 29.6 &   -7.9 \\
  349 & Par\,1772          & B2  & 21.7 &  0.44 &  2.58 &\nodata&  51.9 &  51.2 &\nodata&  1.04 (  13) & 28.4 & 28.9 &   -7.9 \\
 1116 & JW\,660            & B3  & 21.7 &  0.97 &  3.33 &\nodata&  53.8 &  54.3 &\nodata&  1.85 ( 263) & 31.2 & 31.4 &   -5.3 \\
  778 & $\theta^1$\,Ori\,BE& B3  & 21.4 &  0.63 &  2.26 &\nodata&  53.0 &  52.7 &\nodata&  0.70 (  16) & 29.9 & 30.1 &   -6.7 \\
&&&&&&&&&&&&& \\
\multicolumn{14}{c}{`Weak-Wind' Sample} \\
 1473 & $\theta^2$\,Ori\,C & B5  & 21.4 &  0.86 &  2.86 &\nodata&  53.4 &  53.9 &\nodata&  1.88 ( 166) & 30.9 & 31.0 &   -5.4 \\
  732 & $\theta^1$\,Ori\,E & B5  & 21.1 &  0.64 &  1.58 & $>$15 &  54.8 &  55.3 &  54.2 &  2.58 ( 190) & 32.3 & 32.4 &   -3.6 \\
  142 & JW\,197            & B9  & 21.0 &  0.81 &  3.06 &\nodata&  52.6 &  53.0 &\nodata&  1.08 (  61) & 30.1 & 30.2 &   -5.1 \\
  100 & JW\,153            & B9  & 20.8 &  0.66 &  1.39 &\nodata&  52.1 &  52.2 &\nodata&  0.83 (  29) & 29.3 & 29.4 &   -6.1 \\
 1415 & JW\,831            & A0  & 21.3 &  0.26 &  1.00 &\nodata&  52.6 &  52.2 &\nodata&  1.11 (  31) & 29.2 & 29.6 &   -5.6 \\
  113 & JW\,165            & A7  & 21.4 &  0.87 &  2.41 &\nodata&  53.3 &  53.7 &\nodata&  1.54 (  80) & 30.7 & 30.9 &   -4.2 \\
  995 & JW\,599            & A9  & 22.0 &  1.89 &\nodata&\nodata&  52.2 &\nodata&\nodata&  0.71 (  25) & 28.8 & 29.2 &   -6.3 \\
\enddata

\end{deluxetable}

\clearpage
\newpage

\begin{deluxetable}{lllcccrr}
\tabletypesize{\small}  \tablewidth{0pt} \tablecolumns{8}

\tablecaption{X-ray variability of strong-wind (SW) and weak-wind (WW) stars in the ONC\label{tab:variab}}

\tablehead{ \colhead{COUP \#} & \colhead{ID} & \colhead{SpTy} &
\colhead{Prob KS} & \colhead{Prob MLB} & \colhead{$N_{\rm F}$} &
\colhead{$\frac{t_{\rm char}}{t_{\rm total}}$} &
\colhead{$\frac{R_{\rm peak}}{R_{\rm char}}$}}

\startdata
\multicolumn{8}{c}{`Strong-Wind' Sample} \\
     809 & $\theta^1$\,Ori\,C  & O7  & $\surd$ & $\surd$ & 0 & 1.0 & 1.2 \\
    1232 & $\theta^2$\,Ori\,A  & O9.5& $\surd$ & $\surd$ & 0 & 1.0 & 1.8 \\
     745 & $\theta^1$\,Ori\,A  & B0  & $\surd$ & $\surd$ & 2 & 0.9 & 2.9 \\
     869 & $\theta^1$\,Ori\,D  & B0.5& $-$     & $-$     & 0 & 1.0 & 1.0 \\
    1468 & NU\,Ori             & B1  & $-$     & $-$     & 0 & 1.0 & 1.0 \\
    1360 & $\theta^2$\,Ori\,B  & B1  & $\surd$ & $\surd$ & 1 & 0.9 & 4.2 \\
     349 & Par\,1772           & B2  & $\surd$ & $\surd$ & 0 & 1.0 & 1.4 \\
    1116 & JW\,660             & B3  & $\surd$ & $\surd$ & 0 & 0.9 & 1.5 \\
     778 & $\theta^1$\,Ori\,BE & B3  & $-$     & $\surd$ & 1 & 1.0 & 3.3 \\
\multicolumn{8}{c}{`Weak-Wind' Sample} \\
    1473 & $\theta^2$\,Ori\,C  & B5  & $\surd$ & $\surd$ & 3 & 0.7 & 3.0 \\
     732 & $\theta^1$\,Ori\,E  & B5  & $\surd$ & $\surd$ & 1 & 0.9 & 1.9 \\
     142 & JW\,197             & B9  & $\surd$ & $\surd$ & 4 & 0.8 &51.3 \\
     100 & JW\,153             & B9  & $\surd$ & $\surd$ & 2 & 0.9 &11.8 \\
    1415 & JW\,831             & A0  & $\surd$ & $\surd$ & 0 & 1.0 & 1.6 \\
     113 & JW\,165             & A7  & $\surd$ & $\surd$ & 1 & 0.9 & 3.8 \\
     995 & JW\,599             & A9  & $-$     & $\surd$ & 0 & 1.0 & 1.6 \\
\enddata

\end{deluxetable}

\newpage

\begin{deluxetable}{lccc}
\tabletypesize{\small}  \tablewidth{0pt} \tablecolumns{4}

\tablecaption{Comparison of flares on SW, WW and solar-type stars in the ONC \label{tab:flare_comp}}

\tablehead{ Parameter & \multicolumn{3}{c}{Sample} \\ \cline{2-4}
& SW+WW & WW & 1 M$_\odot$ }

\startdata
$N_{\rm *}$                             &  9~~  &  7~~ & 28~~ \\
$N_{\rm f,total}$                       & 15~~  & 11~~ & 41~~ \\
$N_{\rm f,spec}$                        & 12~~  &  9~~ & 37~~ \\
Median duration          (ks)           & 13~~  & 18~~ & 65~~ \\
Median $\log{L_{\rm F}}$ (erg s$^{-1}$) & 32.3  & 31.2 & 30.8 \\
Median $\log{E_{\rm F}}$ (erg)          & 36.2  & 35.4 & 35.5
\enddata

\tablecomments{Numbers for $1$\,M$_\odot$ stars from \citet{Wolk05.1};
$N_{\rm *}$ = number of stars; $N_{\rm f,total}$ = total number of
flares; $N_{\rm f,spec}$ = number of flares with spectral
information}

\end{deluxetable}

\newpage

\begin{deluxetable}{lrrrrr}
\tabletypesize{\small}  \tablewidth{0pt} \tablecolumns{6}

\tablecaption{Periodic variability of SW and WW stars\label{tab:rotmod}}

\tablehead{
\colhead{Name} & \colhead{$P_{\rm opt}$} &
\colhead{$P_{\rm X}$} & \colhead{FAP} & \colhead{Median} &
\colhead{Ampl} \\

& \colhead{d} & \colhead{d} && \colhead{cts ks$^{-1}$} &
\colhead{cts ks$^{-1}$} }

\startdata
$\theta^1$\,Ori\,C & 15.42 & 15.51 & 0.000 & 72.6~~~ & $93.9 \pm 4.2$ \\
JW\,660            &  6.15 &  6.67 & 0.001 & 61.3~~~ & $30.7 \pm 6.6$ \\
JW\,165            &  5.77 &  4.61 & 0.001 &  7.8~~~ & $ 6.1 \pm 3.8$ \\
\enddata
\end{deluxetable}

\begin{deluxetable}{lcccrrccccc}
\tabletypesize{\small}  \tablewidth{0pt} \tablecolumns{11}

\tablecaption{Multiple systems with early-type primary in the COUP field of the ONC \label{tab:binaries}}

\tablehead{
\colhead{Name} & \colhead{Type} & \multicolumn{4}{c}{Component separation} &&
\multicolumn{2}{c}{Masses (M$_\odot$)} & \colhead{Ref.} & \colhead{COUP} \\

&& \colhead{Comp} & \colhead{Sep (mas)} & \colhead{PA ($^\circ$)} & \colhead{Epoch} &&
\colhead{Pri} & \colhead{Sec} && \#
}

\startdata
$\theta^1$\,Ori\,C & VB  & C1-C2  & $38 \pm 2$   & 208   & 2001  && 45 & $\geq 6$ & \tablenotemark{a} &  809 \\
&&&&&&&&&& \\
$\theta^2$\,Ori\,A & VB  & A1-A2  & $383 \pm 10$ & 291   & 1997  && 25 & $\sim 7$ & \tablenotemark{b} & 1232 \\
$\theta^2$\,Ori\,A & SB  & A1-A3  & 100          &\nodata&\nodata&& 25 & $\sim 9$ & \tablenotemark{b} &      \\
&&&&&&&&&& \\
$\theta^1$\,Ori\,A & VB  & A1-A2  & $215 \pm  3$ & 357   & 2001  && 16 & $\sim 4$ & \tablenotemark{a} &  745 \\
$\theta^1$\,Ori\,A & ecl & A1-A3  &              &       &       && 16 & $\sim 3$ & \tablenotemark{c} &      \\
&&&&&&&&&& \\
NU\,Ori             & VB  & A-B    & $471 \pm 17$ &  98   & 1997  && 14 & $\sim 1$ & \tablenotemark{b} & 1468 \\
NU\,Ori             & SB  & A-C    &  80          &\nodata&\nodata&& 14 & $\sim 3$ & \tablenotemark{b} &      \\
&&&&&&&&&& \\
$\theta^1$\,Ori\,B & VB  & B1-B2  & $942 \pm 10$ & 255   & 2001  && $7$      & $\sim 4;1.6$ & \tablenotemark{a} \tablenotemark{b} & 778+766 \\
$\theta^1$\,Ori\,B & VB  & B2-B3  & $117 \pm  4$ & 210   & 2001  && $\sim 4$ & $\sim 3;0.7$ & \tablenotemark{a} \tablenotemark{b} &  \\
$\theta^1$\,Ori\,B & VB  & B1-B4  & $609 \pm  8$ & 298   & 2001  && $7$      & $\sim 0.2$   & \tablenotemark{a} \tablenotemark{b} &  \\
$\theta^1$\,Ori\,B & SB  & B1-B5  & 30           &\nodata&\nodata&& $7$      & $\sim 2.7$   & \tablenotemark{b} \tablenotemark{d} &  \\
&&&&&&&&&& \\
$\theta^2$\,Ori\,C & SB  & C1-C2  & \nodata      &\nodata&\nodata&&\nodata   & \nodata      & \tablenotemark{e} & 1473
\enddata

\tablecomments{VB = visual binary, SB = spectroscopic binary, Comp = component,
Pri = primary, Sec = secondary, PA = position angle (degrees East of North)}

\tablenotetext{a}{\citet{Schertl03.1}}
\tablenotetext{b}{\citet{Preibisch99.1}}
\tablenotetext{c}{\citet{Lloyd99.1}}
\tablenotetext{d}{\citet{Palla01.1}}
\tablenotetext{e}{\citet{Corporon99.1}}

\end{deluxetable}

\newpage

\begin{deluxetable}{lllccccc}
\tabletypesize{\small}  \tablewidth{0pt} \tablecolumns{7}

\tablecaption{Tentative classification of early-type ONC X-ray sources\label{tab:class}}

\tablehead{\colhead{COUP \#} & \colhead{Name} & \colhead{SpTy} &
\colhead{Luminosity} & \colhead{Spec} & \colhead{Var} &
\colhead{Disk} & \colhead{Magn}}

\startdata
\hline
\multicolumn{8}{c}{`Strong-Wind' Sample} \\
\hline
 809   & $\theta^1$Ori\,C & O7   & $>$W  &\nodata& S     &\nodata & RM,B \\
1232   & $\theta^2$Ori\,A & O9.5 & W     & W     & W\tablenotemark{a}&\nodata & \\
 745   & $\theta^1$Ori\,A & B0   & $>$W  &\nodata& T     &\nodata & NT \\
 869   & $\theta^1$Ori\,D & B0.5 & W     & W     & W     &\nodata & \\
1468   & NU Ori           & B1   & W     & W     & W     & fIR    & \\
1360   & $\theta^2$Ori\,B & B1   & W     &\nodata& T     &\nodata & \\
 349   & Par 1772         & B2   & W     &\nodata&\nodata& fIR    & CP? \\
1116   & JW 660           & B3   & $>$W  &\nodata& S     &\nodata & RM \\
 778   &$\theta^1$Ori\,BE & B3   & W     &\nodata&\nodata& nIR    & \\
\hline
\multicolumn{8}{c}{`Weak-Wind' Sample} \\
\hline
 766   &$\theta^1$Ori\,BW & ?    &\nodata&\nodata& T     &\nodata & \\
1473   & $\theta^2$Ori\,C & B5   & T     &\nodata& T     & fIR    & \\
 732   & $\theta^1$Ori\,E & B5   & T     &\nodata& T     &\nodata & NT \\
\nodata& $\theta^1$Ori\,F & B8   & $-$   & $-$   & $-$   &\nodata & \\
 142   & JW 197           & B9   & T     &\nodata& T     &\nodata & \\
 100   & JW 153           & B9   & T     &\nodata& T     & fIR    & \\
1415   & JW 831           & A0   & T     &\nodata&\nodata&\nodata & \\
\nodata& JW 531           & A1   & $-$   & $-$     & $-$ & fIR    & \\
\nodata& JW 108           & A3   & $-$   & $-$     & $-$ & nIR    & \\
\nodata& JW 608           & A5   & $-$   & $-$     & $-$ &\nodata & \\
 113   & JW 165           & A7   & T     &\nodata& T/S   &\nodata & RM \\
 995   & JW 599           & A9   & T     &\nodata&\nodata& nIR    & \\
\enddata
\tablecomments{
Classes: W = hydrodynamic wind (many small shocks), S = smoothly variable lightcurve,
T = T Tauri type emission, nIR = near-infrared excess,  fIR = far-infrared excess, RM= rotational modulation,
B = detected magnetic field, CP = chemically peculiar, NT = non-thermal cm emission.}

\tablenotetext{a}{This star exhibited powerful flares during
the earlier $Chandra$ observation described by
\citet{Feigelson02.1}, and thus might also be classified "T"
here.}

\end{deluxetable}

\clearpage
\newpage

%

%

%

%

%
\begin{figure}
\begin{center}
\resizebox{15cm}{!}{\includegraphics{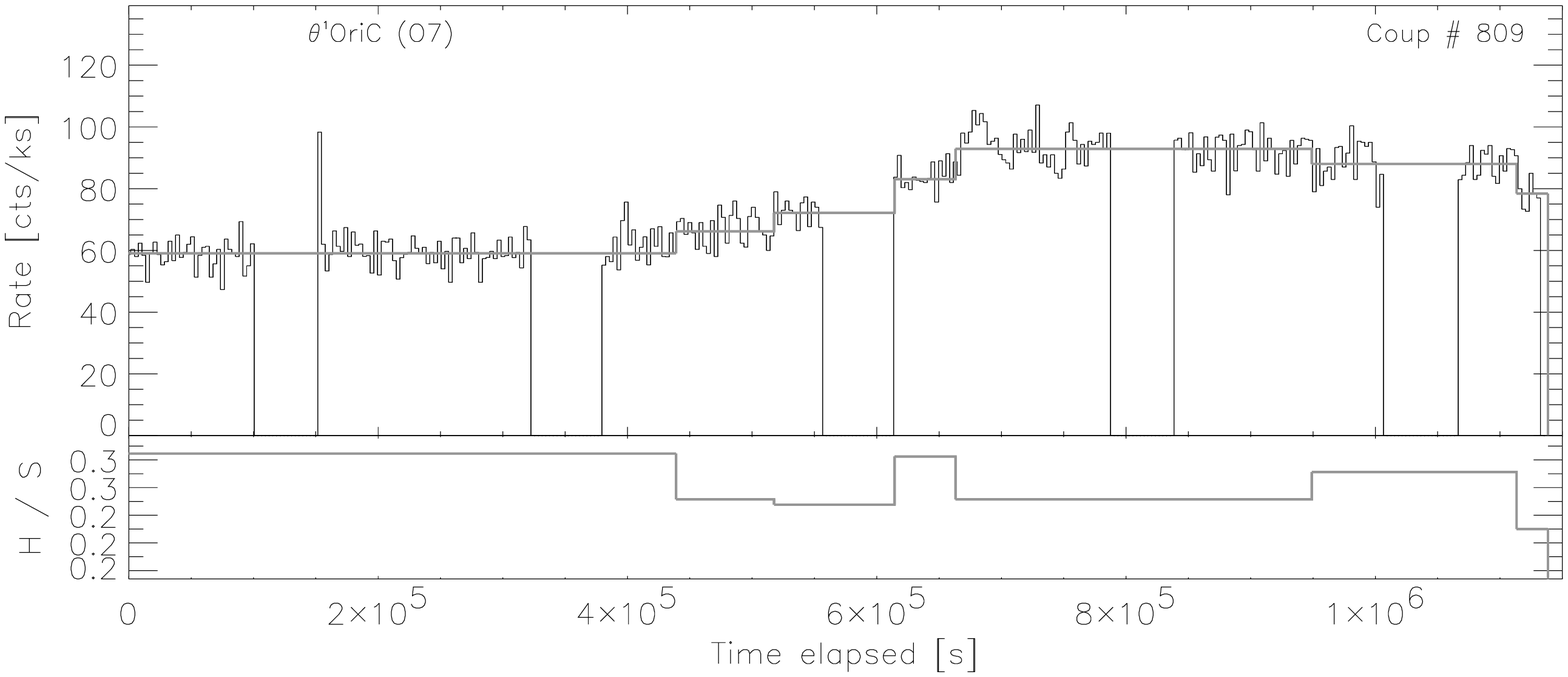}}
\resizebox{15cm}{!}{\includegraphics{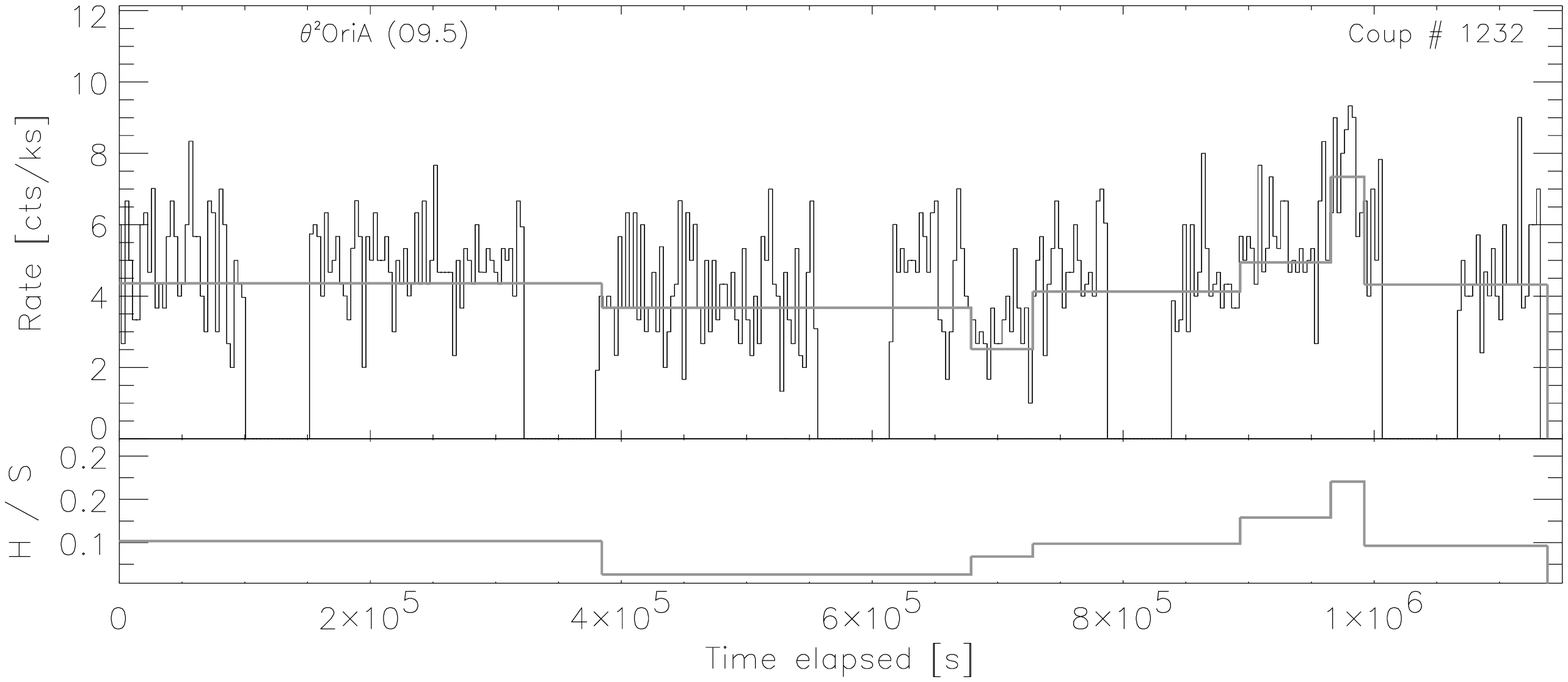}}
\resizebox{15cm}{!}{\includegraphics{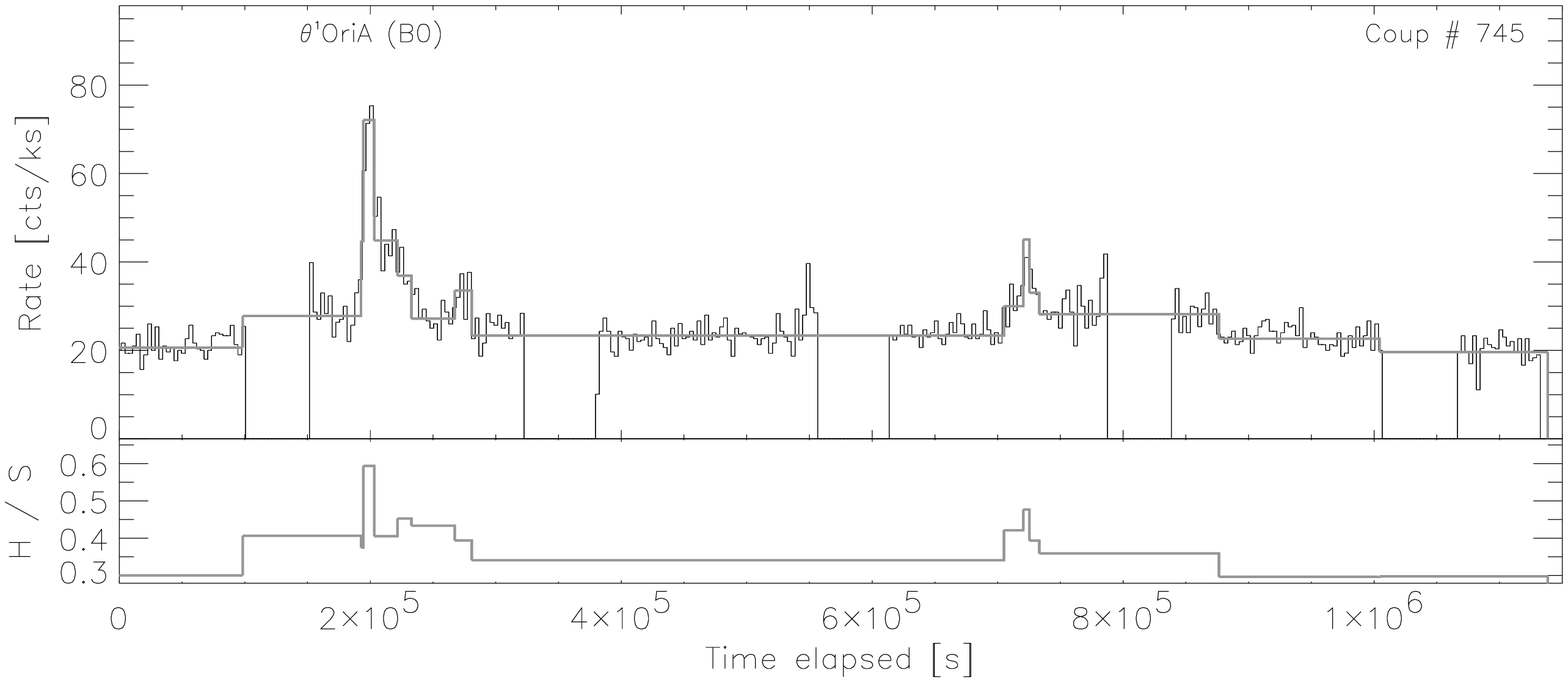}} \caption{Lightcurves 
of COUP count rate and hardness ratio for detected SW and WW stars 
in order of decreasing spectral type. 
Segments derived from the MLB analysis are overplotted with horizontal lines
onto the binned lightcurve. The hardness ratios have been 
evaluated in individual segments.} \label{fig:lcs}
\end{center}
\end{figure}

\addtocounter{figure}{-1}

\begin{figure}
\begin{center}
\resizebox{15cm}{!}{\includegraphics{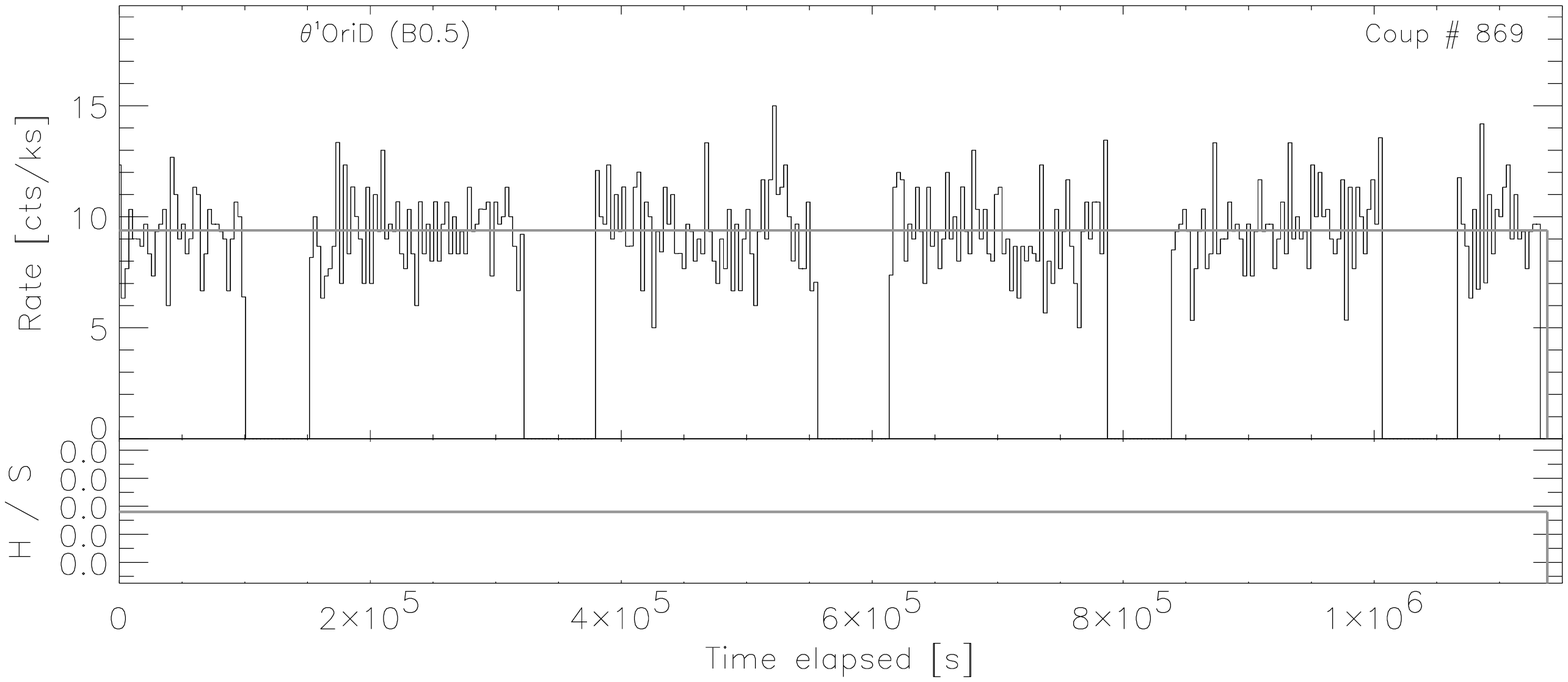}}
\resizebox{15cm}{!}{\includegraphics{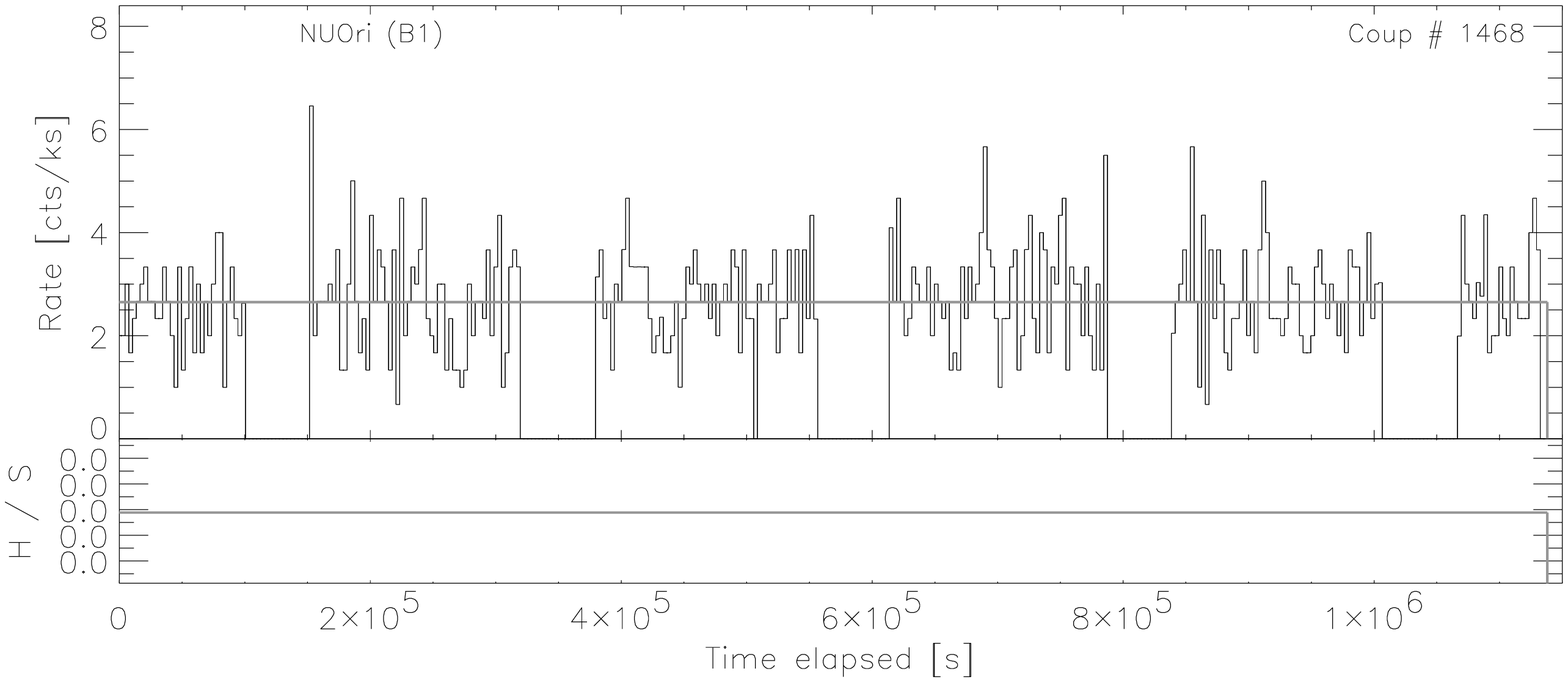}}
\resizebox{15cm}{!}{\includegraphics{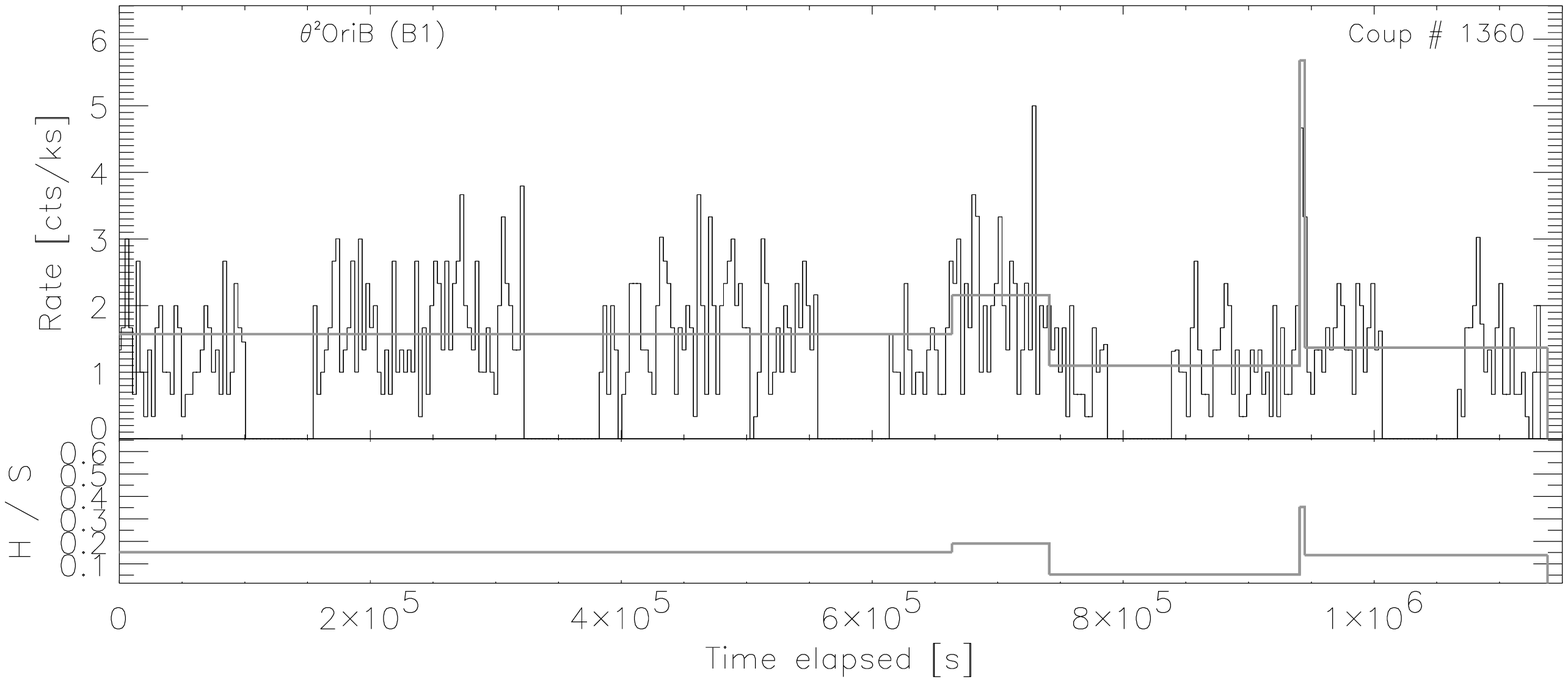}}
\resizebox{15cm}{!}{\includegraphics{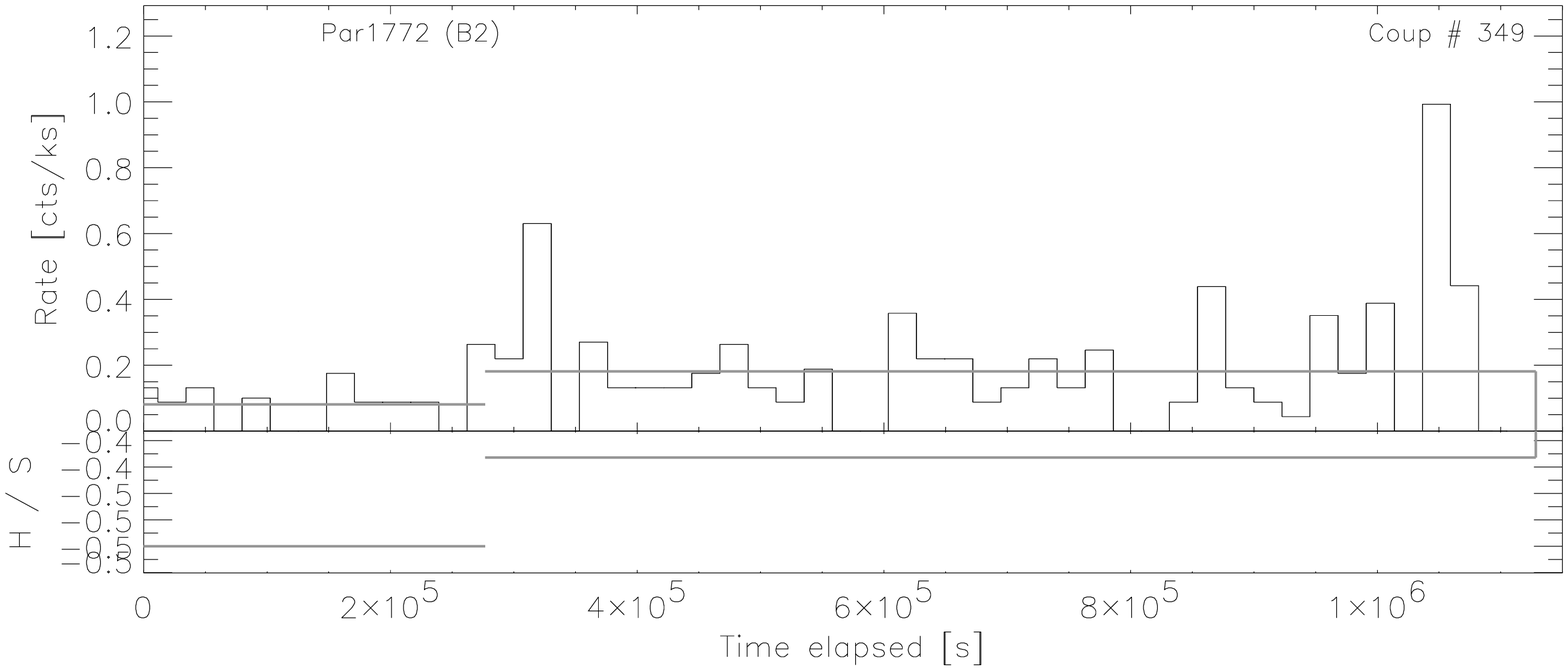}} \caption{{\em
continued}}
\end{center}
\end{figure}

\clearpage

\addtocounter{figure}{-1}

\begin{figure}
\begin{center}
\resizebox{15cm}{!}{\includegraphics{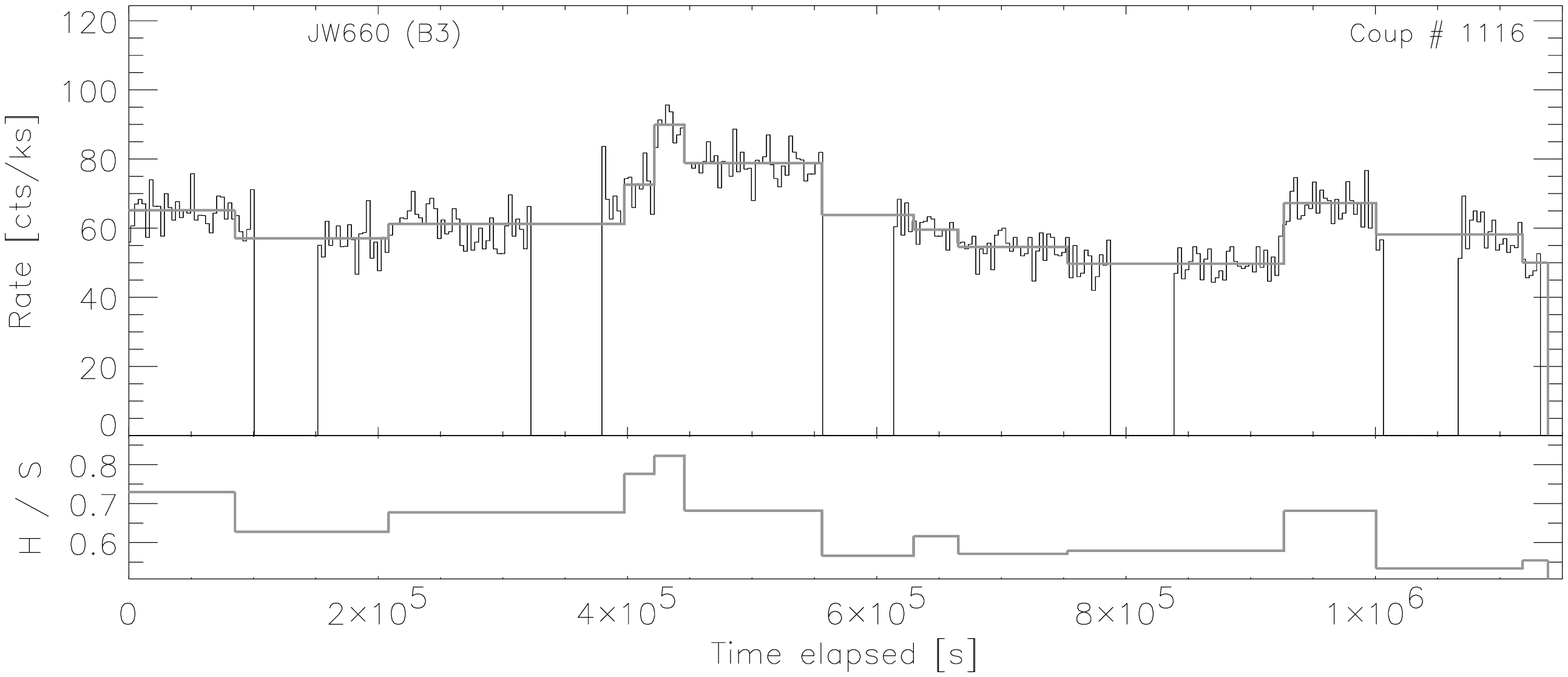}}
\resizebox{15cm}{!}{\includegraphics{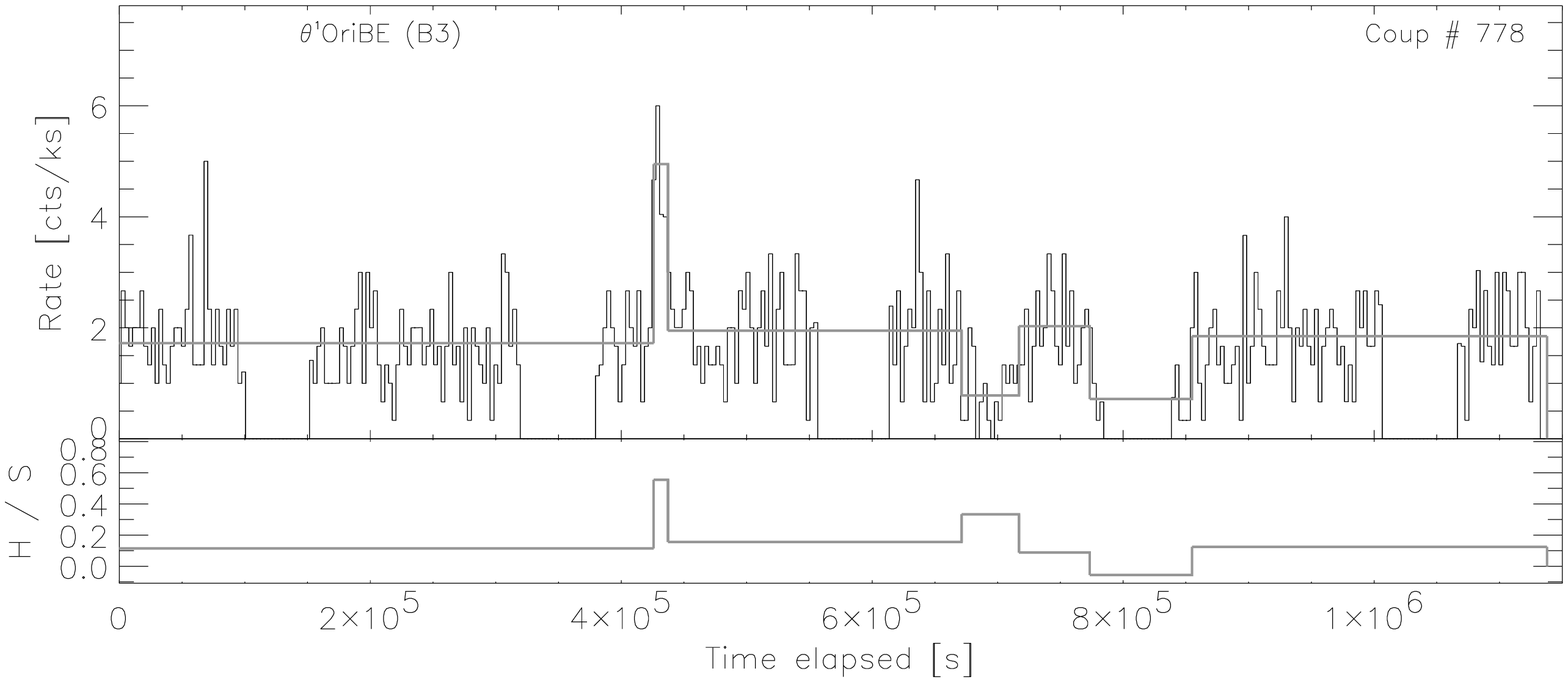}}
\resizebox{15cm}{!}{\includegraphics{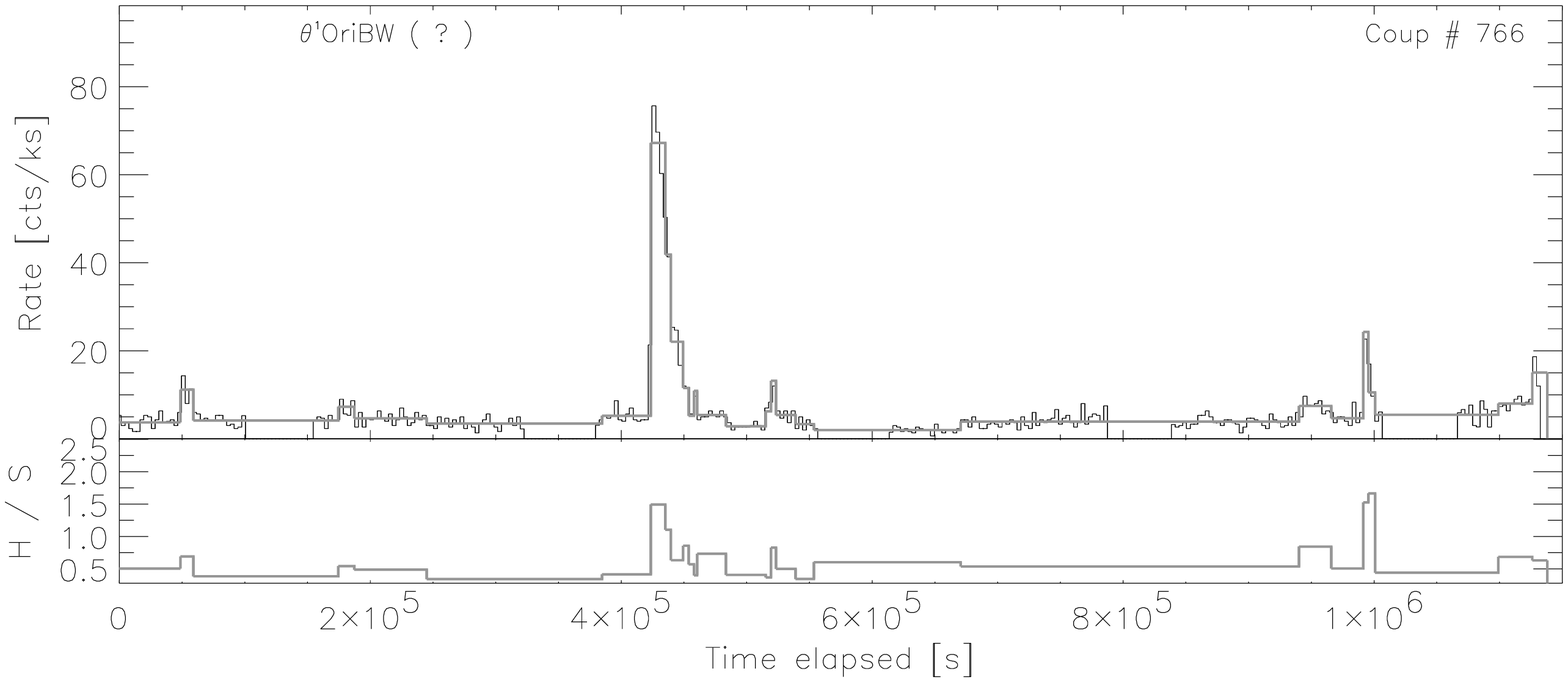}}
\resizebox{15cm}{!}{\includegraphics{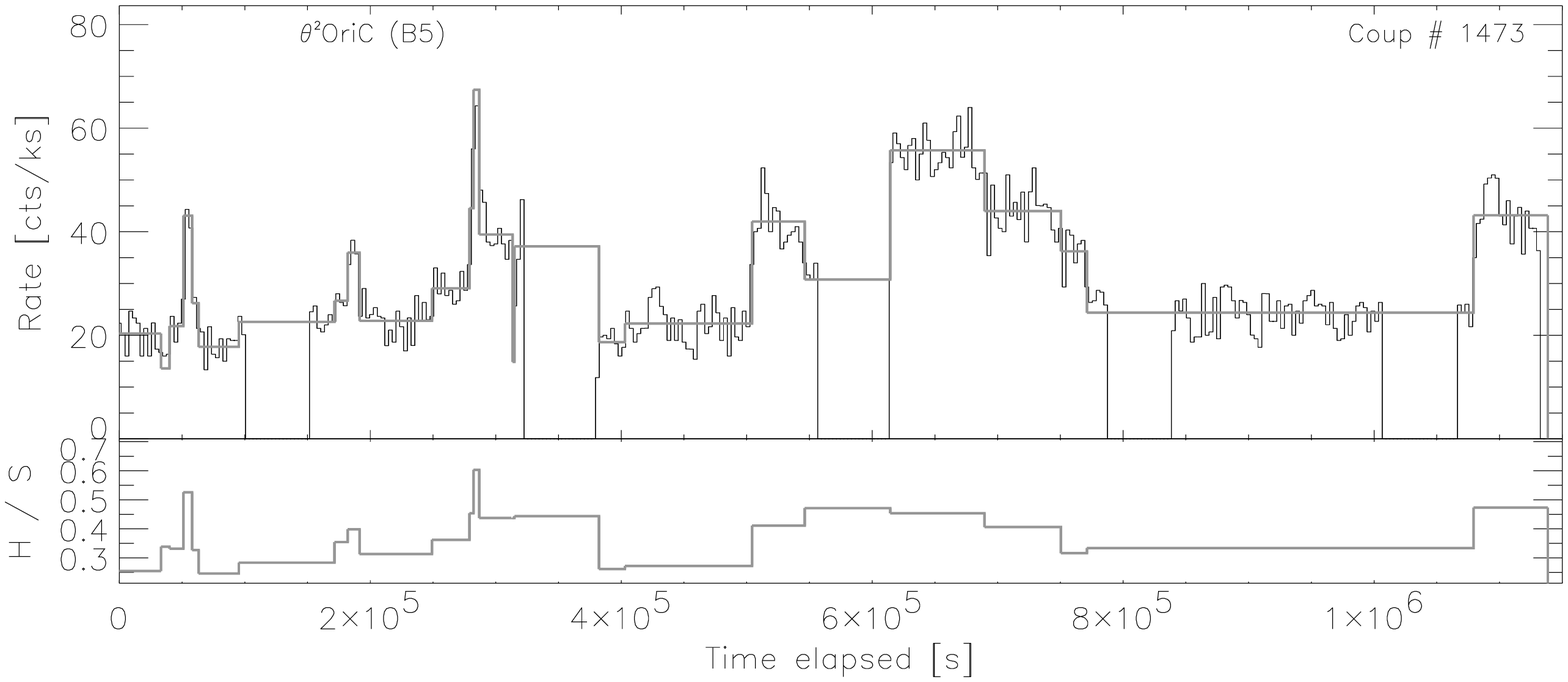}} \caption{{\em
continued}}
\end{center}
\end{figure}

\addtocounter{figure}{-1}

\begin{figure}
\begin{center}
\resizebox{15cm}{!}{\includegraphics{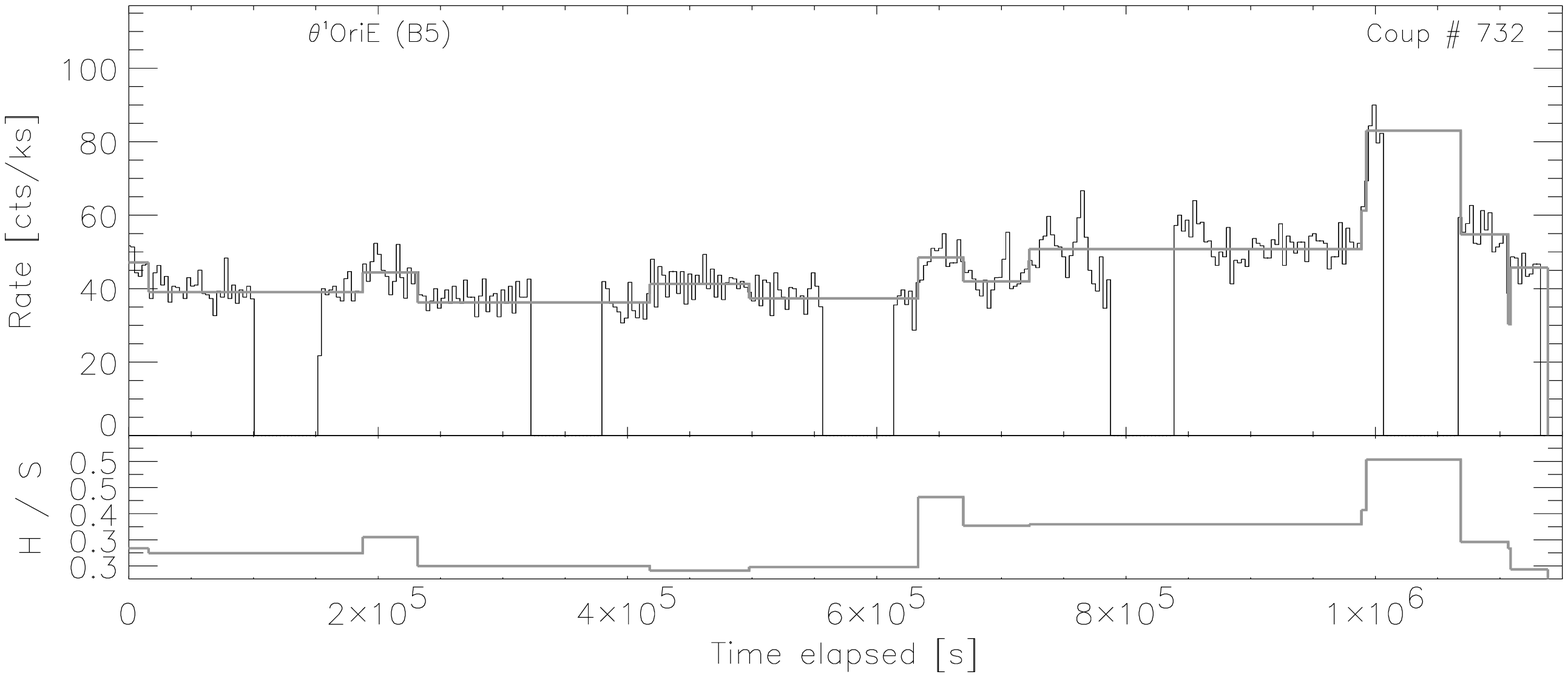}}
\resizebox{15cm}{!}{\includegraphics{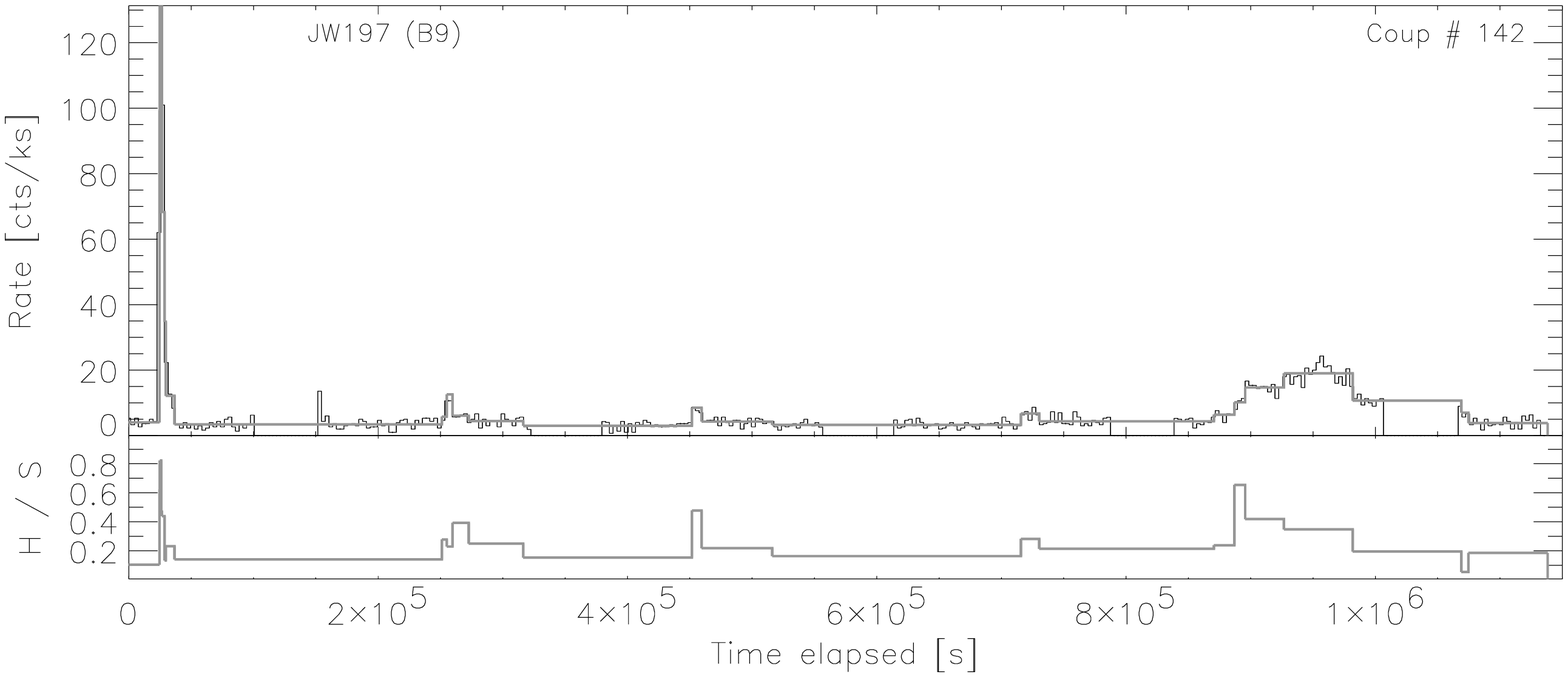}}
\resizebox{15cm}{!}{\includegraphics{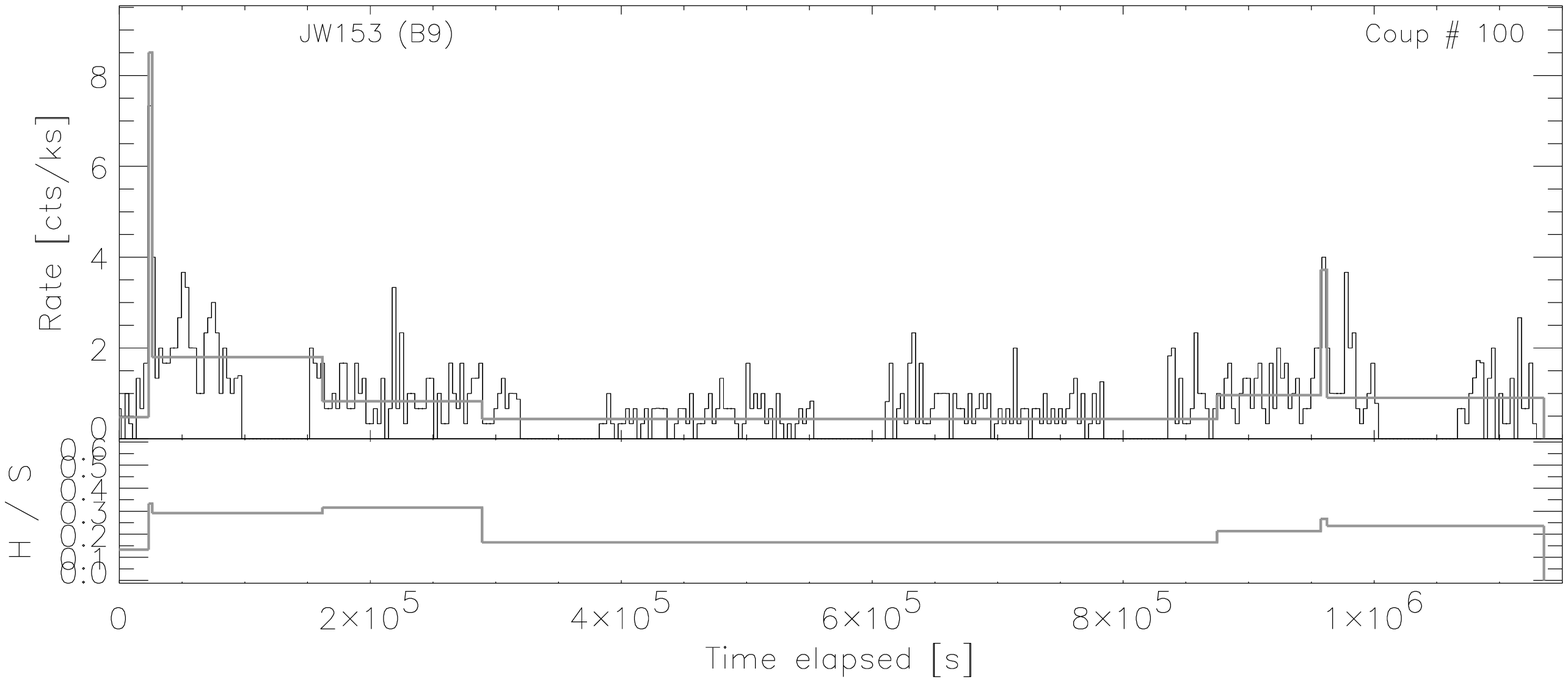}}
\resizebox{15cm}{!}{\includegraphics{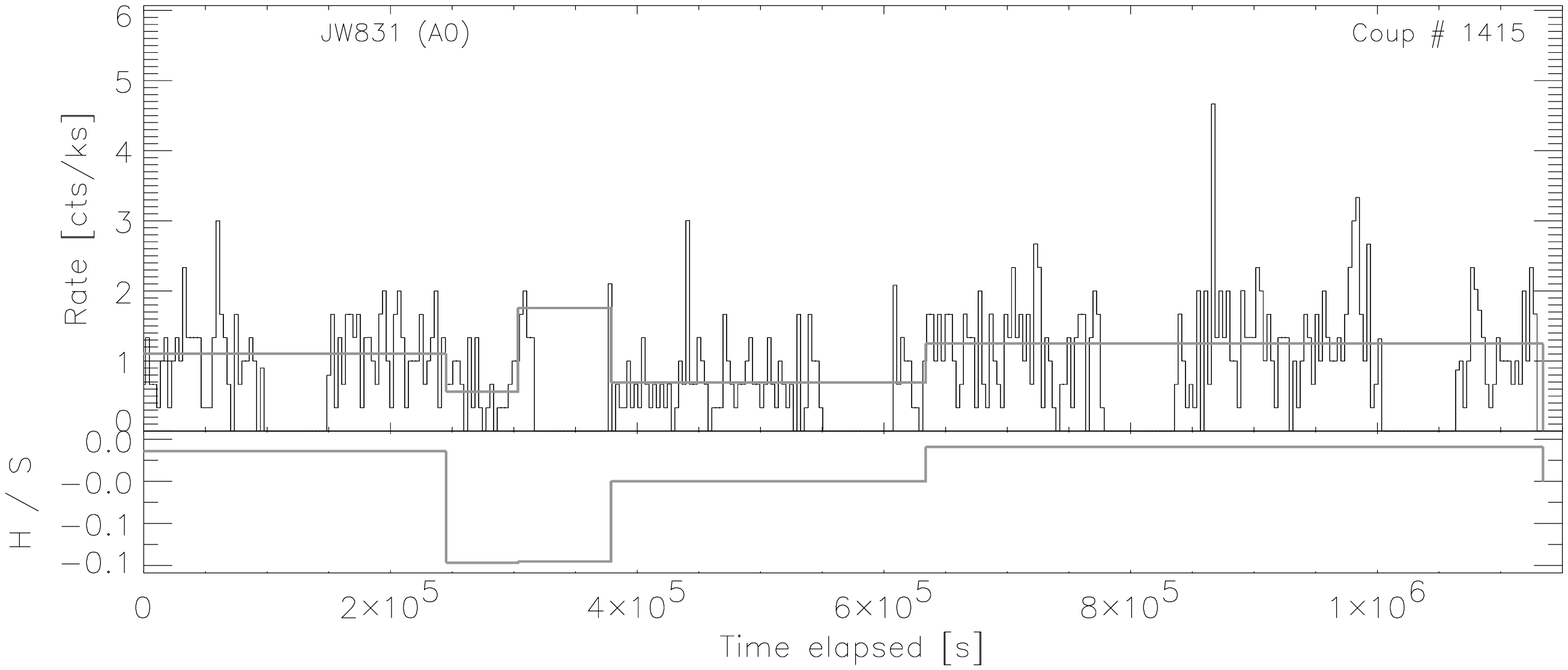}} \caption{{\em
continued}}
\end{center}
\end{figure}

\addtocounter{figure}{-1}

\begin{figure}
\begin{center}
\resizebox{15cm}{!}{\includegraphics{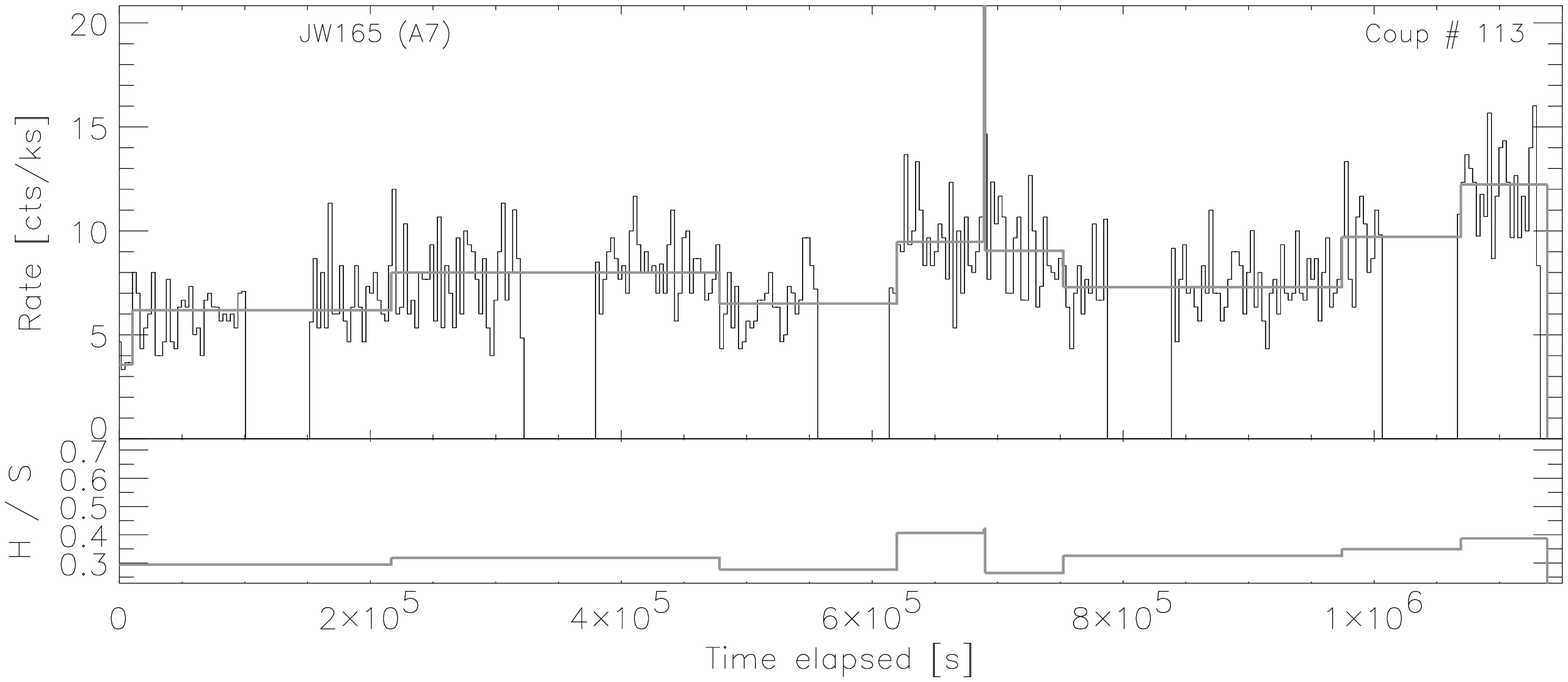}}
\resizebox{15cm}{!}{\includegraphics{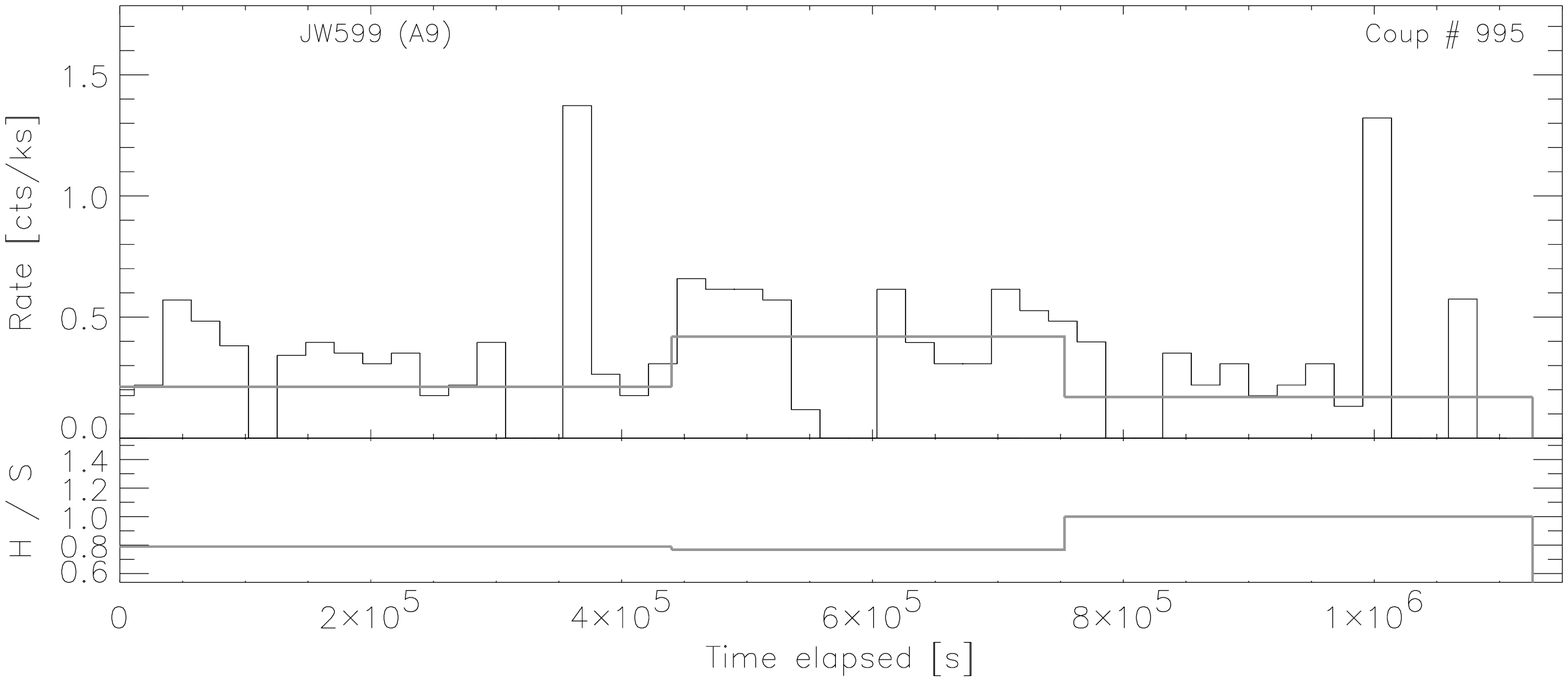}} \caption{{\em
continued}}
\end{center}
\end{figure}

%
\begin{figure}
\begin{center}
\resizebox{16cm}{!}{\includegraphics{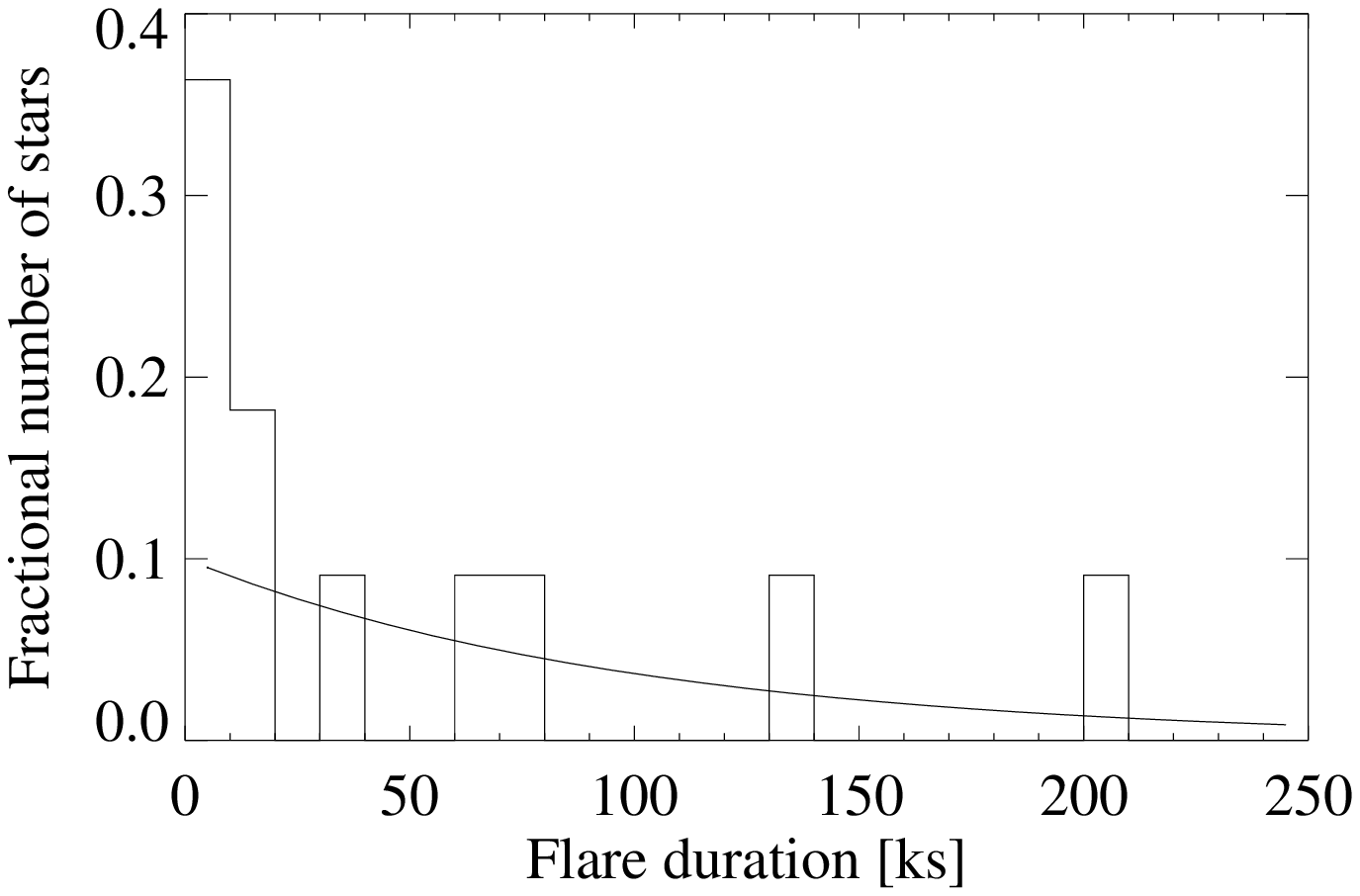}} \caption{Histogram 
showing the distribution of flare durations for the $11$ flares on 
WW stars. The 
exponential $0.1 \times e^{(-t/100\,ks)}$ from \citet{Wolk05.1} is 
overplotted.} \label{fig:histo_flareduration}
\end{center}
\end{figure}

%
\begin{figure}
\begin{center}
\parbox{16cm}{
\parbox{8cm}{\resizebox{8cm}{!}{\includegraphics{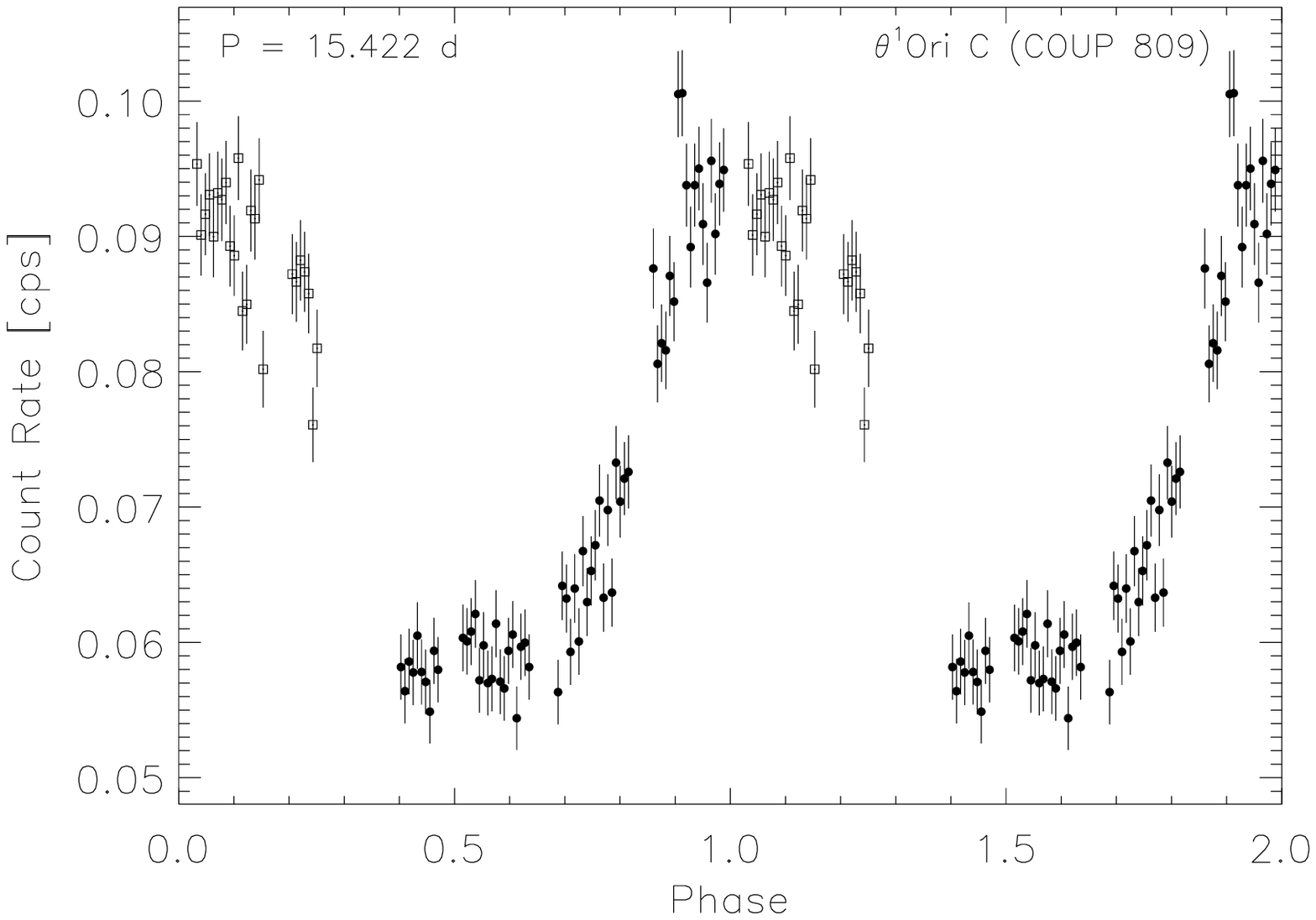}}}
\parbox{8cm}{\resizebox{8cm}{!}{\includegraphics{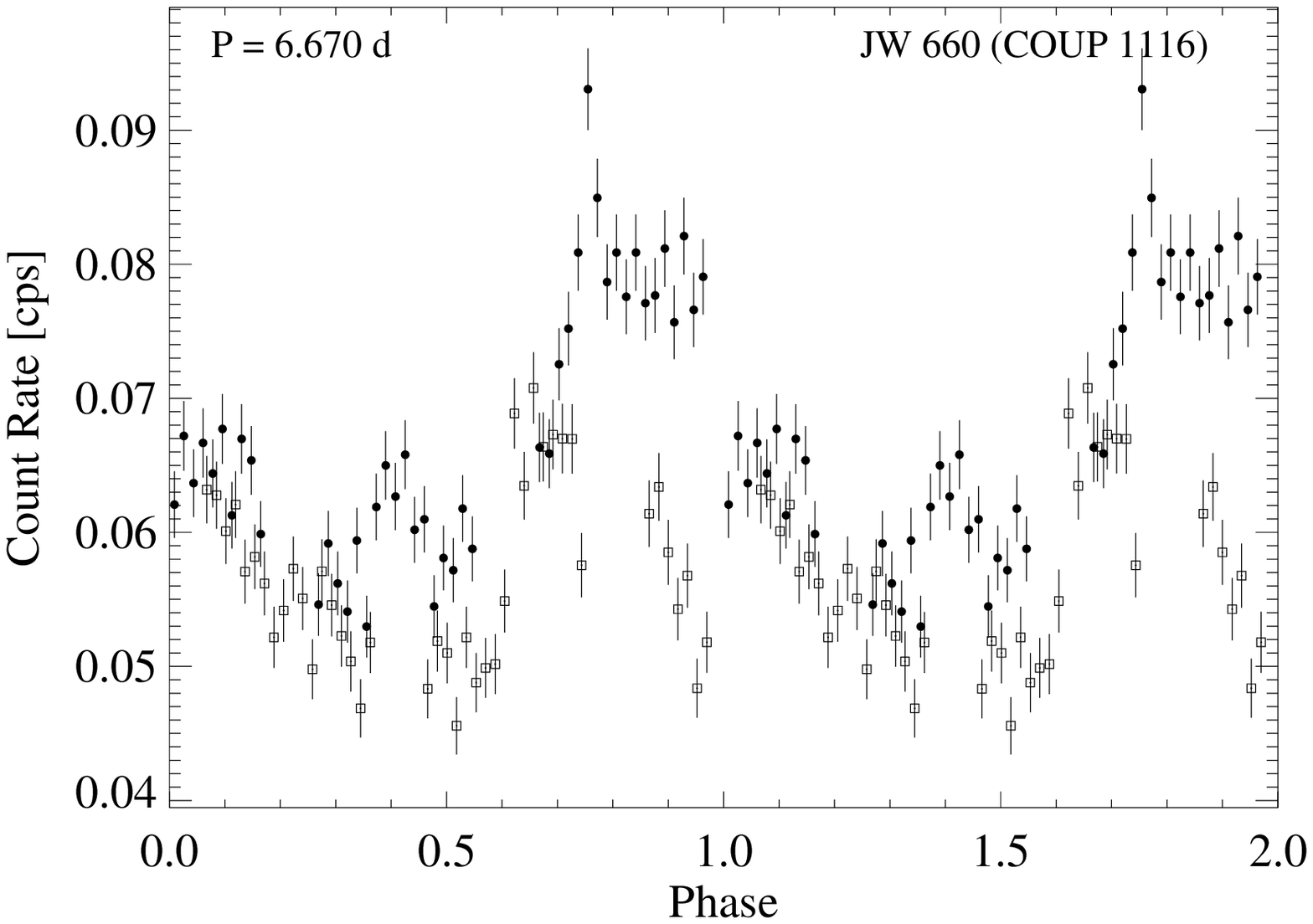}}}
}
\parbox{16cm}{
\parbox{8cm}{\resizebox{8cm}{!}{\includegraphics{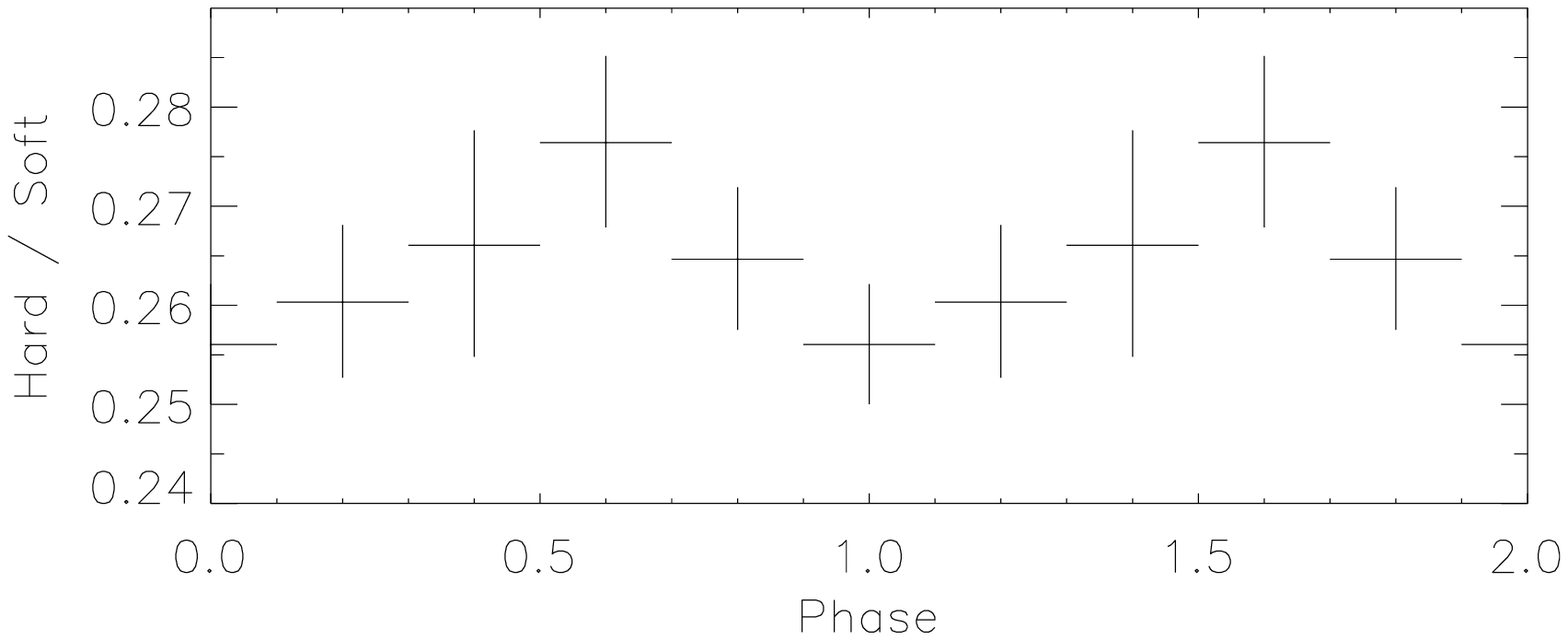}}}
\parbox{8cm}{\resizebox{8cm}{!}{\includegraphics{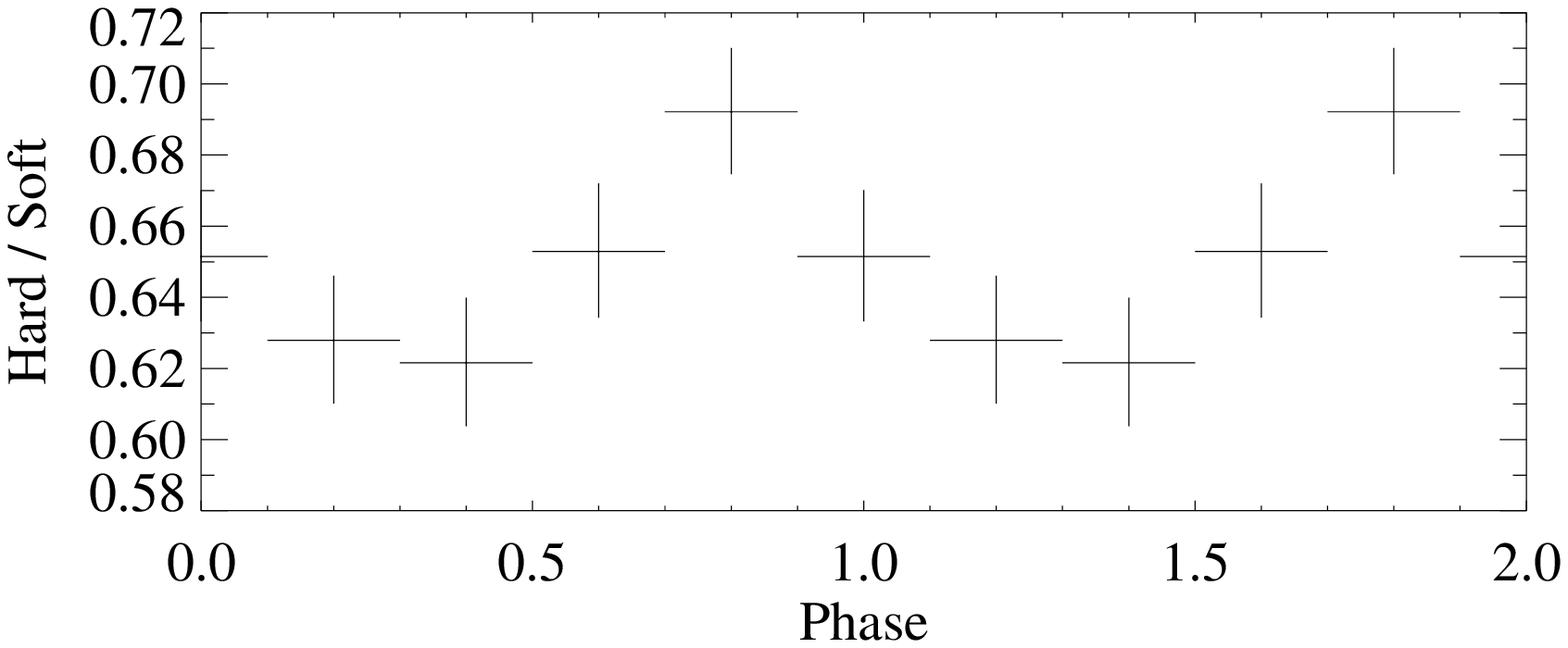}}}
}
\parbox{16cm}{
\parbox{8cm}{\resizebox{8cm}{!}{\includegraphics{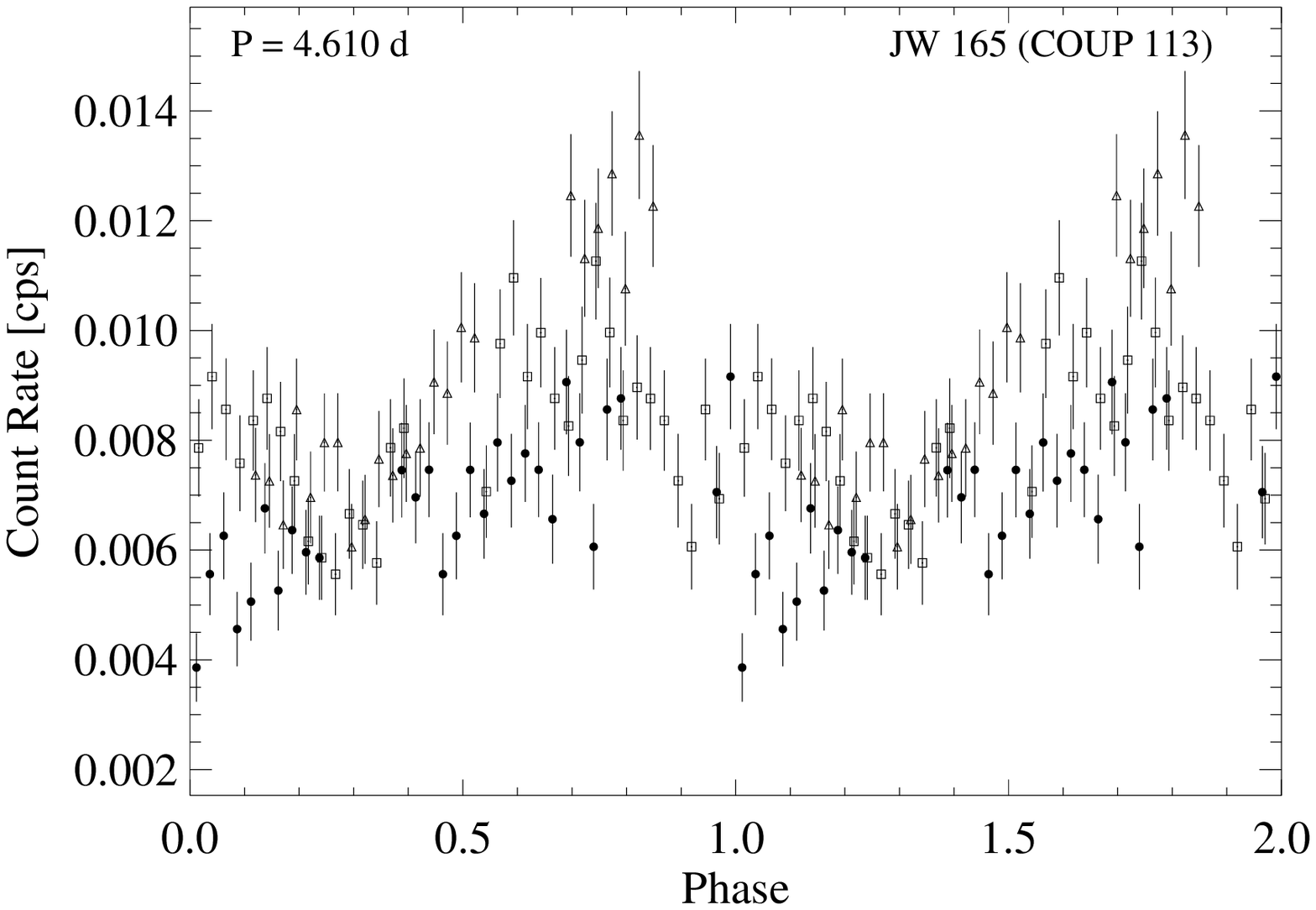}}}
\parbox{8cm}{\resizebox{8cm}{!}{\includegraphics{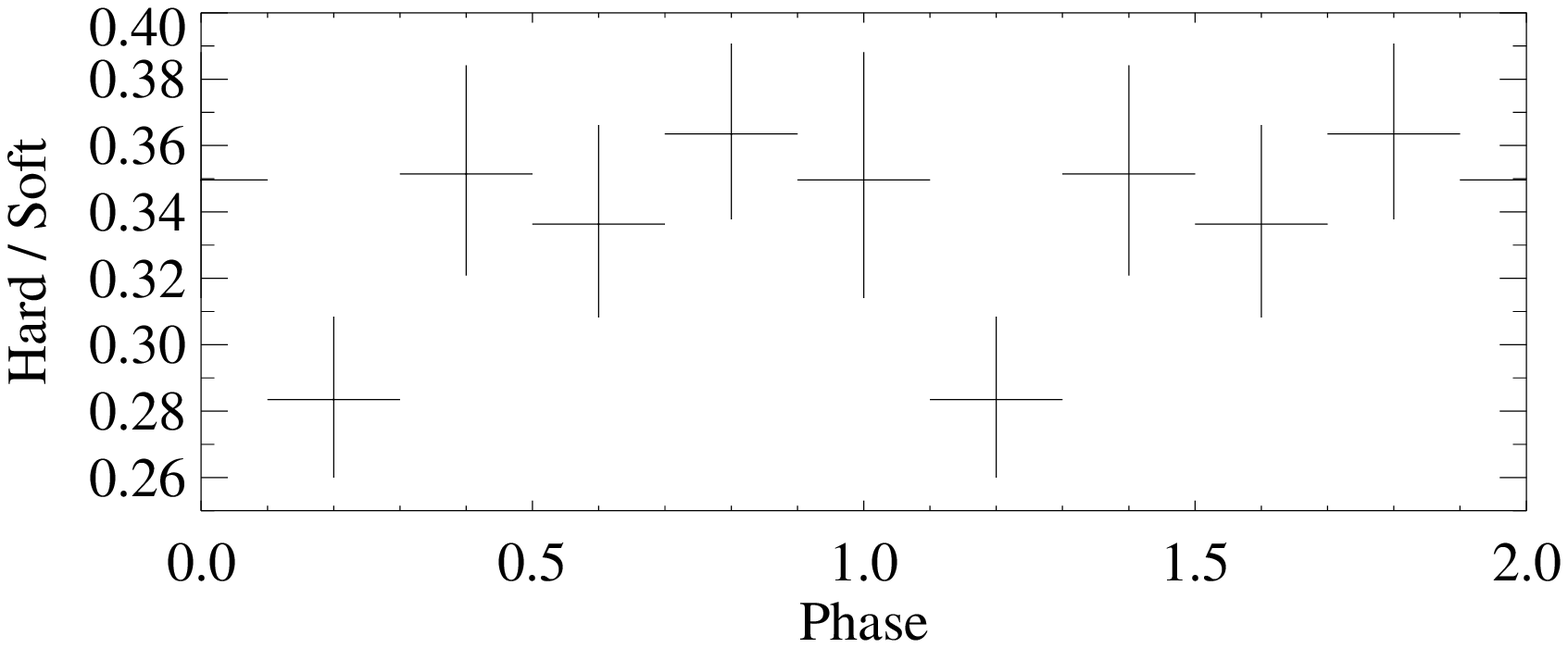}}}
}
\parbox{16cm}{
\parbox{8cm}{
\caption{Folded X-ray lightcurves and hardness ratios for the
three stars with optical photometric periodicities.
$\theta^1$\,Ori\,C is folded with the optical period using the
ephemeris by \protect\citet{Stahl96.1}. JW\,660 and JW\,165 are
folded using the period derived from the X-ray data. Data from
consecutive cycles are shown with different plotting symbols. 
The hardness ratio is examined in $5$ phase bins.}
\label{fig:lcs_folded}
}
}
\end{center}
\end{figure}

%
\begin{figure*}
\begin{center}
\resizebox{8cm}{!}{\includegraphics{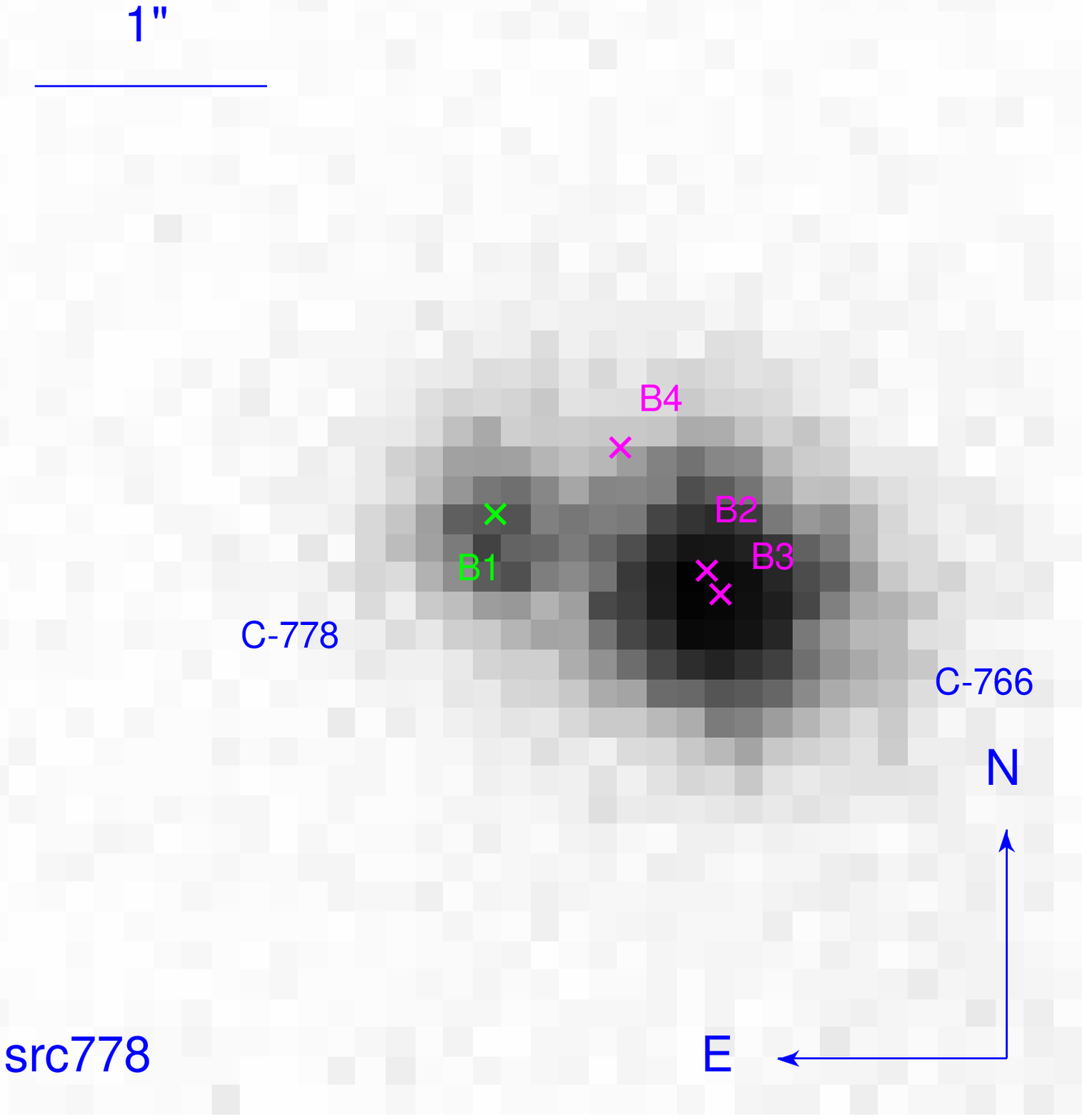}} \caption{COUP image of 
the region near the $\theta^1$\,Ori\,B quintet displayed at very 
high resolution ($0.125^{\prime\prime}$ pixel size). The yellow and 
magenta $\times$ symbols show the positions of the optical 
components.} \label{fig:binary_images}
\end{center}
\end{figure*}

%
\begin{figure}
\begin{center}
\resizebox{8cm}{!}{\includegraphics{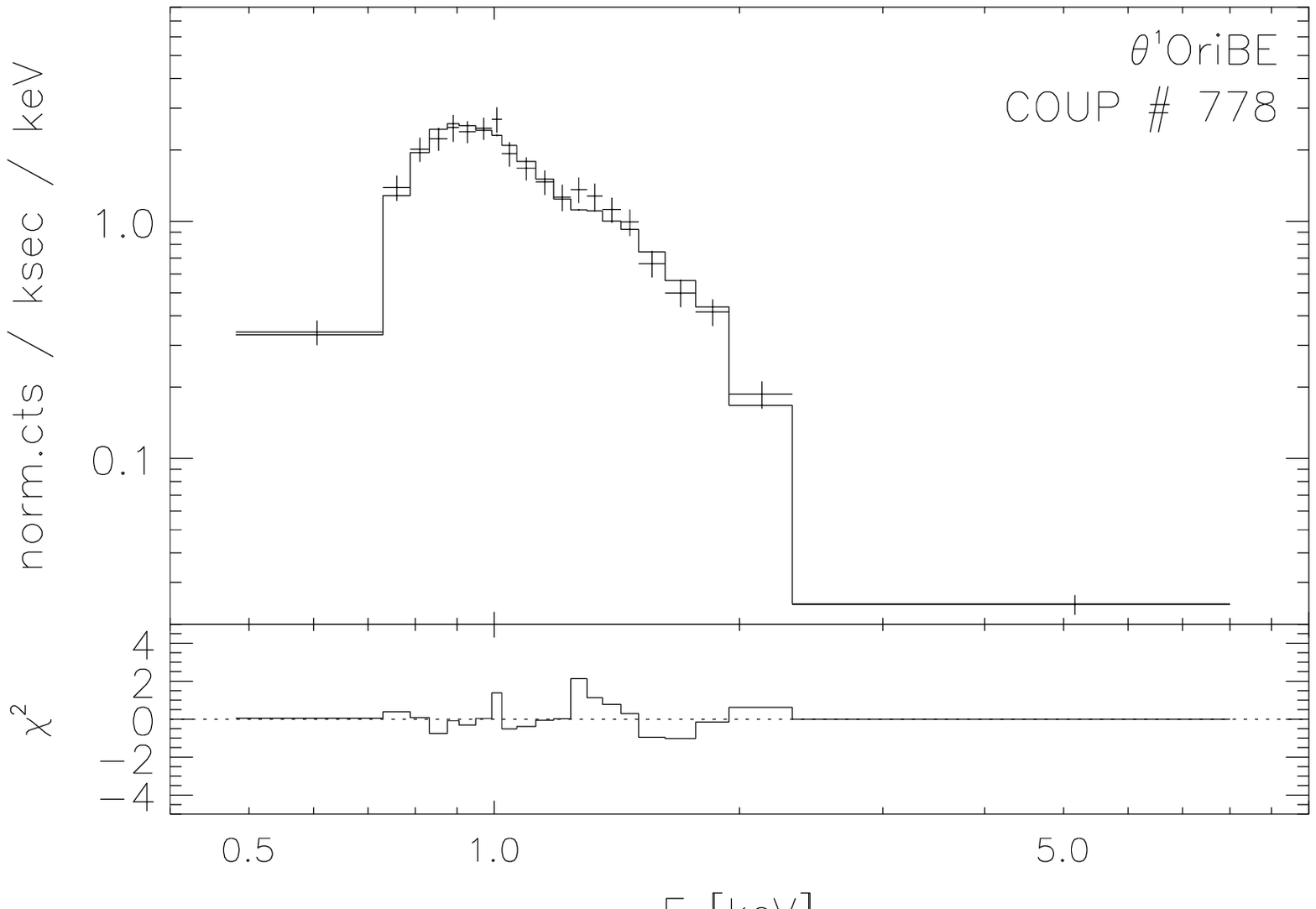}}
\resizebox{8cm}{!}{\includegraphics{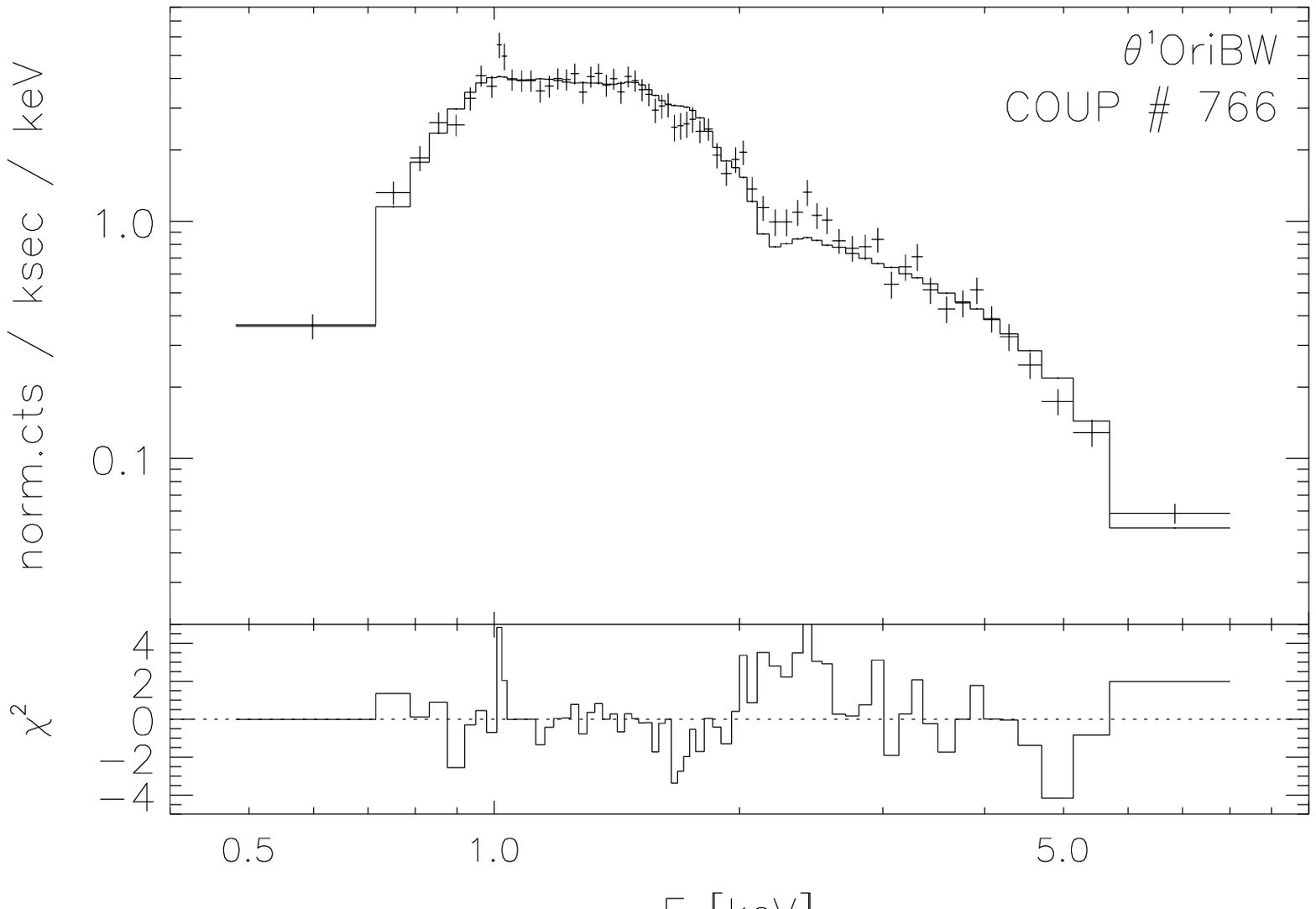}} \caption{ACIS spectra 
of $\theta^1$\,Ori\,B\,East (SpT B3; COUP\,778) and 
$\theta^1$\,Ori\,B\,West (?; COUP\,766).  The data are shown with 
error bars, the continuous histogram is the best-fit model from {\em 
XSPEC}, and the bottom panels show residuals between the data and 
models.} \label{fig:spec_thetaorib}
\end{center}
\end{figure}

%
\begin{figure*}
\begin{center}
\resizebox{8cm}{!}{\includegraphics{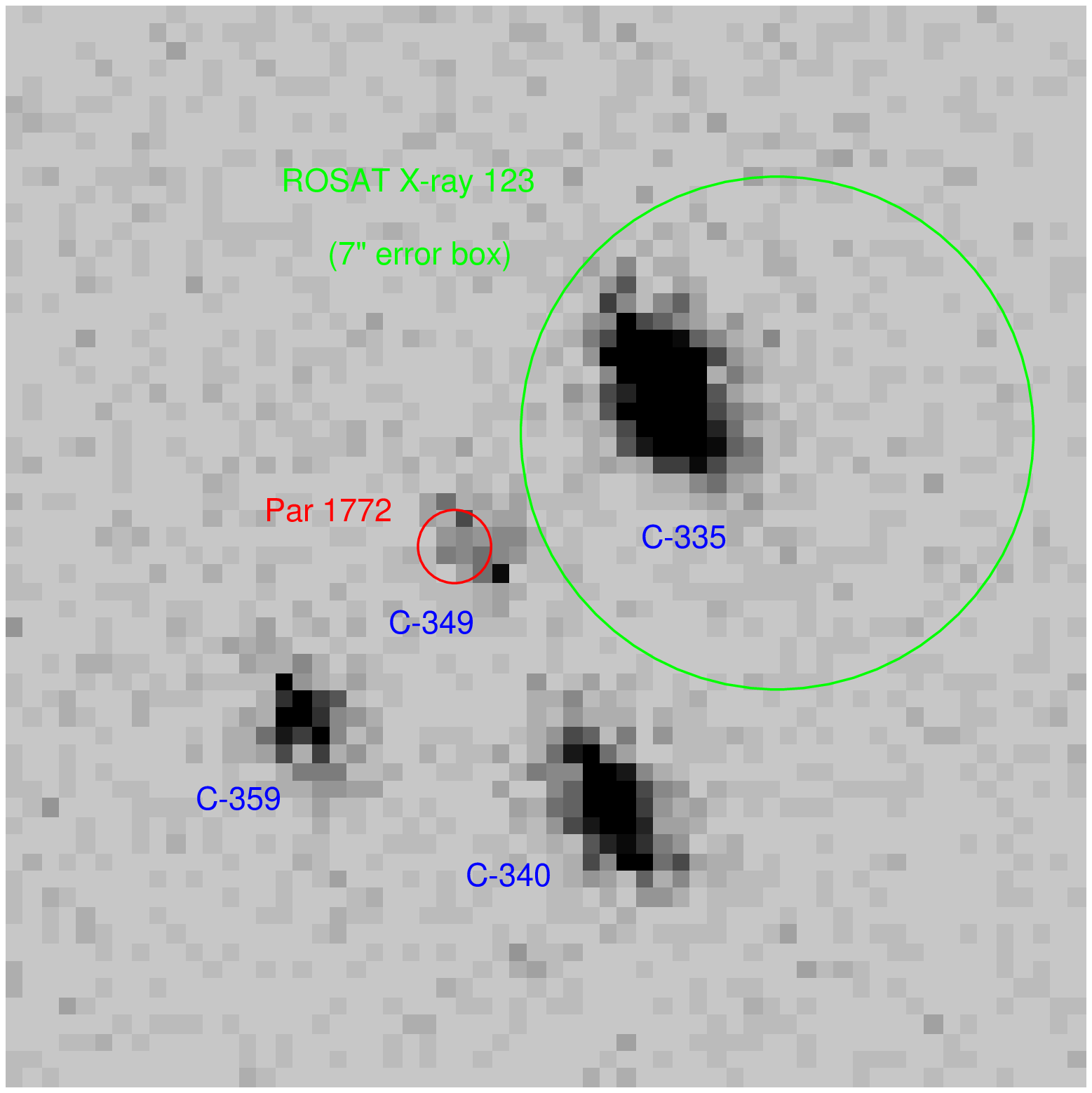}} \caption{COUP image 
showing the X-ray detection of the B1.5 star Par\,1772 (red circle), 
resolved for the first time in X-rays from adjacent brighter X-ray 
sources. `C-XXX' denotes the COUP source number, and the green 
circle shows the $ROSAT$ HRC error circle previously associated to 
Par 1772.} \label{fig:crowded_images}
\end{center}
\end{figure*}

%
\begin{figure}
\begin{center}
\resizebox{8cm}{!}{\includegraphics{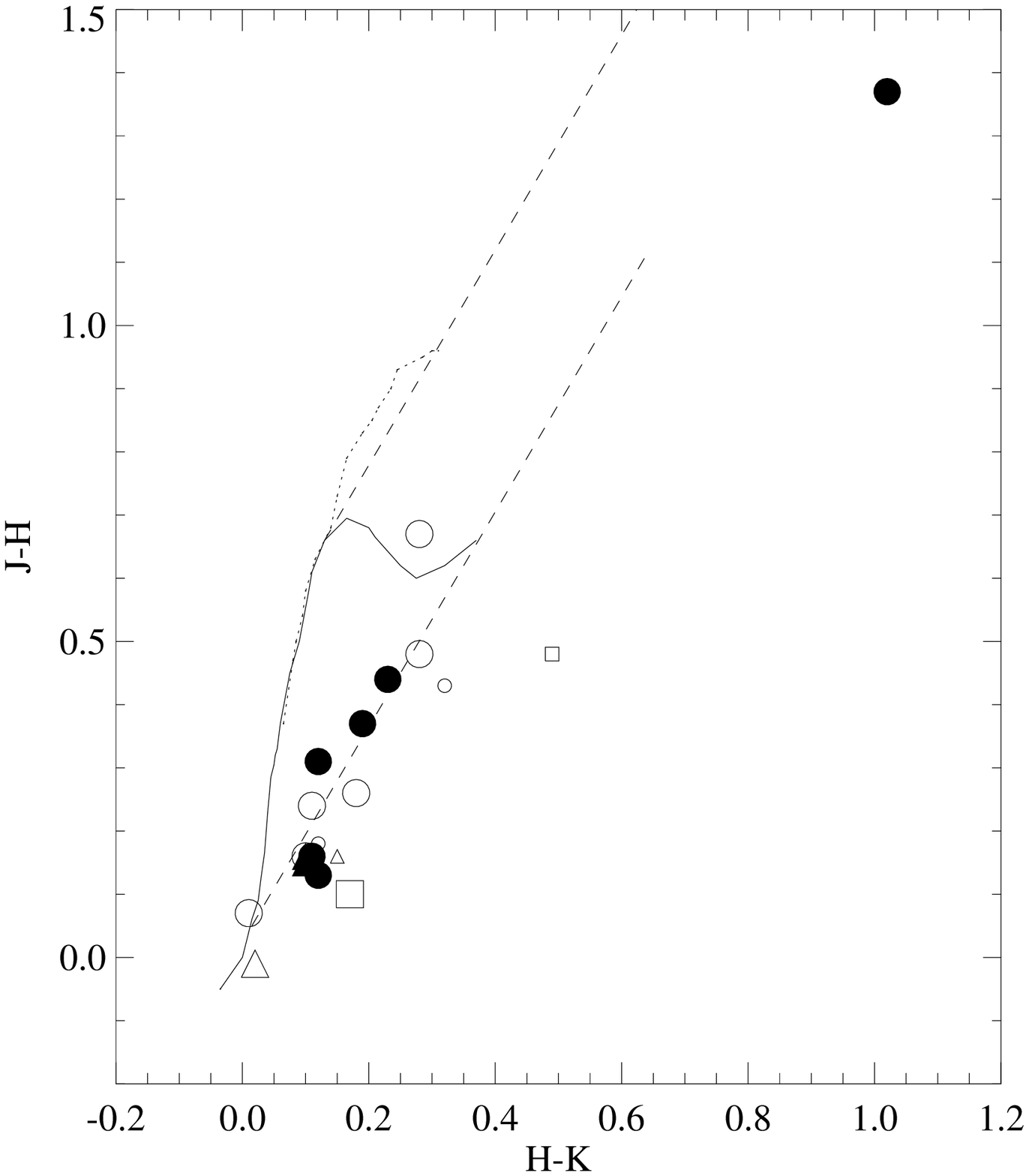}}
\resizebox{8cm}{!}{\includegraphics{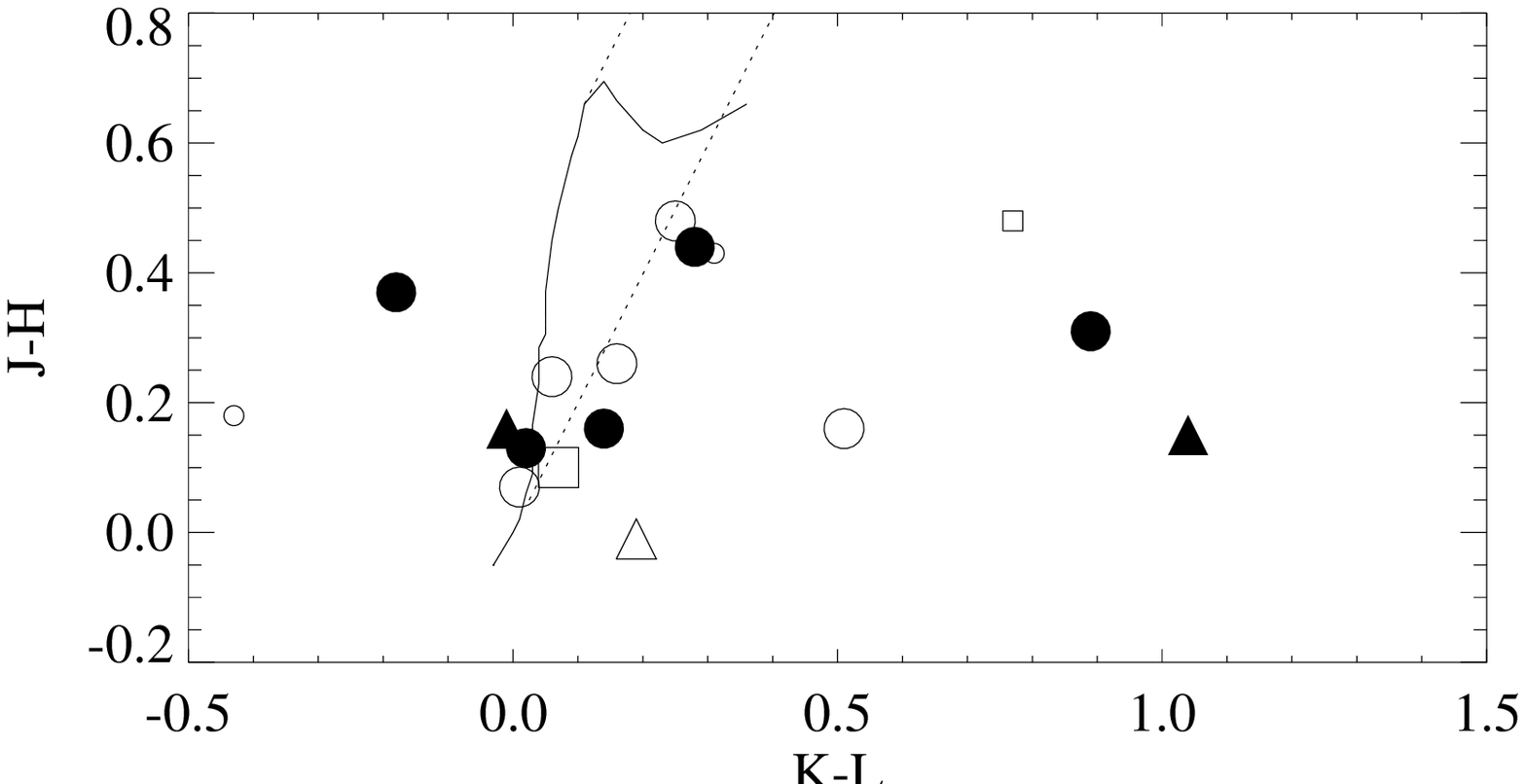}}
\caption{Near-IR color-color diagrams for SW and WW stars in the COUP field. Smaller symbols 
are for stars not detected in the COUP, and filled symbols for known 
or suspected binaries. Plotting symbols indicate the following 
properties: {\em triangles} are $IRAS$-excess stars from 
\protect\citet{The94.1}, {\em squares} are chemically peculiar (CP) 
star candidates from \protect\citet{Renson91.1}, and {\em circles} 
are the remaining normal stars. The solid and the dotted curves are 
the loci of dwarf and giant stars from \protect\citet{Bessel88.1}, 
and the dashed lines indicate a reddening corresponding to $A_{\rm 
V}=10$\,mag using the extinction law by \protect\citet{Rieke85.1}.} 
\label{fig:jhk}
\end{center}
\end{figure}

%
\begin{figure}
\begin{center}
\resizebox{16cm}{!}{\includegraphics{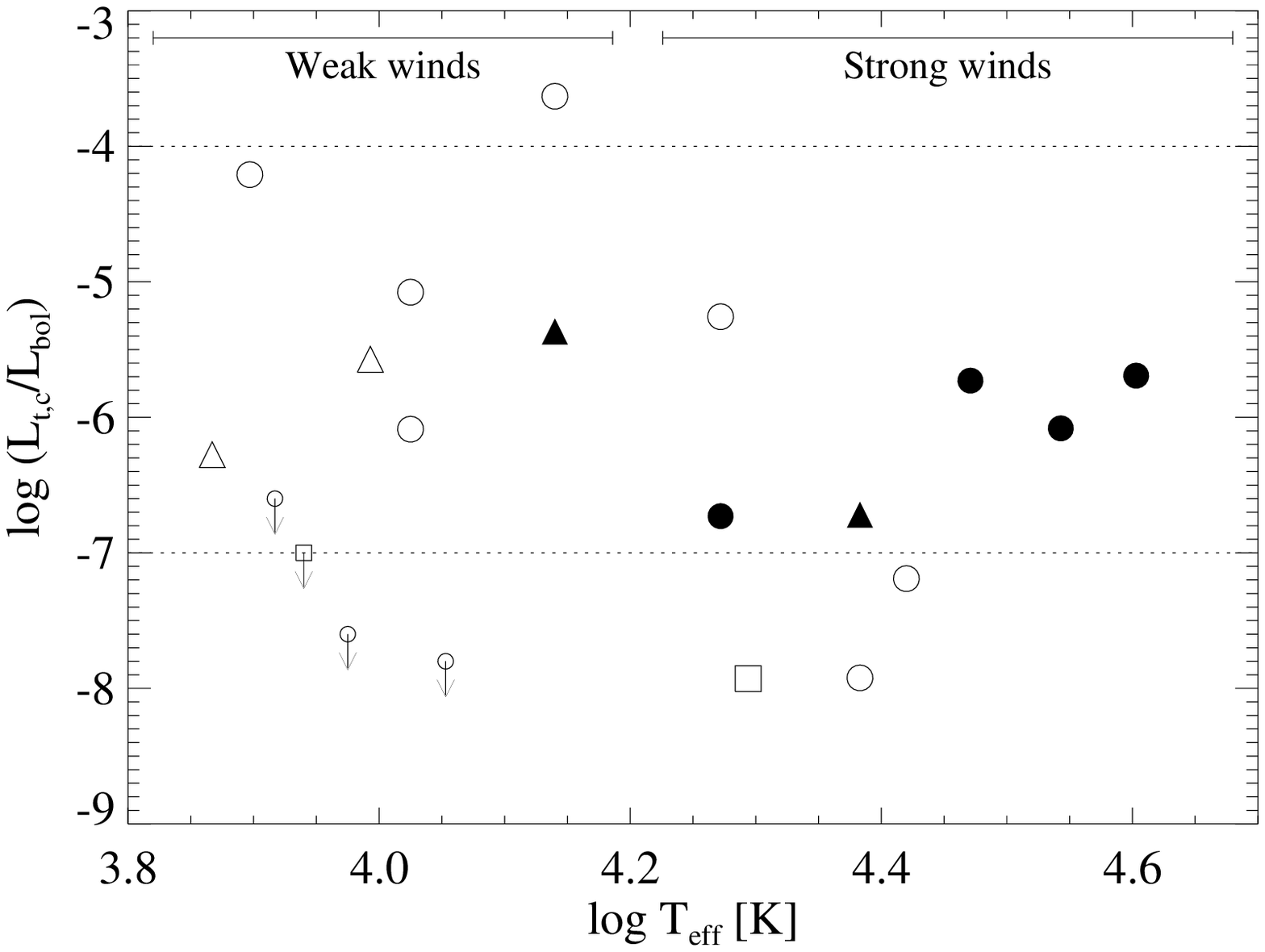}} \caption{X-ray
efficiency as a function of effective temperature. The dotted lines 
indicate the threshold values defined in 
Sect.~\ref{subsect:disc_coup}. The plotting symbols are defined in 
Figure~\ref{fig:jhk}. } \label{fig:lxlbol_lbol}
\end{center}
\end{figure}


\clearpage
\newpage

\begin{thebibliography}{}

\bibitem[\protect\astroncite{Abbott et~al.}{1979}]{Abbott79.1}
Abbott D. C., 1979, in Mass loss and evolution of O-type stars, P.
S. Conti \& C. de Loore (eds.), IAU Symp. 83, Dordrecht: Reidel,
p.237

\bibitem[\protect\astroncite{Abbott et~al.}{1984}]{Abbott84.1}
Abbott D. C., Bieging J. H. \& Churchwell E., 1984, ApJ 280, 671

\bibitem[\protect\astroncite{Abt et~al.}{1991}]{Abt91.1}
Abt H. A., Wang R. \& Cardona O., 1991, ApJ 367, 155

\bibitem[\protect\astroncite{Aikman et~al.}{1974}]{Aikman74.1}
Aikman G. C. L. \& Goldberg B. A., 1974, J.\ Roy.\ Astron.\ Soc.\
Canada, 68, 205

\bibitem[\protect\astroncite{Anders \& Grevesse}{1989}]{Anders89.1}
Anders E. \& Grevesse N.,  1989, Geochim.\ Cosmochim.\ Acta 53,
197

\bibitem[\protect\astroncite{Antokhina et~al.}{1989}]{Antokhina89.1}
Antokhina E. A., Ismailov N. Z. \& Cherpashchuk A. M., 1989, Sov.A.L. 15, 362

\bibitem[\protect\astroncite{Babel}{1996}]{Babel96.1}
Babel J., 1996, A\&A 309, 867

\bibitem[\protect\astroncite{Babel \& Montmerle}{1997a}]{Babel97.1}
Babel J. \& Montmerle T., 1997a, A\&A 323, 121

\bibitem[\protect\astroncite{Babel \& Montmerle}{1997b}]{Babel97.2}
Babel J. \& Montmerle T., 1997b, ApJ 485, L29

\bibitem[\protect\astroncite{Bergh\"ofer et~al.}{1997}]{Berghoefer97.1}
Bergh\"ofer T. W., Schmitt J. H. M. M., Danner R. \& Cassinelli J. P.,
1996, A\&A 322, 167

\bibitem[Behar et al.(2004)]{Behar04} Behar, E., Leutenegger,
M., Doron, R., G{\" u}del, M., Feldman, U., Audard, M., \& Kahn,
S.~M.\ 2004, \apjl, 612, L65

\bibitem[\protect\astroncite{Bernabeu et~al.}{1989}]{Bernabeu89.1}
Bernabeu G., Magazz\'u A. \& Stalio R., 1989, A\&A 226, 215
 
\bibitem[\protect\astroncite{Bessel \& Brett}{1988}]{Bessel88.1}
Bessel M. S. \& Brett J. M., 1988, PASP 100, 1134

\bibitem[\protect\astroncite{B\"ohm \& Catala}{1995}]{Boehm95.1}
B\"ohm T. \& Catala C., 1995, A\&A 301, 155

\bibitem[\protect\astroncite{Bouvier et~al.}{1999}]{Bouvier99.1}
Bouvier J., Chelli A., Allain S., et al., 1999, A\&A 349, 619

\bibitem[\protect\astroncite{Caillault et~al.}{1994}]{Caillault94.1}
Caillault J.-P., Gagn\'e M. \& Stauffer J. R., 1994, ApJ 432, 386

\bibitem[\protect\astroncite{Caillault \& Zoonematkermani}{1989}]{Caillault89.1} 
Caillault J.-P. \& Zoonematkermani S., 1989, ApJ 338, L57

\bibitem[\protect\astroncite{Carpenter et~al.}{2001}]{Carpenter01.1}
Carpenter J. M., Hillenbrand, L. A. \& Skrutskie M. F., 2001, AJ 121, 3160

\bibitem[\protect\astroncite{Castor et~al.}{1975}]{Castor75.1}
Castor J., Abbott D. \& Klein R., 1975, ApJ 195, 157

\bibitem[\protect\astroncite{Catala et~al.}{1999}]{Catala99.1}
Catala C., Donati J. F., B\"ohm, T., et al., 1999, A\&A 345, 884

\bibitem[\protect\astroncite{Charbonneau \& MacGregor}{2001}]{Charbonneau01.1}
Charbonneau P. \& MacGregor K. B., 2001, ApJ 559, 1094

\bibitem[\protect\astroncite{Choi \& Herbst}{1996}]{Choi96.1}
Choi P. I. \& Herbst W., 1996, AJ 111, 283

\bibitem[\protect\astroncite{Cohen et~al.}{1997}]{Cohen97.1}
Cohen D. H., Cassinelli J. P. \& Macfarlane J. J., 1997, ApJ 487, 867

\bibitem[\protect\astroncite{Cohen et~al.}{2003}]{Cohen03.1}
Cohen, D. H.; de Messi\`eres, G. E.; MacFarlane, J. J.; Miller, N. A.; 
Cassinelli, J. P.; et al., 2003, ApJ 586, 495

\bibitem[\protect\astroncite{Corporon \& Lagrange}{1999}]{Corporon99.1}
Corporon P. \& Lagrange A.-M., 1999, A\&AS 136, 429

\bibitem[\protect\astroncite{de Jong et~al.}{2001}]{deJong01.1}
de Jong J. A., Henrichs A. F., Kaper L., et al., 2001, A\&A 368, 601

\bibitem[\protect\astroncite{Donati et~al.}{2002}]{Donati02.1}
Donati J.-F., Babel J., Harries T. J., et al., 2002, MNRAS 333, 55

\bibitem[\protect\astroncite{Favata et~al.}{2005}]{Favata05.1}
Favata F., Flaccomio, E., et al.\ 2005, \apjs, in press

\bibitem[\protect\astroncite{Favata \& Micela}{2003}]{Favata03.1}
Favata F. \& Micela M., 2003, Space Science Reviews 108, 577

\bibitem[\protect\astroncite{Feigelson et~al.}{2005}]{Feigelson05.1}
Feigelson, E.\ D., Getman, K.\ V., et al., \apjs, in press

\bibitem[\protect\astroncite{Feigelson et~al.}{2003}]{Feigelson03.1}
Feigelson E. D., Lawson W. A. \& Garmire G. P., 2003, ApJ 599, 1207

\bibitem[\protect\astroncite{Feigelson et~al.}{2002}]{Feigelson02.1}
Feigelson E. D., Broos P., Gaffney J. A. III., et al., 2002, ApJ 574, 258

\bibitem[\protect\astroncite{Feigelson \& Montmerle}{1999}]{Feigelson99.1}
Feigelson E. D. \& Montmerle T., 1999, ARA\&A 37, 363

\bibitem[\protect\astroncite{Feigelson \& Nelson}{1985}]{Feigelson85.1}
Feigelson E. D. \& Nelson P. I., 1985, ApJ 293, 192

\bibitem[\protect\astroncite{Felli et~al.}{1993}]{Felli93.1}
Felli M., Taylor G. B., Catarzi M., Churchwell E. \& Kurtz S., 1993,
A\&AS 101, 127

\bibitem[\protect\astroncite{Flaccomio et~al.}{2003}]{Flaccomio03.1}
Flaccomio E., Damiani F., Micela G., et al., 2002, ApJ 582, 382

\bibitem[\protect\astroncite{Flaccomio et~al.}{2005}]{Flaccomio05.1}
Flaccomio E., et al., 2005, \apjs, in press 

\bibitem[\protect\astroncite{Frost et~al.}{1926}]{Frost26.1}
Frost E. B., Barrett S. B. \& Struve O., 1926, ApJ 64, 1

\bibitem[\protect\astroncite{Gagn\'e et~al.}{2005}]{Gagne05.1}
Gagn\'e M., Oksala M. E., Cohen D. H., Tonnesen S. K., Ud-Doula A., et al., 2005, 
astro-ph/0504296

\bibitem[\protect\astroncite{Gagn\'e et~al.}{1997}]{Gagne97.1}
Gagn\'e M., Caillault J.-P., Stauffer J. R. \& Linsky J. L., 1997, ApJ 478, L87

\bibitem[\protect\astroncite{Getman et~al.}{2005a}]{Getman05.1}
Getman K., Flaccomio E., Broos P., et al., 2005a, ApJ in press

\bibitem[\protect\astroncite{Getman et~al.}{2005b}]{Getman05.2}
Getman K., et al., 2005b, \apjs, in press

\bibitem[\protect\astroncite{Grosso et~al.}{2005}]{Grosso05.1}
Grosso N., Feigelson, E.\ D., et al., 2005, \apjs, in press

\bibitem[\protect\astroncite{Grosso et~al.}{2004}]{Grosso04.1}
Grosso N., Montmerle T., Feigelson E. D. \& Forbes T. G., 2004,
A\&A 419, 653

\bibitem[\protect\astroncite{Hamaguchi et~al.}{2005}]{Hamaguchi05.1}
Hamaguchi K., Yamauchi S. \& Koyama K., 2005, ApJ 618, 360

\bibitem[\protect\astroncite{Harnden et~al.}{1979}]{Harnden79.1}
Harnden F. R. Jr., Branduardi G., Gorenstein P., Grindlay J.,
Rosner R., et al., 1979, ApJ 234, 51

\bibitem[\protect\astroncite{Herbst et~al.}{2000}]{Herbst00.1}
Herbst W., Rhode K.L., Hillenbrand L.A. \& Curran G., 2000, AJ 119, 261

\bibitem[\protect\astroncite{Hillenbrand \& White}{2004}]{Hillenbrand04.1}
Hillenbrand L. A. \& White R. J., 2004, ApJ 604, 741

\bibitem[\protect\astroncite{Hillenbrand et~al.}{1998}]{Hillenbrand98.1}
Hillenbrand L. A., Strom S. E., Calvet N., et al., 1998, AJ 116, 1816

\bibitem[\protect\astroncite{Hillenbrand}{1997}]{Hillenbrand97.1}
Hillenbrand L. A., 1997, AJ 113, 1733

\bibitem[\protect\astroncite{Kaper et~al.}{1999}]{Kaper99.1}
Kaper L., Henrichs H. F., Nichols J. S. \& Telting J. H., 1999, A\&A 344, 231

\bibitem[\protect\astroncite{Kissmann et~al.}{2004}]{Kissmann04.1}
Kissmann R., Fichtner H. \& Ferreira S. E. S., 2004, A\&A 419, 357

\bibitem[\protect\astroncite{Kudritzki \& Puls}{2000}]{Kudritzki00.1}
Kudritzki R.-P. \& Puls J., 2000, ARA\&A 38, 613

\bibitem[\protect\astroncite{Landstreet}{2003}]{Landstreet03.1}
Landstreet J. D., 2003, In: Proc. of `Magnetism and Activity of the Sun and Stars',
EAS Publications Series, vol. 9, 235

\bibitem[\protect\astroncite{Landstreet}{1992}]{Landstreet92.1}
Landstreet J. D., 1992, A\&AR 4, 35

\bibitem[\protect\astroncite{Linsky et~al.}{1992}]{Linsky92.1}
Linsky J. L., Drake S. A. \& Bastian T. S., 1992, ApJ 393, 341

\bibitem[\protect\astroncite{Lloyd \& Strickland}{1999}]{Lloyd99.1}
Lloyd C. \& Strickland D. J., 1999, IBVS 4809

\bibitem[\protect\astroncite{Lucy \& White}{1980}]{Lucy80.1}
Lucy L. B. \& White R. L. 1980, ApJ 241, 300

\bibitem[\protect\astroncite{MacGregor \& Cassinelli}{2003}]{MacGregor03.1}
MacGregor K. B. \& Cassinelli J. P., 2003, ApJ 586, 480

\bibitem[\protect\astroncite{Manoj et~al.}{2002}]{Manoj02.1}
Manoj P., Maheswar G. \& Bhatt H. C., 2002, MNRAS 334, 419

\bibitem[\protect\astroncite{Mason et~al.}{1998}]{Mason98.1}
Mason B. D., Gies D. R., Hartkopf W. I., et al., 1998, AJ 115, 821

\bibitem[\protect\astroncite{Mewe et~al.}{1985}]{Mewe85.1}
Mewe R., Gronenschild E. H. B. M. \& van den Oord G. H. J., 1985, A\&AS 62, 197

\bibitem[\protect\astroncite{Mewe et~al.}{1995}]{Mewe95.1}
Mewe R., Kaastra J., S., Schrijver C. J., van den Oord G. H. J. \& Alkemade
F. J. M., 1995, A\&A 296, 477

\bibitem[\protect\astroncite{Miller et~al.}{2002}]{Miller02.1}
Miller N. A., Cassinelli J. P., Waldron W. L., MacFarlane J. J. 
\& Cohen D. H., 2002, ApJ 577, 951

\bibitem[\protect\astroncite{Montmerle et~al.}{2000}]{Montmerle00.1}
Montmerle Th., Grosso N., Tsuboi Y. \& Koyama K., 2000, ApJ 532, 1097

\bibitem[\protect\astroncite{Montmerle et~al.}{1983}]{Montmerle83.1}
Montmerle Th., Koch-Miramond L., Falgarone E. \& Grindlay J. E., 1983, ApJ 269, 182

\bibitem[\protect\astroncite{Morrell \& Levato}{1991}]{Morrell91.1}
Morrell N. \& Levato H., 1991, ApJS 75, 965

\bibitem[\protect\astroncite{Muench et~al.}{2002}]{Muench02.1}
M\"unch A. A., Lada E. A., Lada C. J. \& Alves Joao, 2002, ApJ 573, 366

\bibitem[\protect\astroncite{Nakajima et~al.}{2003}]{Nakajima03.1}
Nakajima H., Imanishi K., Takagi S.-I., Koyama K. \& Tsujimoto M.,
2002, PASJ 55, 635

\bibitem[\protect\astroncite{Oksala et al.}{2004}]{Oksala04.1} Oksala, M.,
Gagne, M., Cohen, D., Tonnesen, S., ud-Doula, A., Owocki, S., \&
MacFarlane, J.\ 2004, Bull.\ AAS, 204, \#62.14

\bibitem[\protect\astroncite{Owocki \& Cohen}{1999}]{Owocki99.1}
Owocki S. P. \& Cohen D. H., 1999, ApJ 520, 833

\bibitem[\protect\astroncite{Palla \& Stahler}{2001}]{Palla01.1}
Palla F. \& Stahler S. W., 2001, ApJ 553, 299

\bibitem[\protect\astroncite{Pallavicini et~al.}{1981}]{Pallavicini81.1}
Pallavicini R., Golub L., Rosner R., Vaiana G. S., Ayres T., et al.,
1981, ApJ 248, 279

\bibitem[\protect\astroncite{Petr et~al.}{1998}]{Petr98.1}
Petr M. G., Coud\'e du Foresto V., Beckwith S. V. W., et al., 1998, ApJ 500, 825

\bibitem[\protect\astroncite{Popper \& Pavec}{1976}]{Popper76.1}
Popper D. M. \& Plavec M., 1976, ApJ 205, 462

\bibitem[\protect\astroncite{Prato et~al.}{2002}]{Prato02.1}
Prato L., Simon M., Mazeh T., et al., 2002, ApJ 569, 863

\bibitem[\protect\astroncite{Preibisch et~al.}{2005}]{Preibisch05.1}
Preibisch Th., et al., 2005, \apjs, in press

\bibitem[\protect\astroncite{Preibisch}{2003}]{Preibisch03.1}
Preibisch Th., 2003, A\&A 401, 543

\bibitem[\protect\astroncite{Preibisch \& Zinnecker}{2001}]{Preibisch01.1}
Preibisch Th. \& Zinnecker H., 2001, AJ 123, 1613

\bibitem[\protect\astroncite{Preibisch et~al.}{1999}]{Preibisch99.1}
Preibisch Th., Balega Y., Hofmann K.-H., Weigelt G. \& Zinnecker H., 1999, 
New Astronomy 4, 531

\bibitem[\protect\astroncite{Preibisch}{1997}]{Preibisch97.1}
Preibisch Th., 1997, A\&A 320, 525

\bibitem[\protect\astroncite{Renson et~al.}{1991}]{Renson91.1}
Renson P., Gerbaldi M. \& Catalano F. A., 1991, A\&AS 89, 429

\bibitem[\protect\astroncite{Rieke \& Lebofsky}{1985}]{Rieke85.1}
Rieke G. H. \& Lebofsky M. J., 1985, ApJ 288, 618 (RL85)

\bibitem[\protect\astroncite{Scargle}{1982}]{Scargle82.1}
Scargle J. D., 1982, ApJ 263, 835

\bibitem[\protect\astroncite{Scargle}{1998}]{Scargle98.1}
Scargle J. D., 1998, ApJ 504, 405

\bibitem[\protect\astroncite{Schertl et~al.}{2003}]{Schertl03.1}
Schertl D., Balega Y. Y., Preibisch Th. \& Weigelt G., 2003, A\&A 402, 267

\bibitem[\protect\astroncite{Sciortino et~al.}{1990}]{Sciortino90.1}
Sciortino S., Vaiana G. S., Harnden F. R. Jr., Ramella M., Morossi C., et al.,
1990, ApJ 361, 621

\bibitem[\protect\astroncite{Schmitt et~al.}{1985}]{Schmitt85.1}
Schmitt J. H. M. M., Golub L., Harnden F. R. Jr., Maxson C. W., Rosner R., et al.,
1985, ApJ 290, 307

\bibitem[\protect\astroncite{Schulz et~al.}{2001}]{Schulz01.1}
Schulz N. S. Canizares C., Huenemoerder D., et al., 2001, ApJ 549, 441

\bibitem[\protect\astroncite{Schulz et al.}{2003}]{Schulz03.1} 
Schulz, N.~S., Canizares, C., Huenemoerder, D., \& Tibbets, K.\ 2003, \apj, 595,
365

\bibitem[\protect\astroncite{Siess et~al.}{2000}]{Siess00.1}
Siess L., Dufour E. \& Forestini M., 2000, A\&A 358, 598

\bibitem[\protect\astroncite{Simon et~al.}{1999}]{Simon99.1}
Simon M., Close L. M. \& Beck T. L., 1999, AJ 117, 1375

\bibitem[\protect\astroncite{Simon et~al.}{1995}]{Simon95.1}
Simon Th., Drake S. A. \& Kim P. D., 1995, PASP 107, 1034

\bibitem[\protect\astroncite{Smith \& Fullerton}{2005}]{Smith05.1}
Smith, M.~A., \& Fullerton, A.~W.\ 2005, \pasp, 117, 13

\bibitem[\protect\astroncite{Snow \& Morton}{1976}]{Snow76.1}
Snow Th. P. \& Morton D. C., 1976, ApJ 32, 429
 
\bibitem[\protect\astroncite{Stahl et~al.}{1996}]{Stahl96.1}
Stahl O., Kaufer A., Rivinius Th., et al., 1996, A\&A 312, 539

\bibitem[\protect\astroncite{Stahl et~al.}{1993}]{Stahl93.1}
Stahl O., Wolf B., G\"ang Th., et al., 1993, A\&A 274, L29

\bibitem[\protect\astroncite{Stelzer et~al.}{2005}]{Stelzer04.1}
Stelzer B., Hu\'elamo,N., Hubrig S., Micela G., Zinnecker H. \&
Guenther E., 2005, in 13$^{th}$ Cambridge Workshop on Cool Stars,
Stellar Systems, and the Sun, ESA SP Ser., in press

\bibitem[\protect\astroncite{Stelzer et~al.}{2003}]{Stelzer03.1}
Stelzer B., Hu\'elamo,N., Hubrig S., Zinnecker H. \& Micela G., 2003, A\&A 407, 1067

\bibitem[\protect\astroncite{Stelzer et~al.}{1999}]{Stelzer99.1}
Stelzer B., Neuh\"auser R., Casanova S \& Montmerle T., 1999, A\&A 344, 154

\bibitem[\protect\astroncite{Th\'e et~al.}{1994}]{The94.1}
Th\'e P. S., de Winter D. \& Perez M. R., 1994, A\&AS 104, 315

\bibitem[\protect\astroncite{Tsuboi et~al.}{1998}]{Tsuboi98.1}
Tsuboi Y., Koyama K. \& Murakami H., 1998, ApJ 503, 894

\bibitem[\protect\astroncite{Tovmassian et~al.}{1997}]{Tovmassian97.1}
Tovmassian H. M., Navarro S. G., Tovmassian G. H., et al., 1997, AJ 113, 1888

\bibitem[\protect\astroncite{Trigilio et~al.}{2004}]{Trigilio04.1}
Trigilio C., Leto P., Umana G., Leone F. \& Buemi C. S., 2004, A\&A 418, 593

\bibitem[\protect\astroncite{Ud'Doula \& Owocki}{2002}]{UdDoula02.1}
Ud'Doula A. \& Owocki S. P., 2002, ApJ 576, 413

\bibitem[\protect\astroncite{Usov \& Melrose}{1992}]{Usov92.1}
Usov V. V. \& Melrose D. B., 1992, ApJ 395, 575

\bibitem[\protect\astroncite{Van Loo et~al.}{2004}]{VanLoo04.1}
van Loo S., Runacres M. C. \& Blomme R., 2004, A\&A 418, 717

\bibitem[\protect\astroncite{Vitrichenko \& Plachinda}{2000}]{Vitrichenko00.1}
Vitrichenko E. A. \& Plachinda S. I., 2000, AstL 26, 390

\bibitem[\protect\astroncite{Vuong et~al.}{2003}]{Vuong03.1}
Vuong M. H., Montmerle T., Grosso N., et al., 2003, A\&A 408, 581

\bibitem[\protect\astroncite{Weigelt et~al.}{1999}]{Weigelt99.1}
Weigelt G., Balega Y., Preibisch Th., et al., 1999, A\&A 347, L15

\bibitem[\protect\astroncite{Wolk et~al.}{2005}]{Wolk05.1}
Wolk S. J., et al., 2005, \apjs, in press

\end{thebibliography}
\end{document}